\documentclass[aps,prd,onecolumn,showpacs,superscriptaddress,groupedaddress,nofootinbib]{revtex4-2}  
\usepackage{cancel}
\usepackage{graphicx} 
\usepackage[T1]{fontenc}
\usepackage{lmodern} 
\usepackage{graphicx}  
\usepackage{dcolumn}   
\usepackage{bm}        
\usepackage{amssymb}   
\usepackage{amsmath}
\usepackage{dsfont}
\usepackage{bbm}
\usepackage{bbold}
\usepackage{array}
\usepackage{float}
\usepackage{makecell}
\usepackage{braket}
\usepackage{longtable}
\usepackage{supertabular,booktabs}
\usepackage{soul}

\usepackage{titlesec} 
\usepackage[colorlinks,citecolor=blue,urlcolor=blue,hypertexnames=true]{hyperref}
\usepackage{subfigure}

\usepackage[usenames,dvipsnames,svgnames,table]{xcolor} 

\renewcommand{\arraystretch}{2}

\newcommand{\be}{\begin{equation}}
\newcommand{\ee}{\end{equation}}
\newcommand{\bea}{\begin{eqnarray}}
\newcommand{\eea}{\end{eqnarray}}
\newcommand{\bml}{\begin{subequations}}
\newcommand{\eml}{\end{subequations}}
\newcommand{\bfig}{\begin{figure}}
\newcommand{\efig}{\end{figure}}

\newcommand{\slashed}{\hspace{-1.1ex}/}

\newcommand{\bmat}{\begin{pmatrix}}
\newcommand{\emat}{\end{pmatrix}}
\usepackage{graphicx, slashed,booktabs, color, multirow, float,
amsfonts, bbold, mathtools, sidecap, tikz, bm,enumitem}
\usepackage{multirow}
\usepackage{bbding}
\usepackage{titlesec}
\usepackage{hyperref}
\usepackage{wasysym}
\usepackage{amssymb}
\usepackage{pifont}
\newcommand{\grad}{\nabla}

\renewcommand{\leq}{\leqslant}
\renewcommand{\geq}{\geqslant}

\usepackage[dvipsnames, usenames]{xcolor}

\definecolor{linkcolor}{rgb}{0.55, 0.13, .32}

\definecolor{oucrimsonred}{rgb}{0.6, 0.0, 0.0}
\definecolor{persianblue}{rgb}{0.11, 0.22, 0.73}
\definecolor{forestgreen}{rgb}{0.13,0.35,0.13}
\definecolor{lightgray}{rgb}{0.83, 0.83, 0.83}
 \hypersetup{colorlinks, citecolor=oucrimsonred, linkcolor=persianblue, urlcolor=oucrimsonred}
\definecolor{cornellred}{rgb}{0.7, 0.11, 0.11}
\definecolor{navyblue}{rgb}{0.0, 0.0, 0.5}
\definecolor{amethyst}{rgb}{0.6, 0.4, 0.8}
\definecolor{yellow}{rgb}{1.0, 1.0, 0.0}
\definecolor{firebrick}{rgb}{0.7, 0.13, 0.13}
\definecolor{tangerineyellow}{rgb}{1.0, 0.8, 0.0}
\definecolor{deepfuchsia}{rgb}{0.76, 0.33, 0.76}
\definecolor{amber}{rgb}{1.0, 0.75, 0.0}
\definecolor{VioletRed4}{rgb}{0.55, 0.13, .32}
\definecolor{indiagreen}{rgb}{0.07, 0.53, 0.03}
\definecolor{VioletRed4}{rgb}{0.55, 0.13, .32}

\usepackage{hyperref}
\usepackage{graphics, appendix, afterpage, makecell} 
\usepackage{bbold}
\usepackage{tikz}
\usepackage{adjustbox}

\usepackage{tcolorbox}

\definecolor{oucrimsonred}{rgb}{0.6, 0.0, 0.0}
\definecolor{persianblue}{rgb}{0.11, 0.22, 0.73}
\definecolor{forestgreen}{rgb}{0.13,0.35,0.13}
\definecolor{lightgray}{rgb}{0.83, 0.83, 0.83}
 \hypersetup{colorlinks, citecolor=oucrimsonred, linkcolor=persianblue, urlcolor=oucrimsonred}
\definecolor{cornellred}{rgb}{0.7, 0.11, 0.11}
\definecolor{navyblue}{rgb}{0.0, 0.0, 0.5}
\definecolor{amethyst}{rgb}{0.6, 0.4, 0.8}
\definecolor{yellow}{rgb}{1.0, 1.0, 0.0}
\definecolor{firebrick}{rgb}{0.7, 0.13, 0.13}
\definecolor{tangerineyellow}{rgb}{1.0, 0.8, 0.0}
\definecolor{deepfuchsia}{rgb}{0.76, 0.33, 0.76}
\definecolor{amber}{rgb}{1.0, 0.75, 0.0}
\definecolor{VioletRed4}{rgb}{0.55, 0.13, .32}
\definecolor{indiagreen}{rgb}{0.07, 0.53, 0.03}
\definecolor{VioletRed4}{rgb}{0.55, 0.13, .32}
\newcommand{\nn}{\nonumber}

\definecolor{oucrimsonred}{rgb}{0.6, 0.0, 0.0}
\newcommand\vertarrowbox[3][6ex]{%
  \begin{array}[t]{@{}c@{}} #2 \\
  \left\uparrow\vcenter{\hrule height #1}\right.\kern-\nulldelimiterspace\\
  \makebox[0pt]{\scriptsize#3}
  \end{array}%
}

\definecolor{mtcolor}{rgb}{.8,.3,.1}

\definecolor{violachiaro}{rgb}{1,0.6,1}

\definecolor{gbcolor}{rgb}{.43,.22,.12}
 
\definecolor{gbcolor2}{rgb}{.9,.2,.6}
\definecolor{gbcolor3}{rgb}{.3,.2,.6}

\definecolor{verdechiaro}{rgb}{0.6,1,0.6}
\definecolor{giallochiaro}{rgb}{1,1,0.6}
\definecolor{bluscuro}{rgb}{0.15, 0.2, 0.9}
\definecolor{verdes}{rgb}{0.1, 0.5, 0.1}%
\definecolor{tangerineyellow}{rgb}{1.0, 0.8, 0.0}
\definecolor{smokyblack}{rgb}{0.06, 0.05, 0.03}

\definecolor{americanrose}{rgb}{1.0, 0.01, 0.24}
\definecolor{cobalt}{rgb}{0.0, 0.28, 0.67}
\definecolor{brandeisblue}{rgb}{0.0, 0.44, 1.0}
\definecolor{mycolor}{rgb}{0.0, 0.0, 0.5}
\definecolor{oxfordblue}{rgb}{0.0, 0.13, 0.28}
\definecolor{azure}{rgb}{0.0, 0.5, 1.0}
\definecolor{turquoiseblue}{rgb}{0.0, 1.0, 0.94}
\newtcolorbox{mynewbox}[1]{colback=white!5!white,colframe=azure!75!black,fonttitle=\bfseries,title=#1}
\newtcolorbox{mybox}{colback=mycolor!5!white,colframe=azure!75!black}
\newtcolorbox{mynamedbox}[1]{colback=mycolor!5!white,colframe=azure!75!black,title=#1}
\definecolor{venetianred}{rgb}{0.78, 0.03, 0.08}
\newtcolorbox{mynamedbox1}[1]{colback=venetianred!5!white,colframe=venetianred!80!black,title=#1}
\newtcolorbox{mynamedbox2}[1]{colback=azure!5!white,colframe=azure!80!black,title=#1}

\definecolor{rossocorsa}{rgb}{0.83, 0.0, 0.0}

\tikzset{->-/.style={decoration={
  markings,
  mark=at position #1 with {\arrow{>}}},postaction={decorate}}}
\tikzset{-<-/.style={decoration={
  markings,
  mark=at position #1 with {\arrow{<}}},postaction={decorate}}} 

\def\be{\begin{equation}}
\def\ee{\end{equation}}
\def\ba{\begin{eqnarray}}
\def\ea{\end{eqnarray}}

\def\L*{{\cal L}_*}
\def\L{\mathcal{L}}
\def\({\left(}
\def\){\right)}

\def\nn{\nonumber}

\def\<{\langle}
\def\>{\rangle}

\def\cs2{c_{s}^{2}}

 \def\be   {\begin{equation}}   \def\ee   {\end{equation}}
 \def\ba   {\begin{array}}      \def\ea   {\end{array}}
 \def\bea  {\begin{eqnarray}}   \def\eea  {\end{eqnarray}}
 \def\bean {\begin{eqnarray*}}  \def\eean {\end{eqnarray*}}

\newcommand{\fnl}{f_{\mathrm{NL}}}

\titleclass{\subsubsubsection}{straight}[\subsection]

\newcounter{subsubsubsection}[subsubsection]
\renewcommand\thesubsubsubsection{\thesubsubsection.\arabic{subsubsubsection}}

\titleformat{\subsubsubsection}
  {\normalfont\normalsize\bfseries}{\thesubsubsubsection}{1em}{}
\titlespacing*{\subsubsubsection}
{0pt}{3.25ex plus 1ex minus .2ex}{1.5ex plus .2ex}

\makeatletter
\renewcommand\paragraph{\@startsection{paragraph}{5}{\z@}%
  {3.25ex \@plus1ex \@minus.2ex}%
  {-1em}%
  {\normalfont\normalsize\bfseries}}
\renewcommand\subparagraph{\@startsection{subparagraph}{6}{\parindent}%
  {3.25ex \@plus1ex \@minus .2ex}%
  {-1em}%
  {\normalfont\normalsize\bfseries}}
\def\toclevel@subsubsubsection{4}
\def\toclevel@paragraph{5}
\def\toclevel@paragraph{6}
\def\l@subsubsubsection{\@dottedtocline{4}{7em}{4em}}
\def\l@paragraph{\@dottedtocline{5}{10em}{5em}}
\def\l@subparagraph{\@dottedtocline{6}{14em}{6em}}
\makeatother

\usepackage{titlesec} 
\usepackage[colorlinks,citecolor=blue,urlcolor=blue,hypertexnames=true]{hyperref}
\usepackage{subfigure}

\usepackage[usenames,dvipsnames,svgnames,table]{xcolor}  
\usepackage{tikzsymbols}
\usepackage{natbib}
\usepackage{float}

\usepackage{tikz,xcolor,hyperref}

\usepackage{slashed}

\definecolor{lime}{HTML}{A6CE39}
\DeclareRobustCommand{\orcidicon}{
	\begin{tikzpicture}
	\draw[lime, fill=lime] (0,0) 
	circle [radius=0.2] 
	node[white] {{\fontfamily{qag}\selectfont \tiny ID}};
	\draw[white, fill=white] (-0.0625,0.095) 
	circle [radius=0.007];
	\end{tikzpicture}
	\hspace{-2mm}
}

\foreach \x in {A, ..., Z}{\expandafter\xdef\csname orcid\x\endcsname{\noexpand\href{https://orcid.org/\csname orcidauthor\x\endcsname}
			{\noexpand\orcidicon}}
}
 
\usepackage{bbold}
\usepackage{tikz}
\usepackage{adjustbox}
\usepackage{tcolorbox}
\usepackage{enumitem}
\usepackage{amsfonts}

\setlist[itemize,1]{label=$\times$}
\setlist[itemize,2]{label=$\checkmark$}
\setlist[itemize,3]{label=$\diamond$}
\setlist[itemize,4]{label=$\bullet$}

\begin{document}
\title{\Large \textcolor{Sepia}{Negative non-Gaussianity as a salvager for PBHs with PTAs in bounce
}}
\author{\large Sayantan Choudhury\orcidA{}${}^{1,2}$}
\email{sayanphysicsisi@gmail.com, 
schoudhury@fuw.edu.pl, sayantan.choudhury@nanograv.org (Corresponding author)} 
 \author{\large Kritartha Dey\orcidB{}${}^{2}$}
\email{kritartha09@gmail.com }
\author{\quad\quad\quad\quad\quad\quad\quad\quad\quad\quad\quad\quad\quad\quad\quad\quad\quad\quad\quad\quad\quad\quad\quad\quad\quad\quad\quad\quad\quad\quad\quad\quad\quad\quad\quad\quad\quad\quad\quad\quad\large  Siddhant Ganguly\orcidC{}{}${}^{3}$}\email{ms22100@iisermohali.ac.in }
\author{\large Ahaskar Karde\orcidD{}${}^{2}$}
\email{kardeahaskar@gmail.com }
\author{\large Swapnil Kumar Singh\orcidF{}${}^{2}$}
\email{swapnil.me21@bmsce.ac.in}
\author{\large Pranjal Tiwari\orcidE{}${}^{3}$}
\email{ms22104@iisermohali.ac.in}

\affiliation{ ${}^{1}$Institute of Theoretical Physics, Faculty of Physics,\\
University of Warsaw, ul. Pasteura 5, 02-093 Warsaw, Poland,}
\affiliation{ ${}^{2}$Centre For Cosmology and Science Popularization (CCSP),\\
        SGT University, Gurugram, Delhi- NCR, Haryana- 122505, India,}
 \affiliation{${}^{3}$Indian Institute of Science Education and Research, Mohali-140306, India}

\begin{abstract}

Non-Gaussianity in the primordial curvature perturbation is a crucial element of the early universe due to its significant impact on the primordial black hole (PBH) production. In this work, we focus on the effects of negative non-Gaussianity on PBH abundance through the lens of the compaction function criterion for PBH formation. Our setup utilizes an effective field theory of non-singular bounce, including the standard slow-roll inflation with an ultra-slow roll phase for amplifying the curvature perturbations to form PBHs. We investigate with two separate values of the non-Gaussianity parameter, $f_{\rm NL}=(-39.95,-35/8)$, found within the ekpyrotic contraction and the matter bounce scenarios, respectively, and show that a negatively large amount of $f_{\rm NL}$ can provide sizeable abundance, $10^{-3}\leq f_{\rm PBH}\leq 1$, and completely mitigates the PBH overproduction issue. We also highlight that the case with the effective sound speed $c_{s}=1$, coupled with $f_{\rm NL}=-39.95$, provides an agreement under $1\sigma$ for the scalar-induced gravitational wave explanation of the latest PTA (NANOGrav15 and EPTA) signal. Lastly, we extract an upper bound on the most negative value of, $f_{\rm NL}\sim -60$, below which we show breaching of the underlying perturbativity constraints on the power spectrum amplitude.

\end{abstract}

\maketitle
\tableofcontents
\newpage
\section{Introduction}

Recent advancements by the PTA experiments has marked a significant milestone with the detection of the Stochastic Gravitational Wave Background (SGWB). NANOGrav \cite{NANOGrav:2023gor,NANOGrav:2023hde,NANOGrav:2023ctt,NANOGrav:2023hvm,NANOGrav:2023hfp,NANOGrav:2023tcn,NANOGrav:2023pdq,NANOGrav:2023icp}, EPTA \cite{EPTA:2023fyk,EPTA:2023sfo,EPTA:2023akd,EPTA:2023gyr,EPTA:2023xxk}, CPTA \cite{Xu:2023wog}, PPTA \cite{Reardon:2023gzh,Zic:2023gta,Reardon:2023zen}, have provided compelling evidence of the Hellings-Downs correlation pattern, strong signature of a gravitational wave origin. There are several possible sources to explain the origin of the SGWB. Some of the prominent ones include supermassive black holes, cosmic strings, domain walls, cosmic first order phase transitions, and induced Gravitational Waves (GW) from large quantum fluctuations, primarily the Scalar Induced Gravitational Waves (SIGW). For details on the above, please refer to \cite{Choudhury:2023fwk,Choudhury:2023fjs,Choudhury:2024one,Choudhury:2023hfm,Bhattacharya:2023ysp,Franciolini:2023pbf,Inomata:2023zup,Balaji:2023ehk,HosseiniMansoori:2023mqh,Gorji:2023sil,DeLuca:2023tun,Choudhury:2023kam,Yi:2023mbm,Cai:2023dls,Cai:2023uhc,Huang:2023chx,Vagnozzi:2023lwo,Frosina:2023nxu,Zhu:2023faa,Jiang:2023gfe,Cheung:2023ihl,Oikonomou:2023qfz,Liu:2023pau,Liu:2023ymk,Wang:2023len,Zu:2023olm, Abe:2023yrw, Gouttenoire:2023bqy,Salvio:2023ynn, Xue:2021gyq, Nakai:2020oit, Athron:2023mer,Ben-Dayan:2023lwd, Madge:2023cak,Kitajima:2023cek, Babichev:2023pbf, Zhang:2023nrs, Zeng:2023jut, Ferreira:2022zzo, An:2023idh, Li:2023tdx,Blanco-Pillado:2021ygr,Buchmuller:2021mbb,Ellis:2020ena,Buchmuller:2020lbh,Blasi:2020mfx, Madge:2023cak, Liu:2023pau, Yi:2023npi,Vagnozzi:2020gtf,Benetti:2021uea,Inomata:2023drn,Lozanov:2023rcd,Basilakos:2023jvp,Basilakos:2023xof,Li:2023xtl,Domenech:2021ztg,Yuan:2021qgz,Chen:2019xse,Cang:2023ysz,Cang:2022jyc,Konoplya:2023fmh,Huang:2023chx,Ellis:2023oxs,Yu:2023jrs,Nassiri-Rad:2023asg,Heydari:2023rmq,Li:2023xtl,Chang:2023aba,Bernardo:2023jhs,Choi:2023tun,Elizalde:2023rds,Chen:2023bms,Nojiri:2023mbo,Domenech:2023jve,Liu:2023hpw,Huang:2023zvs,Oikonomou:2023bli,Cyr:2023pgw,Fu:2023aab,Kawai:2023nqs,Kawasaki:2023rfx,Maji:2023fhv,Bhaumik:2023wmw,He:2023ado,An:2023jxf,Zhu:2023lbf,Das:2023nmm,Roshan:2024qnv,Chen:2024fir,Chowdhury:2023xvy,Battista:2021rlh,Battista:2022hmv,Battista:2022sci,Battista:2023znv,DeFalco:2023djo,DeFalco:2024ojf,Domenech:2024rks,Iovino:2024uxp,Andres-Carcasona:2024wqk,Vagnozzi:2022qmc,Calza:2024qxn,Vagnozzi:2022moj,Pedrotti:2024znu,Wang:2023ost,Wang:2023sij,Zhao:2022kvz,Zhu:2023gmx,Jin:2023wri,Tan:2024esx,Chen:2024twp,Basilakos:2024diz,Papanikolaou:2024cwr,He:2024luf,Heydari:2023rmq,Solbi:2024zhl,Teimoori:2021thk,Ashrafzadeh:2023ndt,Heydari:2024bxj,Ashrafzadeh:2024oll}. In this paper, we have focused on generating the SIGWs from the curvature perturbation. They do not interact with the surrounding plasma and, therefore, act as a crucial probe of the early universe (typically covering the scales $k\sim (10^7-10^{18}){\rm Mpc^{-1}}$). These SIGWs are tensor perturbations sourced at the second order in cosmological perturbation theory by the non-linear interactions between the scalar perturbations. Because the GWs come in the second order, the corresponding curvature perturbation needs to be significantly enhanced at small scales, compared to the Cosmic Microwave Background (CMB) scales, for possible detection by PTA experiments. This enhancement of the perturbation is also associated with formation of cosmic underdense and overdense patches within an otherwise isotropic and homogeneous FLRW background, which gravitationally collapse beyond a certain critical threshold that leads to the creation of Primordial Black Holes (PBHs)\cite{Zeldovich:1967lct,Hawking:1974rv,Carr:1974nx,Carr:1975qj,Chapline:1975ojl,Carr:1993aq,Choudhury:2011jt,Yokoyama:1998pt,Kawasaki:1998vx,Rubin:2001yw,Khlopov:2002yi,Khlopov:2004sc,Saito:2008em,Khlopov:2008qy,Carr:2009jm,Choudhury:2011jt,Lyth:2011kj,Drees:2011yz,Drees:2011hb,Ezquiaga:2017fvi,Kannike:2017bxn,Hertzberg:2017dkh,Pi:2017gih,Gao:2018pvq,Dalianis:2018frf,Cicoli:2018asa,Ozsoy:2018flq,Byrnes:2018txb,Ballesteros:2018wlw,Belotsky:2018wph,Martin:2019nuw,Ezquiaga:2019ftu,Motohashi:2019rhu,Fu:2019ttf,Ashoorioon:2019xqc,Auclair:2020csm,Vennin:2020kng,Nanopoulos:2020nnh,Inomata:2021uqj,Stamou:2021qdk,Ng:2021hll,Wang:2021kbh,Kawai:2021edk,Solbi:2021rse,Ballesteros:2021fsp,Rigopoulos:2021nhv,Animali:2022otk,Frolovsky:2022ewg,Escriva:2022duf,Ozsoy:2023ryl,Ivanov:1994pa,Afshordi:2003zb,Frampton:2010sw,Carr:2016drx,Kawasaki:2016pql,Inomata:2017okj,Espinosa:2017sgp,Ballesteros:2017fsr,Sasaki:2018dmp,Ballesteros:2019hus,Dalianis:2019asr,Cheong:2019vzl,Green:2020jor,Carr:2020xqk,Ballesteros:2020qam,Carr:2020gox,Ozsoy:2020kat,Baumann:2007zm,Saito:2008jc,Saito:2009jt,Choudhury:2013woa,Sasaki:2016jop,Raidal:2017mfl,Papanikolaou:2020qtd,Ali-Haimoud:2017rtz,Di:2017ndc,Raidal:2018bbj,Cheng:2018yyr,Vaskonen:2019jpv,Drees:2019xpp,Hall:2020daa,Ballesteros:2020qam,Carr:2020gox,Ozsoy:2020kat,Ashoorioon:2020hln,Papanikolaou:2020qtd,Wu:2021zta,Kimura:2021sqz,Solbi:2021wbo,Teimoori:2021pte,Cicoli:2022sih,Ashoorioon:2022raz,Papanikolaou:2022chm,Papanikolaou:2023crz,Wang:2022nml,ZhengRuiFeng:2021zoz,Cohen:2022clv,Cicoli:2022sih,Brown:2017osf,Palma:2020ejf,Geller:2022nkr,Braglia:2022phb,Frolovsky:2023xid,Aldabergenov:2023yrk,Aoki:2022bvj,Frolovsky:2022qpg,Aldabergenov:2022rfc,Ishikawa:2021xya,Gundhi:2020kzm,Aldabergenov:2020bpt,Cai:2018dig,Cheng:2021lif,Balaji:2022rsy,Qin:2023lgo,Riotto:2023hoz,Riotto:2023gpm,Papanikolaou:2022did,Choudhury:2011jt,Choudhury:2023vuj, Choudhury:2023jlt, Choudhury:2023rks,Choudhury:2023hvf,Choudhury:2023kdb,Choudhury:2023hfm,Bhattacharya:2023ysp,Choudhury:2023fwk,Choudhury:2023fjs,Choudhury:2024one,Choudhury:2024ybk,Choudhury:2024jlz,Choudhury:2024ezx,Choudhury:2025kxg,Choudhury:2024dei,Choudhury:2024aji,Harada:2013epa,Harada:2017fjm,Kokubu:2018fxy,Gu:2023mmd,Saburov:2023buy,Stamou:2023vxu,Libanore:2023ovr,Friedlander:2023qmc,Chen:2023lou,Cai:2023uhc,Karam:2023haj,Iacconi:2023slv,Gehrman:2023esa,Padilla:2023lbv,Xie:2023cwi,Meng:2022low,Qiu:2022klm,Mu:2022dku,Fu:2022ypp,Davies:2023hhn,Firouzjahi:2023ahg,Firouzjahi:2023aum, Iacconi:2023ggt,Davies:2023hhn,Jackson:2023obv,Riotto:2024ayo,Ragavendra:2024yfp,Papanikolaou:2024rlq,Papanikolaou:2024kjb,Banerjee:2021lqu,Choudhury:2023kam,Heydari:2021gea,Heydari:2021qsr,Heydari:2023xts,Heydari:2023rmq,Caravano:2024tlp,Banerjee:2022xft,Papanikolaou:2021uhe,Papanikolaou:2022hkg,Calza:2024xdh,Calza:2024fzo,Sharma:2024whg,Papanikolaou:2023nkx,Papanikolaou:2023cku,Papanikolaou:2023oxq,Heydari:2023rmq,Solbi:2024zhl,Teimoori:2021thk,Ashrafzadeh:2023ndt,Heydari:2024bxj,Ashrafzadeh:2024oll}. PBHs come through as an intriguing alternative to explain a sizeable portion of the complete dark matter of the universe \cite{Ivanov:1994pa,Afshordi:2003zb,Frampton:2010sw,Carr:2016drx,Kawasaki:2016pql,Inomata:2017okj,Espinosa:2017sgp,Ballesteros:2017fsr,Sasaki:2018dmp,Ballesteros:2019hus,Dalianis:2019asr,Cheong:2019vzl,Green:2020jor,Carr:2020xqk,Ballesteros:2020qam,Carr:2020gox,Ozsoy:2020kat}. However, there is a caveat that demands thorough scrutiny. Much like the most promising scenarios, the drawback of PBH formation is the risk of their overproduction \cite{Franciolini:2023pbf,Franciolini:2023wun,Inomata:2023zup,LISACosmologyWorkingGroup:2023njw, Inui:2023qsd,Chang:2023aba, Gorji:2023ziy,Li:2023xtl, Li:2023qua,Firouzjahi:2023xke,Gorji:2023sil,Ota:2022xni,Raatikainen:2023bzk,Choudhury:2023fwk,Choudhury:2023fjs,Choudhury:2024one}, a scenario where an abundance of PBHs, denoted by $f_{\rm PBH}$, appear to become more than unity; their energy density exceeds the totality of dark matter. Evading PBHs overproduction is a challenging task, more so in the context of SIGW interpretation of the PTA experimental data. Following are some references that have come up with a solution to this issue - \cite{Franciolini:2023pbf,Franciolini:2023wun,Inomata:2023zup,LISACosmologyWorkingGroup:2023njw, Inui:2023qsd,Chang:2023aba, Gorji:2023ziy,Li:2023xtl, Li:2023qua,Firouzjahi:2023xke,Gorji:2023sil,Ota:2022xni,Raatikainen:2023bzk,Choudhury:2023fwk,Choudhury:2023fjs,Choudhury:2024one}.

PBHs are formed from large amplification of the curvature perturbation at specific scales. To this day, the exact scenario of PBH formation is a largely discussed topic \cite{Yoo:2022mzl,Escriva:2021aeh}. The calculation of PBH abundance can be thought to be indirectly derived from knowledge of either of the two profiles: the density perturbation profile and the curvature perturbation profile. The standard production mechanism involves the density perturbation amplitude averaged over a spherical region, crossing a certain threshold $\delta_c$, before collapsing into PBHs. In the traditional Press-Schechter (PS) formalism \cite{Press:1973iz}, a Gaussian distribution of the density perturbation is assumed and is generally applicable to spherically symmetric models. In other words, the formalism depends on the shape of the perturbation profile. 
A physically more robust procedure to estimate the PBHs abundance has been developed known as the compaction function approach, which we here integrate in our study. The compaction function is similar to the Schwarzschild gravitational potential, since it relies mainly on the functional form of the gravitational potential(or mass excess) within a region. In \cite{Escriva:2019phb}, it was shown that the threshold of the PBHs formation is largely unaffected beyond the critical radius $r_p$, denoting the peak of the compaction function. A crucial takeaway of this approach is there exists a universal threshold value of the volume-averaged compaction function that can easily be related to $\delta_c$, and is independent of the perturbation profile to a high-accuracy estimate \cite{Escriva:2019phb}, unlike the PS method which relies on direct $\delta_c$ value and is heavily dependent on the perturbation shape. Additionally, PBH formation is a nonlinear process and simple consideration of a Gaussian profile of density perturbation as in PS theory is unrealistic. In contrast, the compaction function utilizes a non-linear relationship between the density and curvature perturbation in the long-wavelength approximation, which we discuss later in detail. 

Another important aspect of this work comes from considering the impact of quantum loop effects while studying PBH formation. More precisely, the scenario where huge enhancements in the curvature perturbation can induce significant quantum corrections at large scales whilst contributing to PBH formation was first investigated by Kristiano and Yokoyama in \cite{Kristiano:2022maq,Kristiano:2023scm}. This scenario quickly became a topic of intense discussion, followed by works such as \cite{Riotto:2023hoz,Riotto:2023gpm,Firouzjahi:2023aum,Firouzjahi:2023ahg,Firouzjahi:2023bkt,Motohashi:2023syh,Franciolini:2023lgy,Cheng:2023ikq,Tasinato:2023ukp,Tasinato:2023ioq, Iacconi:2023ggt,Davies:2023hhn} discussing the proper conditions to conclude whether or not PBH formation remains ruled out. Parameters that include the duration of the perturbation amplifying phase, also known as the ultra-slow roll (USR) phase, and the nature (smooth or sharp) and implications of the transition in and out of this phase demand much scrutiny before establishing any firm results. In the midst of this, a series of papers by the authors of \cite{Choudhury:2023vuj, Choudhury:2023rks, Choudhury:2023jlt}, proposed a strong no-go theorem as a constraint on the largest mass of the PBHs, $M_{\rm PBH}\sim{\cal O}({10^2}\rm gm)$, as a result of the USR phase. The calculation mechanism involved the application of regularization, then renormalization, followed by a dynamical group resummation, to arrive at a finite result for the two-point correlation function of importance to us in the late time. Moreover, the underlying effective field theory (EFT) treatment of the no-go theorem propelled these conclusions to cover a wide range of canonical and non-canonical single-field inflationary models. 

Nevertheless, several studies like \cite{Choudhury:2023hvf, Choudhury:2023kdb,Choudhury:2023hfm,Bhattacharya:2023ysp,Choudhury:2023fjs, Choudhury:2024one,Choudhury:2024jlz,Choudhury:2024dei} have demonstrated methods to circumvent this rigid PBH mass constraint using the EFT framework. Among such diverse set-ups, our paper focuses on a specific framework with additional pre-inflationary phases of contraction and an expanding non-singular bounce before following up with the standard SR-USR-SR stages for PBH generation. For our purposes, we utilize the result of performing the same calculation scheme as explained above to obtain the regularized-renormalized-resummed (RRR) scalar power spectrum, \cite{Choudhury:2024dei}, within the EFT setting. The motivation here to introduce such phases in our set-up stems mainly from the fact that other than allowing us to evade the no-go, such phases have been shown to provide non-Gaussianities (NGs) that are orders of magnitude of larger than what is attainable in a standard single-field slow-roll inflation, as first shown from Maldacena's calculations in \cite{Maldacena:2002vr} with $f_{\rm NL}\sim {\cal O}(0.1)$, or even, in scenarios with slow-roll violating phases like the USR with $1<|f_{\rm NL}|< 10$. The matter bounce is such a scenario where the NGs are much larger (and negative) compared to standard inflationary scenarios, primarily because the curvature perturbation here grows on scales beyond the Hubble scale which later contributes to the strength and shape of the NGs \cite{Cai:2009fn}. Likewise, for models focusing on ekpyrotic contraction \cite{Lehners:2008my}, negative NGs are also shown to be larger than the standard SR inflation. At this point, it might be tempting to ask that, with the introduction of all these phases, how large of a non-Gaussianity can be realistically feasible? Before answering that question, let us briefly consider the importance of NGs in the context of PBH formation.

The abundance of PBHs is affected by the Probability Distribution (PDF) of the curvature perturbation, which is generally assumed to be Gaussian. However, for a realistic analysis, the inclusion of NGs in the curvature perturbation, $\zeta$, is necessary to accurately estimate the abundance of PBHs, largely because there is no concrete confirmation that the curvature perturbation itself behaves as a Gaussian random field. They emerge from quantum fluctuations and are responsible for creating a myriad of exotic and large-scale structures in the universe. Therefore, it is highly likely that the perturbation field would inherit intrinsic NGs. Besides, even under an ideal Gaussian assumption, the relationship between the curvature perturbation and the density contrast field is non-linear, which imparts NGs to the density contrast field as an effect due to these non-linear (NL) corrections. Previous case studies in \cite{Franciolini:2023pbf,Choudhury:2024one,Choudhury:2023fjs,Choudhury:2023fwk,Choudhury:2024dzw} demonstrate that incorporating NGs have enabled a reduction in the abundance of PBHs, eliminating the risk of overproduction. Most fiducial cases in the literature treat NGs perturbatively, while there have been some developments in its study through a non-perturbative formalism \cite{Lyth:2005fi,Gow:2022jfb,Ferrante:2022mui}. 

In this paper, we will consider the local form of NGs in the curvature perturbation expressed with the parameter $f_{\rm NL}$ that encodes the first-order NG correction. PBH abundance is highly dependent on the signature of this $f_{\rm NL}$, and most cases show that $f_{\rm NL}<0$ is considered more favourable to suppress the PBH abundance while positive NGs $f_{\rm NL}>0$ invite the risk of overproduction. In our case, such NGs are affected by the USR phase, usually short-lived with $\Delta N_{\rm USR}\sim 2$, and operate to enhance the fluctuations at small scales and violating Gaussianity. One could also envision a curvaton scenario \cite{Franciolini:2023pbf, Ferrante:2023bgz, Gow:2023zzp, Ferrante:2022mui}, where the curvature perturbation acquires NG corrections. Apart from this, scenarios described a while ago with ekpyrotic contraction and matter bounce also invite a large amount of negative non-Gaussianity. The analysis in this paper incorporates the above-mentioned two phases along with the three phases of the usual SR inflation model. Let us revert back to the question we asked a while ago about how large of a negative NG can we dive into. In \cite{Lehners:2008my}, it has been shown through the calculation using the $\delta N$ formalism \cite{Sugiyama:2012tj,Dias:2012qy,Naruko:2012fe,Takamizu:2013gy,Abolhasani:2013zya,Clesse:2013jra,Chen:2013eea,Choudhury:2014uxa,vandeBruck:2014ata,Dias:2014msa,Garriga:2015tea,Choudhury:2015hvr,Choudhury:2016wlj,Choudhury:2017cos,Choudhury:2018glz, Starobinsky:1985ibc,Sasaki:1995aw,Sasaki:1998ug,Lyth:2005fi,Lyth:2004gb,Abolhasani:2018gyz,Passaglia:2018ixg}, that $f_{\rm NL}$ can take values as large as $f_{\rm NL} \sim -166$ from the ekpyrotic phase. While in \cite{Cai:2009fn}, it has been shown that $f_{\rm NL}$ is negative and of the order of $f_{\rm NL}\sim{\cal O}(1)$.
When these phases, along with an added USR phase, are included, a large NG is expected. For our analysis, we adopt two benchmark values of $f_{\rm NL}\in (-39.95,-35/8)$ and present a comparative analysis to conclude with a stringent bound for the leading-order NG parameter $f_{\rm NL}$. We will demonstrate through our analysis that diving below $f_{\rm NL}=-39.95$ for a specific threshold of the compaction function increases the risk of breaking perturbativity conditions. We will confirm that with $f_{\rm NL}=-39.95$, a substantial amplification of the power spectrum is possible that explains the PTA and simultaneously avoids overproduction.  
Throughout this work, we have employed two distinct values of the effective sound speed parameter $c_s \in (0.88,1)$, which lie within the window that preserves the causality and unitarity conditions. The mass spectrum of PBHs examined in this work is between $M_{\rm PBH} \sim (10^{-6}-0.1)M_{\odot}$. Our set-up is successful in generating large mass PBHs while satisfying the inflationary conditions and underlying perturbativity conditions. We also clarify that in this work the equation of state parameter has been fixed to a constant value corresponding to the radiation-dominated epoch. 

This paper is structured as follows. We commence with a quick description of our EFT set-up in sec.(\ref{EFTBasic}). This is followed by the expressions for the tree-level and loop-corrected scalar power spectrum of the curvature perturbation in sec.(\ref{s3}). In sec.(\ref{s4}), the overall dynamics of PBHs formation has been discussed. Special emphasis has been put towards the development of the compaction function approach as it is going to be our prime tool throughout this paper. The formulas for the corresponding joint PDF and PBHs abundance have been laid out along with the domain of integration for the PBHs mass fraction. This is followed by highlighting the main expressions for the SIGW spectrum in sec.(\ref{s5}) for $w=1/3$. We then present the outcomes of our numerical analysis in sec.(\ref{s6}), where motivation for consideration of large negative NGs for the analysis has been largely discussed. To familiarize the reader with a rough idea of the recent studies, we have presented a light comparison in sec.(\ref{s7}) between several models. Various experiments that our analysis has confronted have been enlisted in sec.(\ref{s8}). Based on our examination, we provide a stringent upper bound on the NG parameter $f_{\rm NL}$, with its validity, in sec.(\ref{s9}). Finally, we draw our conclusion and highlight the major findings from this work in sec.(\ref{s10}).

\section{Bounce from EFT framework}\label{EFTBasic}

Our approach for the EFT of bounce framework builds on the foundation discussed extensively in \cite{Choudhury:2024dei,Choudhury:2024dzw} and we thus follow the same construction in this work and elaborate on the pre-inflationary features to make the underlying picture clear. For more details of EFT construction see also refs. \cite{Weinberg:2008hq,Cheung:2007st,Choudhury:2017glj,Choudhury:2024ybk,Choudhury:2024jlz,Choudhury:2021brg,Adhikari:2022oxr,Banerjee:2021lqu,Naskar:2017ekm,Choudhury:2015pqa,Choudhury:2015eua,Choudhury:2015zlc,Choudhury:2015hvr,Choudhury:2014sua,Choudhury:2013iaa,Choudhury:2013jya,Choudhury:2013zna,Choudhury:2011sq,Choudhury:2011jt,Choudhury:2012yh,Choudhury:2012whm,Choudhury:2014sxa,Choudhury:2014uxa,Choudhury:2014kma,Choudhury:2016cso,Choudhury:2016pfr,Choudhury:2017cos,Bohra:2019wxu,Akhtar:2019qdn,Choudhury:2020yaa,Choudhury:2021tuu,Choudhury:2016wlj,Cabass:2022avo,Cai:2016thi,Cai:2017tku,Agarwal:2012mq,Piazza:2013coa,Delacretaz:2016nhw,Salcedo:2024smn,Colas:2023wxa,Senatore:2010wk,Noumi:2012vr,Tong:2017iat,Noumi:2012vr,Arkani-Hamed:2015bza,Kim:2021pbr,Baumann:2018muz,Choudhury:2018glz,Hongo:2018ant,Baumann:2017jvh,An:2017hlx,Gong:2017yih,Liu:2016aaf}.

The basic principles in this EFT framework remain the same as behind the construction of the EFT of single-field inflation. We investigate the theory of scalar field perturbations $\delta\phi(t,{\bf x})$ around a time-evolving spatially flat FLRW background solution $\phi_{0}(t)$, where the spatial diffeomorphisms remain unbroken but the temporal ones are broken non-linearly. Upon choosing to work within the unitary gauge $\delta\phi(t,{\bf x})=0$, the scalar field perturbations vanish and get instead absorbed by the metric perturbations increasing its degrees of freedom by one from the usual two helicities. The aforementioned FLRW background geometry is expressed with the metric structure
\bea
ds^2 = a^2(\tau)(-d\tau^2+d{\bf x}^2),
\eea
where now the scale factor $a(\tau)$ can be chosen in the following two conditions to identify the pre-inflationary phases \cite{Khoury:2001wf,Khoury:2001zk,Khoury:2001bz,Buchbinder:2007ad,Lehners:2007ac,Lehners:2008vx,Raveendran:2018yyh,Brandenberger:2012zb,Raveendran:2017vfx,Chowdhury:2015cma,Cai:2011tc,Brandenberger:2016vhg,Boyle:2004gv,Wands:1998yp,Peter:2002cn,Allen:2004vz,Martin:2003sf,Papanikolaou:2024fzf,Raveendran:2023auh,Raveendran:2019idj,Brustein:1998kq,Starobinsky:1980te,Mukhanov:1991zn,Brandenberger:1993ef,Novello:2008ra,Lilley:2015ksa,Battefeld:2014uga,Peter:2008qz,Biswas:2005qr,Bamba:2013fha,Nojiri:2014zqa,Bajardi:2020fxh,Bhargava:2020fhl,Cai:2009in,Cai:2012ag,Shtanov:2002mb,Ilyas:2020qja,Ilyas:2020zcb,Zhu:2021whu,Banerjee:2016hom,Saridakis:2018fth,Barca:2021qdn,Wilson-Ewing:2012lmx,K:2023gsi,Agullo:2020cvg,Agullo:2020fbw,Agullo:2020wur,Chowdhury:2018blx,Chowdhury:2016aet,Nandi:2019xag,Raveendran:2018why,Raveendran:2018yyh,Stargen:2016cft,Sriramkumar:2015yza,Banerjee:2022gpy,Paul:2022mup,Odintsov:2021yva,Banerjee:2020uil,Das:2017jrl,Pan:2024ydt,Colas:2024xjy,Piao:2003zm,Cai:2017dyi,Cai:2017pga,Cai:2015nya,Cai:2019hge,Silva:2015qna,Silva:2020bnn} of our framework,
\begin{itemize}
    \item[\ding{228}] In the Contraction phase, the scale factor obeys the power law solution
    \bea \label{contractionscale} a(\tau) = a_0\left(\displaystyle\frac{\tau}{\tau_0}\right)^{\frac{1}{\epsilon-1}},
    \eea
    where for the choice of parameter in, $1 < \epsilon < 3$, the value $\epsilon=3/2$ results in a matter contraction phase. For values $\epsilon< 1$, one observes a power law solution for the quasi-dS inflationary regime. Beyond $\epsilon>3$, the solution describes the ekpyrotic contraction phase. The conformal time $\tau_{0}$ sets the reference scale with $a(\tau_{0})=a_{0}$. 
    \item[\ding{228}] In the Bouncing phase, the scale factor obeys the solution 
    \bea \label{bouncescale} a(\tau) = a_0\left[1+\left(\displaystyle\frac{\tau}{\tau_0}\right)^2\right]^{\frac{1}{2(\epsilon-1)}},
    \eea
    where the choice of $\epsilon=3/2$ describes the matter bounce solution. For the similar choice of $\epsilon> 3$, this time this solution realizes an ekpyrotic bounce solution.
\end{itemize}

As for the most general Lagrangian of scalar perturbations constructed from various operators invariant under spatial diffeomorphisms and curvature operators, its structure is written as follows \cite{Cheung:2007st}:
\bea \label{eftaction}
S&=&\int d^4x\sqrt{-g}\bigg[\frac{M_{pl}^2}{2}R+M_{pl}^2\dot{H}g^{00}-M_{pl}^2(3H^{2}+\dot{H})+ \frac{M^{4}_{2}(t)}{2!}\left(\delta g^{00}\right)^2+\frac{M^{4}_{3}(t)}{3!}\left(\delta g^{00}\right)^3\nonumber\\
&&\quad\quad\quad\quad\quad -\frac{\bar{M}^{3}_{1}(t)}{2}\left(\delta g^{00}\right)\delta K^{\mu}_{\mu}-\frac{\bar{M}^{2}_{2}(t)}{2}(\delta K^{\mu}_{\mu})^2-\frac{\bar{M}^{2}_{3}(t)}{2}\delta K^{\mu}_{\nu}\delta K^{\nu}_{\mu} + \cdots\bigg],
\eea 
where perturbation in the temporal metric component and in the extrinsic curvature are defined as follows: \bea
\delta g^{00}=1+g^{00}, \quad\quad \delta K_{\mu\nu}=\left(K_{\mu\nu}-a^2Hh_{\mu\nu}\right), 
\eea
where the induced metric $h_{\mu\nu}$ on the spatial slice at fixed time $(t)$ and its corresponding normal vector $n^{\mu}$ have the following relations in between them:
\bea
&& h_{\mu\nu}=g_{\mu\nu}+n_\mu n_\nu ,\quad 
n_\mu = \frac{\partial_\mu t}{\sqrt{-g_{\mu\nu}\partial_\mu t  \partial_\nu t}}, \quad K_{\mu\nu} = h_\mu ^ \sigma \grad _\sigma n_\nu.
\eea
The quantities $M_1(t)$, $M_3(t)$, $\bar{M}_1(t)$, $\bar{M}_2(t)$ and $\bar{M}_3(t)$ describe the Wilson coefficients of this EFT and capture the distinctions between different models. 

With the help of the generic lagrangian from Eqn. $(\ref{eftaction})$, we ultimately want to deal with the quadratic lagrangian for the scalar perturbations. Such an action can be obtained after we restore the gauge invariance in the EFT action and focus exclusively on the decoupling limit that renders the mixing terms of gravity with the scalar perturbations completely irrelevant above the mixing energy scale \cite{Cheung:2007st}.
Under these conditions, the final quadratic action of interest can now be expressed as follows:
\bea \label{action}
S^{(2)}&&=
\int d^4x\; a^3\bigg[\frac{-M_{pl}^2\dot{H}}{c_s^2}\bigg(\dot{\pi}^2-c_s^2\frac{(\partial_i\pi)^2}{a^2}\bigg)\bigg]\quad \text{with}\quad \frac{1}{c_s^2}\equiv 1-\frac{2M_2^4}{\dot{H}M_{pl}^2},
\eea
where $\pi(t,{\bf x})$ are the Goldstone modes that appear when we restore gauge invariance in the original lagrangian and these now play the role of fluctuations of interest in the decoupling limit. The effective sound speed $c_{s}$ depends on the Wilson coefficient and captures deviation from the canonical single-field inflation. The current constraint on this parameter comes under, $0.024 \leq c_{s}< 1$ \cite{Planck:2018jri}, from imposing the conditions from cosmological observations along with causality and unitarity. 

Now, working in the unitary gauge, we utilize the broken time-diffeomorphism property to obtain the significant linear relationship between the comoving curvature perturbation and the Goldstone modes $\pi$, as $\zeta(t,{\bf x})=-H\pi(t,{\bf x}),$ 
Any correction terms beyond linear order to the same relation become highly suppressed in the limit $\tau_{0} \rightarrow 0$, where $\tau_0$ refers to the conformal time in
the late-time when inflation ends. Moving forward with this relation and after taking the Fourier transform Eqn. (\ref{action}) reads:
\bea
S^{(2)}=\int \frac{d^3{\bf k}}{(2\pi)^3}d\tau \;a^2 \bigg[\frac{M^2_{pl}}{c_s^2}\epsilon(\zeta_{\bf k}'^2 -k^2c_s^2\zeta_{\bf k}^2)\bigg], \quad\quad \text{where}\quad \epsilon=1-\frac{\cal H'}{{\cal H}^2},
\eea
with ${\cal H}=a'/a=aH$ defining the conformal Hubble parameter and a prime denoting a conformal time derivative. The variation of the above action with the curvature perturbation leads us to identify the quadratic equation of motion for the same which is the Mukhanov-Sasaki (MS) equation, represented as follows in the Fourier space:
\bea\label{MSEq}
\zeta_{\bf k}''(\tau)+2\frac{z'(\tau)}{z(\tau)}\zeta_{\bf k}^{'}(\tau)+c_s^2k^2\zeta_{\bf k}(\tau)=0,
\eea
which introduces a new conformal time-dependent variable $z$ that has associated with it the following properties: 
\bea \label{MSvariableprops}
z &=& a(\tau)\frac{\sqrt{2\epsilon}}{c_s},\nonumber\\
\frac{z^{'}(\tau)}{z(\tau)} &=& {\cal H}\left(1-\eta+\epsilon-s\right),\quad\quad \text{where}\quad \eta = \frac{\epsilon'}{\epsilon{\cal H}}, \quad s = \frac{c'_s}{c_s{\cal H}}, \nonumber\\
\frac{z^{''}(\tau)}{z(\tau)} &=& \frac{1}{\tau^2}\left(\nu^2-\frac{1}{4}\right),\quad\quad\text{where}\quad \nu=\left\{
\begin{array}{ll}
    3/2+3\epsilon-\eta-3s & \mbox{for }\quad \epsilon\ll 1\\
    1/2-(1-\eta+3s)/(\epsilon-1), & \mbox{for }\quad \epsilon\geq 1,
\end{array}\right.
\eea
including few other slow-roll parameters such as $\eta$ and $s$.

Our next objective is to construct the scalar power spectrum based on the mode solutions of the MS Eqn.(\ref{MSEq}) across all the regions of interest within the framework. These regions include the contraction phase followed by a bouncing phase that transitions to the inflationary stages of the first slow-roll (SRI), an ultra slow-roll (USR), and the final phase of inflation with a second slow-roll (SRII). The detailed structure of the solutions of the curvature perturbation mode is elaborated in the Appendix \ref{appA}. For an in-depth derivation, we direct the reader to \cite{Choudhury:2024dei, Choudhury:2024aji,Choudhury:2024dzw}. Here, we focus on the structure of the scalar power spectrum and subsequent contributions from quantum loop effects that yield the final power spectrum.

\section{Scalar Power Spectrum}\label{s3}
\subsection{Tree Level Contribution} \label{s31}

From mode solutions of the comoving curvature perturbation, the standard tree-level $2$-point correlation function at late times is written down in Fourier space as:
\bea \label{treecorrl}
\langle \hat{\zeta}_{\bf k}\hat{\zeta}_{{\bf k}^{'}}\rangle_{{\bf Tree}} &=&\lim_{\tau_{0}\rightarrow 0}\langle \hat{\zeta}_{\bf k}(\tau_{0})\hat{\zeta}_{{\bf k}^{'}}(\tau_{0})\rangle_{{\bf Tree}}\nonumber\\
&=&(2\pi)^{3}\;\delta^{3}\left({\bf k}+{\bf k}^{'}\right)P^{\bf Tree}_{\zeta}(k),\quad\eea
where the quantity $P^{\bf Tree}_{\zeta}(k)$ refers to the dimensionful scalar power spectrum. Subsequently, the often useful dimensionless version of the same power spectrum is defined as:
\bea \label{treex} \Delta^{2}_{\zeta,{\bf Tree}}(k)=\frac{k^{3}}{2\pi^{2}}P^{\bf Tree}_{\zeta}(k)=\frac{k^{3}}{2\pi^{2}}|{\zeta}_{\bf k}(\tau_{0})|^{2}_{\tau_{0}\rightarrow 0}.\eea
We need to employ the Fourier space solutions for the curvature perturbation in each phase to construct the full tree-level power spectrum as per the above definition. The resulting version coming from each phase is combined to give us the late-time version $(\tau_{0} \rightarrow 0)$ as follows \cite{Choudhury:2024dei, Choudhury:2024aji, Choudhury:2024dzw}:
\begin{widetext}
  \bea
\Delta^2_{\zeta,{\rm\bf Tree-Total}}(k)
&=&\Delta^2_{\zeta,{\rm\bf SRI}}(k_*
)\times\bigg[1+\bigg(\frac{\epsilon_*}{\epsilon_c}\bigg)\times\bigg(\frac{k}{k_*} \bigg)^{\frac{2\epsilon_c}{\epsilon_c-1}}+\bigg(\frac{\epsilon_*}{\epsilon_b}\bigg)\times\bigg(\frac{k}{k_*}\bigg)^2
\times\bigg[1+\bigg(\frac{k_*}{k}\bigg)^2\bigg]^{-\frac{1}{\epsilon_b-1}}\nonumber\\&&\quad\quad\quad\quad\quad\quad\quad\quad+\Theta(k-k_s)\bigg(\frac{k}{k_s}\bigg)^6\times|\alpha_2-\beta_2|^2+\Theta(k-k_e)\bigg(\frac{k_e}{k_s}\bigg)^6\times|\alpha_3-\beta_3|^2\bigg],
\eea  
\end{widetext}
where the top line introduces contributions from the SRI phase, contraction, and bouncing phase, respectively, while the bottom line shows the contributions from the USR and SRII phases.
The quantities $\{\alpha_{2},\beta_{2},\alpha_{3},\beta_{3}\}$ represent the Bogoliubov coefficients that appear as we pass through each phase in the inflationary stages. These coefficients signal a change in the underlying vacuum structure from the initial conditions at the start of inflation. Notice that the transition from contraction to the bounce and into the SRI is continuous such that the underlying choice of quantum vacuum structure remains the same as the initial Bunch-Davies vacuum. However, we require the implementation of a Heaviside theta function, $\Theta(k-k_i),\;i=\{s,e\}$, which indicates the sharp transition criteria leading to a shift from the initial vacuum structure. Consequently, it is only when we begin to see transitions of the SRI-USR $(k=k_s)$ and USR-SRII $(k=k_e)$, that we need to incorporate the modified quantum vacuum. The wavenumber $k_{*}$ refers to the reference scale which we set here as the pivot scale, $k_{*}=0.02\;{\rm Mpc^{-1}}$. Also important are the quantities, $\epsilon_{c},\epsilon_{b}$, which refer to the choices of the first SR parameter in the contraction and bouncing phases, respectively, and the choice signifies the particular nature whether it is an ekpyrotic kind or a matter kind of phase.

\subsection{Quantum Loop Corrected}\label{s32}

We now examine the quantum loop corrections to the tree-level power spectrum structure from each phase within our framework discussed in the previous section. The task here requires expanding the EFT action to obtain the cubic-order curvature perturbation action and working with the Schwinger-Keldysh or in-in formalism to evaluate the one-loop corrections to the previously mentioned $2$-point function. In the Appendix \ref{appC}, we provide the necessary details concerning the computations and this formalism to the reader. Here, we provide only the relevant expressions that, when combined, give us the final quantum loop corrected version of the scalar power spectrum. To present the procedure in a succinct manner, we divide it into two steps and explain the consequences of each outcome. 

\subsubsection{$\diamond$ Regularization }
\label{s32a}

The primary task with all the relevant one-loop correlation functions, using the terms from the cubic-order action, would be to handle the ultraviolet (UV) and infrared (IR) divergences. The initial technique employed for this purpose is the cut-off regularization, which is to compute the necessary loop integrals. These consist of a finite wavenumber interval for the momentum integration, preferably the modes of significance of the phase of interest, while for the temporal integration the late time limit $(\tau_{0} \rightarrow 0)$ is opted later to gather information at the super-horizon scales. Although we do not perform the said computations explicitly, we do provide the final expressions resulting from these in the Appendix \ref{app:RRR}. After evaluating the loop integrals, the cut-off regularized one-loop scalar power spectrum can be written as follows:
\bea \label{one-loop:Reg} \Delta^{2}_{\zeta, {\bf R}}(k)&=&\bigg[\Delta^{2}_{\zeta,{\bf Tree}}(k)\bigg]_{\bf SRI}\times\bigg(1+ {\bf T}_{\bf C} + {\bf T}_{\bf B} + {\bf T}_{\bf SRI} + {\bf T}_{\bf USR} + {\bf T}_{\bf SRII}\bigg), \eea 
where the bold terms represent the one-loop level corrections coming from the contraction $({\bf T_{C}})$, bounce $({\bf T_{B}})$, SRI $({\bf T_{SRI}})$, USR $({\bf T_{USR}})$, and SRII $({\bf T_{SRII}})$, to the SRI phase. 
We would require here the following information about the power spectrum at tree-level in the SRI region when written for any arbitrary wavenumber $k$:
\bea \bigg[\Delta^{2}_{\zeta,{\bf Tree}}(k)\bigg]_{\bf SRI}
&=&\bigg[\Delta^{2}_{\zeta,{\bf Tree}}(k_*)\bigg]_{\bf SRI}\bigg(1+\bigg(\frac{k}{k_s}\bigg)^2\bigg),\quad\quad\eea
where the coefficient for amplitude at the pivot scale is defined using the following relation:
\bea
\label{pivot} \bigg[\Delta^{2}_{\zeta,{\bf Tree}}(k_*)\bigg]_{\bf SRI}=\left(\frac{2^{2\nu-3}H^{2}}{8\pi^{2}M^{2}_{ pl}\epsilon c_s}\left|\frac{\Gamma(\nu)}{\Gamma\left(\frac{3}{2}\right)}\right|^2\right)_*.\eea
The quantity in Eqn. (\ref{one-loop:Reg}) becomes our first non-trivial result after controlling the presence of divergences from quantum loops. An immediate consequence of such a regularization procedure, however, is the presence of regularization scheme-dependent counterterms that must be adjusted to remove UV divergent contributions, and this is the next topic of discussion.


\subsubsection{$\diamond$ Renormalization}
\label{s32b}

After the regularization of one-loop integrals, our next task is to invoke the renormalization procedure and improve the previous result in Eqn. (\ref{one-loop:Reg}) by making it completely UV divergence free. The source of such harmful quadratic (or power-law) UV contributions appears from scales deep inside the horizon $(-kc_{s}\tau\ll 1)$ due to rapid fluctuations in the quantized scalar field at such short scales. Removing such divergences amounts to choosing a renormalization scheme and evaluating the necessary counter-terms to negate the impact of UV divergent corrections to the one-loop correlations calculated earlier. In \cite{Choudhury:2024dei}, it has been shown via both the adiabatic/wave-function, and late-time renormalization techniques that the final result remains independent of the chosen schemes, and we are left with only the logarithmic IR divergences. The improved regularized-renormalized version of the power spectrum can now be written as follows \cite{Choudhury:2024dei}:
\bea \label{one-loop:RR} \Delta^{2}_{\zeta, {\bf RR}}(k)&=&\bigg[\Delta^{2}_{\zeta,{\bf Tree}}(k)\bigg]_{\bf SRI}\times\bigg(1+\overline{{\bf T}}_{\bf C}+\overline{{\bf T}}_{\bf B}+\overline{{\bf T}}_{\bf SRI}+\overline{{\bf T}}_{\bf USR}+\overline{{\bf T}}_{\bf SRII}\bigg), \eea 
where the now regularized-renormalized one-loop integrals give us the terms, $\overline{{\bf T}}_{\bf C}$ (contraction), $\overline{{\bf T}}_{\bf B}$ (bounce), $\overline{{\bf T}}_{\bf SRI}$ (SRI), $\overline{{\bf T}}_{\bf USR}$ (USR), and $\overline{{\bf T}}_{\bf SRII}$ (SRII), and their explicit forms are given in Appendix \ref{app:RRR}. 

The leftover IR divergences remain irremovable and require further analysis. The first step in this direction is the power spectrum renormalization. This technique helps us control/smooth the IR-divergent corrections of the power spectrum induced from all the pre-inflationary (contraction and bouncing) phases and the SRI, USR, and SRII phases. To execute this technique, we introduce an appropriate IR counter-term to renormalize the correlation functions. The estimation of this counterterm depends on our UV counterterms, evaluated previously to renormalize the power-law divergences, and a renormalization condition accepted at the CMB pivot scale $(k_{*})$. For proper details regarding the various steps in the calculations, we refer the reader to \cite{Choudhury:2024dei}. We now show the consequence of the above explained steps which modify our previous result via the IR counter-term as follows: 
\bea \label{one-loop:RR2}
\Delta_{\zeta,{\bf RR}}^{2}(k)&\xrightarrow[\text{renormalization}]{\text{Power-spectrum}}&\bigg[\Delta_{\zeta,\textbf{Tree}}^{2}(k_{*})\bigg]_{\textbf{SRI}}\times\left(1-\overline{{\bf T}}_{\bf C,*}-\overline{{\bf T}}_{\bf B,*}-\overline{{\bf T}}_{\bf SRI,*}-\overline{{\bf T}}_{\bf USR,*}-\overline{{\bf T}}_{\bf SRII,*}\right)\nonumber\\
&&\quad\quad\quad\quad\quad\quad\quad\quad\quad\quad\quad\quad\quad\quad\quad \times\left(1+\overline{{\bf T}}_{\bf C}+\overline{{\bf T}}_{\bf B}+\overline{{\bf T}}_{\bf SRI}+\overline{{\bf T}}_{\bf USR}+\overline{{\bf T}}_{\bf SRII}\right).
\eea
We notice the use of similar terms, as before in Eqn. (\ref{one-loop:RR}), but this time with them evaluated at the pivot scale, hence the asterisk, as per the CMB renormalization condition. After performing the product between the various terms, we get the important result where the linear-order term is completely devoid of any logarithmic IR corrections, and the remaining divergences are pushed to higher loop orders in the perturbative expansion. In the next section, we discuss the strategy for resumming such contributions.

\subsubsection{$\diamond$ Dynamical RG Resummation}
\label{s32c}

We begin from the results of previous discussions on the power spectrum renormalization technique, which was able to soften the IR divergences by pushing them to the higher loop orders and removing their presence at the linear order. Eventually, we must take care of all the large logarithms, which display a disturbing feature of secular growth with time, to extract the well-needed behavior of these correlations at late times. To complete this task of producing a finite result by carefully handling divergent contributions, we soften the remaining logarithmic IR divergences and perfrom their resummation with the help of the dynamical renormalization group (DRG) technique \cite{Chen:2016nrs,Baumann:2019ghk,Boyanovsky:1998aa,Boyanovsky:2001ty,Boyanovsky:2003ui,Burgess:2015ajz,Burgess:2014eoa,Burgess:2009bs,Dias:2012qy,Chaykov:2022zro,Chaykov:2022pwd}. The primary advantage of DRG here is the ability to resum the softened logarithmic terms by absorbing the divergent or secularly growing terms coming at all orders in the perturbative expansion. At the end we get a finite, convergent result that is resummed at all loop orders. We refer the reader to \cite{Choudhury:2024dei} for the full details on calculations from this procedure and to the Appendix \ref{app:RRR} for the explicit list of expressions used here. Here, we mention the final expression that remains useful for the latter sections of this paper, where we deal with PBH formation, \cite{Choudhury:2024dei}:
\bea \label{RRRspectrum}
\Delta^{2}_{\zeta, {\bf RRR}}(k)
=A\times\left(1+\left(\frac{k}{k_s}\right)^2\right)\times\exp\bigg(6\ln\left(\frac{k_s}{k_e}\right)+{\cal Q}_{c}\bigg).\eea
where the last term describes the quantum corrections as a result of the above procedures:
\bea \label{quantcorr}  &&{\cal Q}_{c}=-\left\{\frac{\bigg[  \Delta_{\zeta,\textbf{Tree}}^{2}(k)\bigg]_{\textbf{SRI}}}{\bigg[  \Delta_{\zeta,\textbf{Tree}}^{2}(k_{*})\bigg]_{\textbf{SRI}}}\right\}\times\bigg[\bigg(\overline{{\bf T}}^2_{\bf C,*}+\overline{{\bf T}}^2_{\bf B,*}+\overline{{\bf T}}^2_{\bf SRI,*}+\overline{{\bf T}}^2_{\bf USR,*}+\overline{{\bf T}}^2_{\bf SRII,*}\bigg)+\cdots\bigg],\eea
The final power spectrum in Eqn. (\ref{RRRspectrum}) contains an amplitude parameter in the USR phase that reads as:
\bea A&=&\bigg[\Delta^{2}_{\zeta,{\bf Tree}}(k_*)\bigg]_{\bf SRI}\times \left(\frac{k_e}{k_s}\right)^6.\eea
and which involves the pivot scale amplitude defined before in Eqn. (\ref{pivot}). We note some crucial facts about the DRG analysis here before moving on to the latter part of this paper.
The foremost outcome of the DRG resummed version of the one-loop corrected power spectrum is that, unlike in the renormalized one-loop power spectrum, the two-point function is reliable without any problems at large times from the uncontrolled growth of amplitude due to IR divergent logarithms. The final amplitude gets exponentiated, which is a crucial outcome of this procedure in general, and this leads to its rapid fall off instead of the growth. Another fact is that, in the computations, we do not demand any exact details of the higher-order loop diagrams during resummation nor it needs the dominance of all chain diagrams over other possible diagrams in the computation, but it will definitely increase the leading contributions from these components with Logarithms.


\section{Primordial Black Hole Formation}
\label{s4}

In the previous section, we presented the development of the regularized-renormalized-resummed scalar power spectrum, where we carefully handled the quantum one-loop divergences that lead to a finite outcome. This result can now be utilized in the subsequent sections to investigate the generation of PBHs. The standard theory behind PBH formation begins with the journey of curvature perturbations stretched on the super-horizon scales where they freeze in amplitude and, as inflation ends, these fluctuations start to re-enter the Hubble horizon. Upon their re-entry, these fluctuations leave their signatures in the primordial plasma that appears in the form of density fluctuations, and, with time, they develop regions of densities larger or smaller than the mean background density of the universe. Due to gravitational attraction, if these overdense regions cross a certain threshold value, they suffer gravitational collapse, leading to the generation of PBH. The standard theory was built considering such density fluctuations in the radiation-dominated (RD) era, where we have the relation, $c_{s}^2=w=1/3$, between the sound speed and the equation of state of this era. The exact mechanism of the collapse of perturbations into PBHs has been a topic of intense study for the past few decades. There have been numerous mechanisms exploring this problem, the most widely investigated of them include the Press-Schechter theory \cite{Press:1973iz,Carr:1975qj}, the peak theory \cite{Bardeen:1985tr,Green:2004wb,Yoo:2018kvb,Young:2020xmk,Kitajima:2021fpq,Young:2022phe}, and the compaction function approach \cite{Musco:2018rwt,Kalaja:2019uju,Young:2019yug,Kehagias:2019eil,Musco:2020jjb,Musco:2021sva,Biagetti:2021eep,DeLuca:2022rfz,Gow:2022jfb,Escriva:2022pnz,Ferrante:2022mui,DeLuca:2023tun,Ferrante:2023bgz,Franciolini:2023pbf,Franciolini:2023wun,Raatikainen:2023bzk,Ianniccari:2024bkh}. In the following sections, we will focus on the details of the compaction function approach and use it as a foundation for our analysis to address the problems related to PBHs production.

\subsection{Dynamics of overdensity field : The Compaction Function approach}
\label{s4a}

Initiating the process of forming PBHs requires significantly increasing the strength of curvature perturbations. In the present setup, the USR phase facilitates this increase in the perturbations, leading to a large power spectrum at small scales. This phase also presents a violation of the slow-roll approximations that result in the generation of large non-Gaussianity. The presence of non-Gaussianity is crucial to the maximum possible mass fraction of the resulting PBHs, and we now elucidate this further.

We denote the overdensity field following standard notation, $\delta(t,{\bf x})$. Recall the fact that as the overdensities in a specific region upon their horizon re-entry exceed a threshold value, say $\delta_{\rm th}$, the perturbations quickly collapse to form PBHs. Now, right before collapse, the curvature perturbation in the super-horizon limit, $\zeta({\bf x})$ (conserved and time-independent), has much larger size than the Hubble horizon size and thus one can invoke the gradient approximation or separate universe approach \cite{Salopek:1990jq, Shibata:1999zs, Lyth:2004gb}. This is a perturbative treatment where the expansion is performed in terms of the spatial gradients and requires introducing a relevant small parameter to construct our power series expansion, which is here the ratio of the Hubble radius to the perturbation length scale. The benefit of this in the context of PBH formation comes in being able to also incorporate the non-linear features in the curvature perturbations over large distances where it is assumed to be sufficiently smooth. As a result of this approximation, the overdensity observable is written in terms of the curvature perturbation following the gradient expansion \cite{Harada:2015yda, Musco:2018rwt}:
\bea \label{gradexpansion}
\delta(t,{\bf x}) &=& -\frac{2f(w)}{3}\frac{1}{(aH)^2}e^{-5\zeta({\bf x})/2}\nabla^{2}e^{\zeta({\bf x})/2}\approx -\frac{2f(w)}{3}\frac{1}{(aH)^2}e^{-2\zeta({\bf x})} \bigg[\nabla^{2}\zeta({\bf x}) + \frac{1}{2}\partial_{i}\zeta({\bf x})\partial^{i}\zeta({\bf x})\bigg],
\eea
where \bea f(w)=\frac{3(1+w)}{(5+3w)},\eea refers to the parameter that carries the equation of state (EoS) dependence in the above relation. Throughout the discussions in this paper of the numerical outcomes, we will work with the EoS choice in the RD era that gives $f(w=1/3)=2/3$.  
The most important fact about the expression in Eqn. (\ref{gradexpansion}) is that the overdensity field boasts an inherently non-Gaussian character even though if we had assumed an a priori Gaussian profile for the curvature perturbation field. Only in the case where we work at the linear order approximation for the exponential factors in Eqn. (\ref{gradexpansion}), would we be left with a purely Gaussian picture for the rest of our analysis, but instead, we would now have to deal with a completely non-Gaussian relationship. The scale factor $a(t)$ and Hubble parameter $H(t)$ in the denominator provide the time dependence on superhorizon scales to the overdensity on the left. The first numerical study of black hole formation from non-linear metric perturbations in a FLRW universe was carried out by Shibata and Sasaki in \cite{Shibata:1999zs}, where they first introduced the compaction function method, and since then, it has been a largely powerful tool to ascertain the formation of PBHs.



Considering the spherical symmetry of the curvature perturbation profile, the compaction function provides a criterion to assess whether a given profile can collapse into PBH by identifying the maximum or peak value condition of the same function that occurs at some radial distance. It is defined in the literature as twice the excess mass contained within a certain spherical volume \cite{Shibata:1999zs, Harada:2015yda, Musco:2018rwt}: 
\bea \label{compaction}
{\cal C}(t, r) &=& 2\times\frac{[M(t,r)-M_{o}(t,r)]}{R(t,r)}=\frac{2}{R(r,t)}\underbrace{\int_{S_{R}^{2}} d^{3}\Vec{x} \rho_o(t)\delta(t,{\bf x})}_{\delta M(t,r)},
\eea
where \bea  \delta(t,{\bf x}) = \frac{\rho(t,{\bf x})-\rho_{o}(t)}{\rho_{o}(t)},\eea describes the overdensity or density contrast field.
Here we use: \bea \label{bcgquant}
M_{o}(t,r)=\rho_{0}(t)V_{o}(t,r),\quad\quad {\rm with}\quad\quad V_{o}(t,r)=\left(\frac{4\pi}{3}\right)R^{3}(t,r),\eea denotes the background mass with respect to the background energy density:
\bea\rho_{o}(t)=\frac{3H^2(t)}{8\pi}.\eea Here $M(t,r)$ denotes the Misner-Sharp mass inside a spherical volume of areal radius, 
\bea R(t,r)\equiv a(t)re^{\zeta(r)},
\eea
written in the comoving radial coordinates. In the second equality of Eqn. (\ref{compaction}), we represent the mass excess using the volume integral of the density over the sphere $S^2_{R}$ and using the definition of the density contrast variable.  Now, under the above spherical symmetry arguments, the expression from the first equality in Eqn. (\ref{gradexpansion}) can also be modified after acting the spherical Laplacian operator to give us the following version:
\bea \label{gradradial}
\delta(t,r) &=& -\frac{2f(w)}{3}\frac{1}{(aH)^2}e^{-5\zeta(r)/2}\bigg(\frac{d^{2}}{dr^{2}} + \frac{2}{r}\frac{d}{dr}\bigg)e^{\zeta(r)/2},\nonumber\\
&=& -\frac{2f(w)}{3}\frac{1}{(aH)^2}e^{-2\zeta(r)} \bigg[\zeta''(r) + \frac{2}{r}\zeta'(r) + \frac{1}{2}\zeta'(r)^{2}\bigg].
\eea
where in this case prime denotes, $' \equiv d/dr$.
We now observe that on using the Eqs. (\ref{compaction}, \ref{gradradial}), we can write the new expression after a spherical volume integration:
\bea \label{radcompaction}
{\cal C}(r) = -f(w)r\zeta'(r)(2+r\zeta'(r)).
\eea
Notice that the time dependence on the left has vanished due to the right-hand side getting evaluated on the super-horizon scales. We can see here how the compaction function takes into account the intrinsic non-linear terms in the curvature perturbation when estimating the conditions for collapse into PBH.

As stated earlier, the scale at which the compaction function achieves its maximum defines the condition for PBH formation and thus using Eqn. (\ref{radcompaction}) we determine this condition as follows:
\bea \label{maximacondition}
{\cal C}'(r=r_{p})=0 \implies \zeta'(r_{p}) + r_{p}\zeta''(r_{p}) = 0,
\eea
where $r_{p}$ denotes the scale corresponding to the peak in compaction. Furthermore, we require help from another fact concerning the horizon crossing time $(t_H)$ where the perturbations with their scale, $R(r_p,t_H)$, become equal to the Hubble horizon, $1/H(t_H)$, upon their re-entry. Soon after this, if the peak compaction happens to exceed a certain threshold value, say ${\cal C}_{\rm th}$, then the perturbation quickly collapses to form PBH. This horizon crossing condition can be written as:
\bea \label{horizoncross}
R(t_H,r_p)H(t_H) = a(t_H)H(t_H)r_{p}e^{\zeta(r_p)} = 1.
\eea
We are now in the position to underline the key relation which will be extremely helpful in the numerical analysis of mass fraction and abundance of PBHs conducted in the later sections of this paper. 

The amplitude of the perturbation is defined \cite{Musco:2018rwt} as the excess mass $(\delta M)$ averaged over a volume of areal radius $R(t,r)$. Expanding the spherical volume integral and using Eqs. (\ref{compaction}-\ref{bcgquant}), we can write this amplitude measure as:
\bea
\delta_R \equiv \frac{\delta M}{M_{o}} = \frac{1}{V_o}\int_{0}^{R}4\pi R^{2}\bigg(\frac{\rho-\rho_{o}}{\rho_{o}}\bigg)dR,
\eea
and, since we are interested with the perturbation amplitude at the radial scale of peak compaction, we utilize in the above definition the Eqn. (\ref{gradradial}) along with the horizon crossing condition in Eqn. (\ref{horizoncross}) to give us the \textit{key relation} \cite{Musco:2018rwt}: 
\bea \label{deltacompaction}
\delta_p := \frac{\delta M(t_H,r_p)}{M_{o}(t_H,r_p)} = {\cal C}(r_p), 
\eea
which tells us that the peak in compaction function equals the density contrast whose volume average is taken over the sphere of areal radius $R(t_H,r_p)$. The length scale, $r_{p}\exp{(\zeta(r_p))}$, signifies the characteristic scale over which the perturbations contribute most significantly to the collapse process. By virtue of the Eqn. (\ref{deltacompaction}), we can translate directly the condition to form PBHs from the compaction function, ${\cal C}(r_p)> {\cal C}_{\rm th}$, to being equally well calculated using the density contrast threshold, $\delta_{\rm th}$. Thus, we use this identification for the threshold condition, which works only at the horizon crossing instant, in the later part of this paper to study PBH formation.

We end this discussion on the compaction function by highlighting the presence of primordial non-Gaussianity (PNG), which is embedded into our choice of the curvature perturbation profile, through this function as developed above. We begin with the local in position-space expansion of the curvature perturbation: 
\bea \label{zetaPNG}
\zeta({\bf x}) = \zeta_{G}({\bf x}) + \frac{3}{5}f_{\rm NL} \left(\zeta_{G}^2({\bf x})-\langle\zeta_{G}^2({\bf x})\rangle\right) + \frac{9}{25}g_{\rm NL} \zeta_{G}^3({\bf x}) + \cdots,
\eea
where the factor $3/5$ comes from the relation, $\Phi=(3/5)\zeta$, followed in between the gravitational potential $\Phi$ and $\zeta$ on superhorizon scales. The expression in Eqn. (\ref{zetaPNG}) involves various, model-dependent, local-type estimators quantifying the deviation from pure Gaussian statistics via $f_{\rm NL},g_{\rm NL},\cdots$, where the ellipses contain other estimators present at the higher-orders. However, we must exercise caution as the above expression is more of an ansatz than a proper Taylor expansion about the Gaussian component since there is no guarantee about how large these non-Gaussian estimators can actually become. Although, in principle, we must work with the general expansion above to fully grasp the impact of PNGs, for this work, we truncate this expansion to the often-used quadratic version and qualitatively study this impact. A useful fact about this expansion is the way it manages to describe the short-scale perturbation behavior sitting on a background altered by the long-scale component of the perturbations. To observe this, one would have to split the perturbation variable into the two components, \bea &&\zeta({\bf x}) = \zeta_{s}({\bf x}) + \zeta_{L}({\bf x}),\\ &&\zeta_{G}({\bf x}) = \zeta_{G,s}({\bf x}) + \zeta_{G,L}({\bf x}).\eea After plugging this short-long component decomposition into Eqn. (\ref{zetaPNG}), we get:
\bea \label{short-longzeta}
\zeta_{s}({\bf x}) &=& \zeta_{G,s}({\bf x})\left(1 + \frac{6}{5}\fnl\zeta_{G,L}({\bf x})\right) + \frac{3}{5}f_{\text{NL}}\zeta^{2}_{G,S}({\bf x}),\\
\zeta_{L}({\bf x}) &=& \zeta_{G,L}({\bf x}) + \frac{3}{5}\fnl\zeta^{2}_{G,L}({\bf x}).
\eea
where Eqn. (\ref{short-longzeta}) clearly highlights the intermixing between the short-scale and long-scale components.

Consequently, from the above-discussed version of quadratic NGs, the entire Eqn. (\ref{radcompaction}) can be converted by now using derivatives of the Gaussian curvature perturbation component in radial coordinate space, $\zeta_{G}( r)$, yielding at last:
\bea \label{compactfnl}
{\cal C}(r) = \underbrace{{\cal C}_{G}(r)\frac{d\zeta(r)}{d\zeta_{G}}}_{\rm Linear} -  \frac{1}{4f(w)}\underbrace{\bigg({\cal C}_{G}(r)\frac{d\zeta(r)}{d\zeta_{G}}\bigg)^2}_{\rm Non-Linear}, \quad\quad \text{where}\quad {\cal C}_{G} = -2f(w)r\zeta'_{G}(r).
\eea
The above equation already considers both the effect of non-linear (quadratic) corrections to the density contrast and the PNGs from our choice of curvature perturbation profile. With PNGs, even the first linear term becomes non-Gaussian. The non-linearity highlights the intrinsic non-Gaussian nature, which we also pointed out before concisely following the discussions of Eqn. (\ref{gradexpansion}). The ${\cal C}_{G}$ component behaves like a Gaussian random variable, since $\zeta_{G}$ is defined as the Gaussian component, and then there is the influence of non-linear correction terms. Furthermore, by utilizing the quadratic expansion from Eqn. (\ref{zetaPNG}), we can also write the above Eqn. (\ref{compactfnl}) as follows:
\bea \label{compactfnl2}
{\cal C}(r) = {\cal C}_{G}\left(1+\frac{6}{5}f_{\rm NL}\zeta_{G}(r)\right)\bigg[1 - \frac{{\cal C}_{G}(r)}{4f(w)}\left(1+\frac{6}{5}f_{\rm NL}\zeta_{G}(r)\right)\bigg].
\eea
From the above relation, we notice that for $f_{\rm NL}=0$, it reduces to show clearly the intrinsic non-Gaussian character coming from the resulting fully non-linear relation. However, with either $f_{\rm NL}> 0$ or $f_{\rm NL}< 0$, the actual size and behaviour of the compaction depend on the term in the parenthesis, coming from $d\zeta/d\zeta_{G}$, which is strictly model-dependent. We will show in the subsequent section how this term becomes crucial in getting the allowed values of the variables ${\cal C}_{G},\zeta_{G}$ after executing the compaction maxima condition (see Eqn. (\ref{maximacondition})) to determine PBH formation. 

Exercising the true non-Gaussian relationship for the density contrast is a significant step to rigorously discern how PBH formation undergoes, with non-Gaussianity having a significant impact on this process which we will explore in the rest of our paper in detail. In the next section, we will see how these two different random variables ${\cal C}_{G},\zeta_{G}$ develop correlations among them that will help to construct their joint probability distribution function necessary to attain our goal of calculating the PBH mass fraction.

\subsection{PBH Mass Fraction from Distribution Function}
\label{s4b}

Having concluded our preparatory discussion on the compaction function threshold statistics, we turn our attention to computing the mass fraction for PBHs and their corresponding present-day abundance. We begin this section by exploring the statistics of a Gaussian curvature perturbation variable, leveraging the results from this analysis to apply them within the compaction function approach, and finally obtaining the distribution function relevant to our PBH formation.

\subsubsection{$\diamond$ Gaussian Random Field Distribution and its Statistics}
\label{s4b1}

The standard approach towards obtaining the mass fraction has been to exploit the statistics of the random variables in our theory, here the Gaussian curvature perturbation $\zeta_{G}$, that gives rise to the overdensities. Numerous studies have been conducted assuming Gaussian statistics for the fluctuations \cite{Yoo:2018kvb,Germani:2019zez,Germani:2018jgr}. However, since we are extending this simplistic picture to incorporate PNGs into our analysis, via Eqn. (\ref{zetaPNG}), it requires an improved statistical distribution in contrast to the standard Gaussian one.
To further motivate this development, consider the case of the relation in Eqn. (\ref{gradexpansion}) where only the first linear term is of importance and which implies a one-to-one correspondence between the curvature perturbation and density contrast variables. Assuming Gaussian statistics, the mass fraction, $\beta_{G}$, then gets calculated via integrating the Gaussian distribution after obeying the threshold statistics for the density contrast:
\bea
\beta_{G} = \frac{1}{\sigma\sqrt{2\pi}}\int_{\delta > \delta_{\rm th}}^{\infty}\exp{\left(-\frac{(\delta-\mu)^2}{2\sigma^2}\right)}d\delta, 
\eea
where the mean $(\mu)$ and variance $(\sigma^{2})$ altogether determine the statistics at this level of approximation. However, since the beginning we are focusing on the intrinsic non-linear relation for the density contrast, we now encounter multiple Gaussian random variables, namely, $\zeta_{G}$ and its spatial derivatives $\partial_{i}\zeta_{G}\equiv f_{i}$ and $\partial_{i}\partial_{j}\zeta_{G}\equiv \zeta_{ij}$, to supply us with the total spatial distribution information of density contrast. Due to the symmetric nature of the double partial derivatives, there will be $6$ independent components coming from them, $ij\in \{xx,yy,zz,xy,yz,xz\}$. Combining the rest of the independent terms from $\zeta_{G}$, and its first derivative, we are left to construct a $1+3+6=10$ dimensional joint probability distribution function(PDF).  For the sake of completeness, lets write down such a PDF. 

The set of possible correlations between the $10$ Gaussian random fields can be written following \cite{Ferrante:2022mui} as:
\bea \label{10dPDFcorrelations}
\langle\zeta_{G}\;\zeta_{G}\rangle &=& \sigma_{0}^{2}, \quad\quad\quad\quad\quad\quad \langle\zeta_{G}\;f_{i}\rangle = 0,\nonumber\\
\langle\zeta_{G}\;\zeta_{ij}\rangle &=& -\frac{\sigma_{1}^{2}}{2}\delta_{ij}, \quad\quad\quad\quad \langle f_{i}\;f_{j}\rangle = \frac{\sigma_{1}^{2}}{2}\delta_{ij},\nonumber\\
\langle f_{i}\;\zeta_{jk}\rangle &=& 0, \quad\quad\quad\quad\quad\quad\; \langle \zeta_{ij}\;\zeta_{kl}\rangle = \frac{\sigma_{2}^{2}}{8}(\delta_{ij}\delta_{kl} + \delta_{ik}\delta_{jl} + \delta_{il}\delta_{jk}).
\eea
Notice from the above expressions that the auto-correlation terms are different from each other, indicating that they capture different aspects of the primordial perturbation, each with their own scale of fluctuation. This is important from the perspective that each component of the Gaussian random field $\zeta_{G}$ and its derivatives have unique contribution on the overall statistical properties of the density contrast that later translates to the calculation of the mass fraction of PBHs. Now, based on these various auto and cross correlators, we can write the Gaussian PDFs as follows:
\bea
&&\mathbbm{P}_{\rm G}(f_{i}) = 
 \frac{1}{\sigma_{1}\sqrt{\pi}}\exp\left(-\frac{f_{i}^2}{\sigma_1^2}\right),\quad i=x,y,z, \quad \mathbbm{P}_{\rm G}(\zeta_{ij}) = 
 \frac{2}{\sigma_{2}\sqrt{\pi}}\exp\left(-\frac{4\zeta_{ij}^2}{\sigma_{2}^2}\right),\quad ij=xy,yz,xz,\nonumber \eea

 where the multivariate form in terms of the explicit elements of the spatial derivatives of $\zeta_G$ with the covariance matrix $\bf C$ is given by:
 \bea
&& 
  \mathbbm{P}_{\rm G}(\zeta_{\rm G},\zeta_{xx},\zeta_{yy},\zeta_{zz}) = 
  \frac{1}{(2\pi)^{3/2}\sqrt{|{\rm det}{\bf C}|}}\exp\left(
 -\frac{1}{2}{\bf Z}_{G}^{\rm T} {\bf C}^{-1}{\bf Z}_{G}
  \right)\,,\hspace{0.3cm}
{\bf C} \equiv  
\begin{pmatrix}     
     \sigma_0^2 & -\sigma_1^2/2 & -\sigma_1^2/2  &  -\sigma_1^2/2 \\
    -\sigma_1^2/2 & 3\sigma_2^2/8 & \sigma_2^2/8 & \sigma_2^2/8 \\
    -\sigma_1^2/2 & \sigma_2^2/8 & 3\sigma_2^2/8 & \sigma_2^2/8 \\
    -\sigma_1^2/2 & \sigma_2^2/8 & \sigma_2^2/8 & 3\sigma_2^2/8
\end{pmatrix}, 
\eea 
where we have the vector ${\bf Z}_{G}  \equiv (\zeta_{G},\zeta_{xx},\zeta_{yy},\zeta_{zz})^{\rm T}$.
The determinant of the covariance matrix $\bf C$ is given by:

\bea \label{detC}
{\rm det}{\bf C} = -\frac{\sigma_{2}^4}{128}\left(6\sigma_{1}^2 - 5\sigma_0^2 \sigma_{2}^4\right)\ = \left\{
\begin{array}{ll}
    +ve & \mbox{for } 5\sigma_0^2 \sigma_{2}^4 > 6\sigma_{1}^2  \\
    -ve & \mbox{for }  5\sigma_0^2 \sigma_{2}^4 < 6\sigma_{1}^2
\end{array}\right.
\eea




Additionally, the symmetrical property of covariance matrix $\bf C$ is a reflection of the isotropic and homogeneous nature of the underlying Gaussian fields. The presence of negative cross correlations in the matrix suggest that large values of $\zeta_{G}$ are less probable in regions characterized by  extreme spatial concavity or convexity. Moving on, the final product of the above contributions gives us in return the $10$ dimensional joint Gaussian PDF:
\bea
\label{10Dgaussian}
\mathbbm{P}_{\rm G}(\zeta_{G},f_{i},\zeta_{ij})
 d\zeta_{\rm G}\;df_{i}\;d\zeta_{ij} = 
\bigg[\prod_{i=x,y,z}^{}\mathbbm{P}_{\rm G}(f_{i})\bigg]
\bigg[\prod_{ij=xy,yz,xz}\mathbbm{P}_{\rm G}(\zeta_{ij})\bigg]
\mathbbm{P}_{\rm G}(\zeta_{\rm G},\zeta_{xx},\zeta_{yy},\zeta_{zz})
d\zeta_{G}\;df_{i}\;d\zeta_{ij}.
\eea
 In the presence of these many Gaussian random fields together, the statistical features of the combined random field get encoded inside the variance elements $\sigma_{a}^{2}$, where the index $a=(0,1,2)$ denotes the various moments and which also could be understood as counting the spatial derivatives in Fourier space. We require the scalar power spectrum to evaluate these and their general expression has the structure:
\bea \label{moments}
\sigma_a^{2} = \int \frac{dk}{k}W^2(k,R){\cal T}^2(k,\tau) k^{2a} \Delta^{2}_{\zeta}(k).
\eea
A few comments about the above expression are in order. In the Fourier space, a filter or window function $W(k,R)$ gets multiplied by the power spectrum of the Gaussian random field. This filter function helps smoothen any discontinuous features in the field at the small scales. Some popular choices for this include the top-hat and the Gaussian function. Further appearing is the radiation transfer function $T(k,\tau)$ \cite{Blais:2002gw}:
\bea \label{radtransfer}
{\cal T}(k,R)= 3\bigg[
\frac{\sin(k\tau/\sqrt{3}) - (k\tau/\sqrt{3})\cos(k\tau/\sqrt{3})}{(k\tau/\sqrt{3})^3}\bigg] = \left\{
\begin{array}{ll}
    1 & \mbox{for }  k\tau/\sqrt{3}\ll 1 \quad {\bf super-horizon}, \\
    0 & \mbox{for }  k\tau/\sqrt{3}\gg 1 \quad {\bf sub-horizon},
\end{array}\right.
\eea
which effectively describes how the curvature on the sub-horizon scales evolves, at linear order, following the re-entry of the super-horizon modes before collapse. The transfer function also exhibits the property where it dies down in the sub-horizon limit telling us that it naturally provides a damping feature at only the small-scale regime while it does not impact any features of interest on the super-horizon scales. The explicit conformal time-dependence inside the transfer function will be made clear in the upcoming discussion centered around using the compaction function and how it affects our future analysis.

\subsubsection{$\diamond$ Distribution and Threshold Statistics with Compaction Function }
\label{s4b2}

In light of the compaction function approach, the relation of importance for our analysis is present in Eqn. (\ref{compactfnl}) that includes the $f_{\rm NL}$ parameter. Here, we observe two Gaussian random variables, $\zeta_{G}$, and its derivative, $\zeta'_{G}$, signaling the need to account for all possible correlations between these to specify the distribution function and later apply the threshold statistics.
As previously discussed, in order to properly determine the abundance of the PBHs, the NGs need to be carefully accounted for into the computation of the PDF. Thereby, we commence by writing the joint PDF with the Gaussian variables as follows:
 \vspace{-0.2cm}
   \bea
   \label{PDF}
 \mathbbm{P}_G({\cal C}_G,\zeta_G) = \frac{1}{2\pi\sigma_c \sigma_{r}\sqrt{1-\gamma_{\rm cr}^2}}\exp{\bigg[
\left(- \frac{\zeta_G ^2}{2\sigma_r ^2}\right)-
    \bigg(\frac{1}{2(1-\gamma_{\rm cr}^2)}\bigg(\frac{{\cal C}_G}{\sigma_c}-\frac{\gamma_{\rm cr}\zeta_G}{\sigma_r}\bigg)^2\bigg)\bigg]},
   \eea
where we have the dimensionless correlation parameter and the corresponding covariance matrix as: \bea \label{gammacov}\gamma_{cr}=\frac{\sigma^{2}_{cr}}{\sigma_{c}\sigma_{r}},\quad\quad {\bf C}\equiv \bordermatrix{     
  & {\scriptstyle C_{ G}} & {\scriptstyle \zeta_{G}}  \cr
    {\scriptstyle C_{G}}  &  \sigma_c^2 & \sigma_{cr}^2  \cr
    {\scriptstyle \zeta_{G}}  & \sigma_{cr}^2 & \sigma_r^2 \cr
}.\eea    
The dimensionless number $\gamma_{cr}$ retains some horizon mass $M_H$ dependence through the length scale $r_p$ embedded into the different covariance elements $(\sigma_{r},\sigma_{c},\sigma_{cr})$ on which we also elaborate shortly after. This parameter is also a function of the scalar power spectrum and of the scale $r$ used inside the various correlations. These correlations form the two-dimensional covariance matrix and follow a calculation method similar to Eqn. (\ref{moments}), integrating the full quantum-loop corrected scalar power spectrum smoothed via some choice for the window function.

The following are some notable choices for the window functions:
\bea
W_{th}(k,r) = 3\bigg[
\frac{\sin(kr) - (kr)\cos(kr)}{(kr)^3}\bigg],\quad\quad W_g(k,r) =\exp{\left(-\frac{k^2r^2}{2}\right)},\quad\quad W_s(k,r) =\frac{\sin{(kr)}}{kr}.
\eea
The most popular choices for the window function involve the top-hat $(W_{th})$ and the Gaussian $(W_g)$ functions. Based on the current developments in studying PBH collapse, there is no fixed choice between these two functions yet, and it can vary depending on the nature of the analysis. For our purposes, we prefer to stick with the Gaussian choice, as we found that the broad and oscillatory features appearing in the Fourier space version of the top-hat do not work well to smoothen the small-scale fluctuations during our later analysis. As for the spherical window function $W_s$, the reason for its use in the literature comes from the fact that the expression for the linear component ${\cal C}_{G}$, which contains $\zeta'_{G}$, results after using the linear order term in the Eqn. (\ref{gradexpansion}) and perform a surface integral over a sphere of areal radius $R$ following the definition in Eqn. (\ref{compaction}). Though this procedure provides a motivation for the smoothing function over a spherical shell of radius $r$ for the curvature perturbation $\zeta_G$, for this work especially, we found that sticking to the Gaussian choice again proves easier to conduct the analysis for the similar reasons as stated before upon using the top-hat function. With the help of these window functions, the various correlators involved in the
determination of the joint PDF are in general written as \cite{Young:2022phe, Ferrante:2022mui}: 
\bea 
\label{sigc}\langle C_{G}C_{G}\rangle &=& \sigma_{c}^{2} =4\left(\frac{f(w)}{3}\right)^{2}\int_{0}^{\infty}\frac{dk}{k}(k r_{p})^{4}W_{g}^{2}(k,r_{p})\tilde{\Delta}^{2}_{\zeta,{\bf RRR}}(k),
\\
\label{sigr} \langle\zeta_{G}\zeta_{G}\rangle &=& \sigma_{r}^{2} = \int_{0}^{\infty}\frac{dk}{k}W^{2}_{s}(k,r_{p})\tilde{\Delta}^{2}_{\zeta,{\bf RRR}}(k),
\\
\label{sigcr}
 \langle C_{G}\zeta_{G} \rangle  &=& \sigma_{cr}^2 = 2\bigg(\frac{f(w)}{3}\bigg) \int_{0}^{\infty}\frac{dk}{k}(k r_{p})^{2}W_{g}(k,r_{p})W_{s}(k,r_{p})\tilde{\Delta}^{2}_{\zeta,{\bf RRR}}(k),\quad
\eea
where we observe the cumulative contribution to the correlations coming after integration over each phase. Notably, the USR phase $(k_s,k_e)$ yields the largest contribution attributable to the significant amplification of the scalar power spectrum during this interval. Contrary to the modes corresponding to either contraction $(k_c<k<k_b)$, bounce $(k_b<k<k_{*})$, SRI $(k_*<k<k_s)$, or SRII $(k_e<k<k_{\rm end})$, it is for those scales inside the USR where the curvature perturbation attains substantial amplification for generating the PBHs, 
\bea
\int_{0}^{\infty} \equiv \underbrace{\int_{0}^{k_c}}_{\textbf{negligible}} + \underbrace{\int_{{k}_c}^{k_b}}_{\textbf{negligible}} + \underbrace{\int_{{k}_b}^{k_*}}_{\textbf{negligible}}+ \underbrace{\int_{{k}_{*}}^{k_s}}_{\textbf{suppressed}} + \underbrace{\int_{{k}_s}^{k_e}}_{\textbf{Dominant(USR)}} + \underbrace{\int_{{k}_{e}}^{k_{\rm end}}}_{\textbf{suppressed}} + \underbrace{\int_{k_{\rm end}}^{\infty}}_{\textbf{suppressed}}
\eea
where the contributions keep getting more negligible as we consider modes from before the inflationary phases. The cumulative impact of the chosen window and transfer functions through these correlators can be visualized in the PDF later in the numerical analysis section. We now take a moment to comment on the behaviour of the joint PDF on the basis of the possible correlations. 
\begin{itemize}
    \item[\textbullet] \underline{\textit{No correlation}}    
     ($\gamma_{\rm cr}=0$):  The cross-correlation coefficients of the covariance matrix of Eqn. (\ref{gammacov}) is zero, meaning the joint PDF appears as a bivariate normal distribution with uncorrelated and independent variables $C_G$ and $\zeta_G$; any property for ${\cal C}_{G}$ cannot be enough to derive anything about the behaviour of $\zeta_{G}$. The corresponding ellipse (shape of the PDF) deforms towards concentric circles showing the sign of an isotropic covariance and as one moves away from the center it also indicates decreasing probability density. 
     \item[\textbullet] \underline{\textit{Positive correlation}}    
     ($\gamma_{\rm cr}>0$): The $2$D joint PDF deviates from the circular shape, and as $\gamma_{\rm cr}\rightarrow 1$, the PDF is strongly correlated and gets squeezed along a line, imitating a reduced $1$D type distribution. Now an increasing $\gamma_{cr}$ suggest that few specific configurations of the random variables would contribute to the PBHs formation with the rest lying outside the support of the domain within the PDF. Any property shown in the compaction ${\cal C}_{G}$, at some specific position, can be equally well related to being displayed in ${\zeta}_{G}$ too. As we will display later in Sec \ref{s6}, the integration domain gets pushed toward the boundary along with the squeezing, suggesting that only certain compaction thresholds would be permissible for integrating over the respective domain and obtaining the mass fraction. Additionally, the domain support inside the PDF for likelihood of PBHs formation is sensitive to the sign of $f_{\rm NL}$. For $f_{\rm NL}<0$, the domain is shown to align towards the anti-correlated direction. This implies that $C_{G}$, and $\zeta_{G}$ have opposite signs, reflecting increased range of conditions for probability of PBHs formation. On the contrary, adopting $f_{\rm NL}>0$, leads to the PDF displaying significant support in the correlated quadrants, where the signs of both the gaussian random variables are same for favourable conditions of PBHs formation. A positive $f_{\rm NL}>0$ scenario is not examined in the scope of this paper and we refer the reader to \cite{Ferrante:2022mui} for details.

\end{itemize}
The correlations involve the total power spectrum modified with help of the radiation transfer function ${\cal T}$ (see Eqn. (\ref{radtransfer})) as follows:
\bea 
\tilde{\Delta}^{2}_{\zeta,{\bf RRR}}(k)= {\cal T}^2(k,r_p)\Delta^2_{\zeta,{\bf RRR}}.
\eea
In the above expression, we fix the choice of conformal time to $\tau=r_p$ for the transfer function and it behaves as a damping function for the subhorizon modes while evaluating the integrals at the horizon re-entry scale denoted earlier as $r_p$, where also the peak in the compaction function arises. 
Furthermore, during the instant of horizon re-entry there exists a relation between the peak compaction length scale $r_p = 1/(c_sk_H)$, with the wavenumber $k_{H}$ sourcing the PBH collapse, and the horizon mass contained at the re-entry time \cite{Sasaki:2018dmp}:
 \bea\label{HMass}
M_H \approx 17M_{\odot}\bigg(\frac{g_{*}}{10.75}\bigg)^{-1/6}\bigg(\frac{c_{s}k_{H}}{10^6 \rm Mpc^{-1}}\bigg)^{-2}.
\eea
which will be of importance to us when examining the numerical results later. The quantity $g_{*}=106.75$ above indicates the relativistic degrees of freedom deep inside the RD epoch. The behaviour of the effective sound speed, at the instant of the sharp transition in the present study, lies between the interval, $0.88\leq {c}_{s}\leq 1$. The case of $0< c_s< 1$ characterizes the non-canonical while $c_s=1$ characterizes the canonical EFT frameworks.

Now we would explicitly like to discuss the impact of $\gamma_{\rm cr}$ approaching unity on the likelihood of PBHs formation. When Eqn. (\ref{PDF}) is expanded around $\gamma_{\rm cr}=1$, the PDF can be written as:
\bea \label{taylorPg}
\mathds{P}_{G}({\cal C}_{G},\zeta_{G}) &=& \frac{1}{2 \sqrt{2} \pi \sigma_{c}\sigma_{r} \sqrt{1-\gamma_{cr}}}\left(\displaystyle{1+\frac{1-\gamma_{cr}}{4}+\frac{3 (1-\gamma_{cr})^{2}}{32} +\frac{5 (1-\gamma_{cr})^{3}}{128}+\frac{35 (1-\gamma_{cr})^{4}}{2048}+\text{  HO}}\right)\nonumber\\
   && \quad\quad\quad\quad\quad\quad\quad\quad\quad\quad\quad\times\exp \bigg(\displaystyle{\frac{1}{4 (\gamma_{cr}
   -1)}\left(\frac{{\cal C}_{G}}{\sigma_{c}}-\frac{\zeta_{G}}{\sigma_{r}}\right)^{2}+\sum_{n=3}^{\infty}\left(-\frac{1}{2}\right)^{n}\left(\frac{{\cal C}_{G}}{\sigma_{c}}+\frac{\zeta_{G}}{\sigma_{r}}\right)^{2}(\gamma_{cr}-1)^{n-3}}\bigg), \quad\quad\eea
where we have "HO" representing the higher order terms that are neglected. The above PDF is written using Eqn. (\ref{PDF}) as a product of two truncated power series expansions, one for the inverse square root factor before the exponential and another for the $(1-\gamma_{\rm cr}^2)^{-1}$ containing factor inside the exponential, which when written with the limit $\gamma_{\rm cr} \rightarrow 1$ and multiplied by the extra Gaussian factor for $\zeta_{G}$ reduces to the following expression:
\bea \label{peakpdf}
 \lim_{\gamma_{cr}\rightarrow 1}\mathds{P}_{G}({\cal C}_{G},\zeta_{G}) &=& \frac{1}{\sqrt{2\pi}\sigma_{c}\sigma_{r}}\exp{\left(-\frac{1}{8}\left(\frac{{\cal C}_{G}}{\sigma_{c}}+\frac{\zeta_{G}}{\sigma_{r}}\right)^{2}\right)}
 \lim_{\gamma_{cr}\rightarrow 1}\frac{1}{\sqrt{2\pi(2(1-\gamma_{cr}))}}\exp{\left(\frac{1}{4(\gamma_{cr}-1)}\left(\frac{{\cal C}_{G}}{\sigma_{c}}-\frac{\zeta_{G}}{\sigma_{r}}\right)^{2}\right)},\nonumber \\
&=&\frac{1}{\sqrt{2\pi}\sigma_{c}\sigma_{r}}\exp{\left(-\frac{1}{8}\left(\frac{{\cal C}_{G}}{\sigma_{c}}+\frac{\zeta_{G}}{\sigma_{r}}\right)^{2}\right)}\delta\left(\frac{{\cal C}_{G}}{\sigma_{c}}-\frac{\zeta_{G}}{\sigma_{r}}\right).\quad
\eea
where only the $n=3$ terms remains to contribute and comes outside the limit while the remaining terms vanish. The Dirac delta function imposes the condition, ${\cal C}_{G}/\sigma_c=\zeta_{G}/\sigma_r$, meaning that the PDF effectively squeezes along the diagonal direction indicating that any feature present for the compaction ${\cal C}_{G}$ gets strongly correlated to having a similar effect for $\zeta_{G}$. The final PDF is a peaked and maximally correlated one or a case where the power spectrum is purely monochromatic. In terms of PBHs formation, variation in the value of $\gamma_{\rm cr}$ has significant impact. As previously discussed, in the limit of $\gamma_{\rm cr} \rightarrow 0$, the covariance matrix becomes diagonal and the resulting PDF is the product of the two uncorrelated gaussian random variables. As $\gamma_{\rm cr}$ increases, we observe a sharp PDF, exhibiting a pronounced squeezing. Consequently the parameter space that supports the perturbation crossing the threshold values fails to completely coincide with the regions supported by the PDF, and therefore the probability of obtaining a sizeable mass fraction(for higher mass PBHs) decreases. We mentioned about $\gamma_{\rm cr}$ retaining a mass dependence. It can be seen in sec.(\ref{s6}) that as the mass of PBHs increases, we would observe the PDF getting squeezed indicating higher $\gamma_{\rm cr}$ values. 

Now, utilizing the PDF in Eqn. (\ref{PDF}), we arrive at the main expression of this section for the PBH mass fraction \cite{Ferrante:2022mui}: 
\bea   
\label{massfraction}
\beta_{\rm NG}(M_{H}) = \int_{\cal D} K({\cal C}-{\cal C}_{\rm th})^{\gamma}\mathbbm{P}_{G}({\cal C}_G, \zeta_G) d{\cal C}_G\;d\zeta_{G}. \eea
Let us discuss the various components of the above expression to establish its significance more thoroughly. First, it uses the critical scaling relation, $K({\cal C}-{\cal C}_{\rm th})^{\gamma}$ \cite{Choptuik:1992jv, Niemeyer:1997mt, Musco:2012au}, to estimate the PBH mass spectrum formed at some particular time from the gravitational collapse. This relation is different from the notion where the PBH mass gets approximated by the mass present within the horizon at the formation time but suggests instead a range of PBH masses produced for a given horizon mass $M_{H}$. The variable $K$ depends on the density contrast profile structure and we keep this value within $K\sim {\cal O}(1-10)$, and $\gamma$ here is the critical exponent during the spherical collapse. It is independent of the initial perturbation profile. It generally depends on the equation of state and in the RD era takes the value, $\gamma\simeq 0.36$ \cite{Evans:1994pj}, which is a universal value obtained from numerical simulations and is of interest to us here. For ${\cal C}_{\rm th}$ we can now easily determine this by following the result from numerical analysis for the range of density contrast threshold $\delta_{\rm th}$ values in RD \cite{Musco:2020jjb} and relating them with the help of Eqn. (\ref{deltacompaction}). Aside from the joint PDF in $({\cal C}_{G},\zeta_{G})$, the other crucial component is the domain of integration ${\cal D}$, so let us now discuss how it gets evaluated below. 

The domain ${\cal D}$ is defined as the following region of intersection:
\bea \label{integraldomain}
{\cal D} = \{{\cal C}(r) \geq {\cal C}_{\rm th} \wedge {\cal C}_{G}(d\zeta/d\zeta_{G}) \leq 2f(w)\},
\eea
that is composed of two conditions which we introduced in discussions earlier but now put into action to see the result. $(1).$ The lower limit comes from the very use of the threshold statistics on the compaction function, which suggests that PBH formation begins after the value of compaction ${\cal C}(r)$ signals perturbations beyond the threshold value ${\cal C}_{\rm th}$ at horizon re-entry. $(2).$ The upper limit results as a consequence of the compaction maxima condition in Eqn. (\ref{maximacondition}) applied to the version of the compaction in Eqn. (\ref{compactfnl}):
\bea \label{domain:upperlimit}
\frac{d}{dr}{\cal C}(r) = \frac{d}{dr}{\cal C}_{\rm Linear} - \frac{1}{2f(w)}\bigg({\cal C}_{\rm Linear}\frac{d}{dr}{\cal C}_{\rm Linear}\bigg)=0, \quad\quad \text{where}\quad {\cal C}_{\rm Linear} = {\cal C}_{G}(r)\frac{d\zeta(r)}{d\zeta_{G}},
\eea
from which one obtains the relation, ${\cal C}_{G}(d\zeta/d\zeta_{G}) = 2f(w)$. Since we are concerned with the peak scale at the instant of re-entry $r_{p}$, the condition ${\cal C}_{G}(d\zeta/d\zeta_{G}) \leq 2f(w)$ provides us with the largest possible PBH mass formed at that instant that defines the actual horizon mass $M_{H}$. The combined region from both $(1)$ and $(2)$ within the PDF between $({\cal C}_G,\zeta_G)$ serves as our domain over which we perform the integration to evaluate the PBH mass fraction in our analysis. Although we have explained the conditions needed to establish the domain, we go one step further to work out the true limits one must impose to perform this integral. We go back again to the Eqn. (\ref{compactfnl}) but this time, keeping in mind the equality sign in condition $(1)$, solve the resulting quadratic equation:
\bea
{\cal C}_{\rm th} &=& {\cal C}_{\rm Linear}- \frac{1}{4f(w)}\big({\cal C}_{\rm Linear}\big)^{2}\quad\implies \quad\big({\cal C}_{\rm Linear}\big)^{2} - 4f(w){\cal C}_{\rm Linear} + 4f(w){\cal C}_{\rm th} = 0,
\eea
to obtain the critical condition on ${\cal C}_{G}$:
\bea \label{criticalCG}
({\cal C}_{G})_{\pm} = \bigg[\frac{d\zeta(r)}{d\zeta_{G}}\bigg]^{-1}2f(w)\left(1 \pm \sqrt{1-\frac{{\cal C}_{\rm th}}{f(w)}}\right),
\eea
thus we are able to find the exact limits of integration if we further combine the outcome of condition $(2)$ with Eqn. (\ref{criticalCG}) to give us:
\bea \label{CGcriticalrange}
({\cal C}_{G})_{\pm} \leq {\cal C}_{G} \leq \bigg[\frac{d\zeta(r)}{d\zeta_{G}}\bigg]^{-1}2f(w).
\eea
The two critical values $({\cal C}_{G})_{\pm}$ correspond to different classes of perturbations that lead to the formation of two types of PBH, namely type-I (for $+$) and type-II (for $-$) PBHs \cite{Musco:2020jjb}. Type-I PBHs form with their mass spectrum as per the critical scaling relation, while the type-II PBHs are found to have highly suppressed abundance, and numerically interpreting their formation process has also proven less successful than the type-I scenario, see more on this in \cite{Young:2022phe}. 
Also, about the above limits, we recall the statement made during the discussion of Eqn. (\ref{compactfnl2}), where we first remarked on the importance of the term $d\zeta_{G}/d\zeta$ to determine the two Gaussian variables $({\cal C}_{G},\zeta_{G})$. Since the mentioned term shows no guarantee of always remaining either positive or negative, we observe two separate branches of domain solutions that we highlight in our upcoming numerical analysis section. The strength of PNGs, with either $f_{\rm NL}>0$ or $f_{\rm NL}< 0$, clearly plays a role in determining the values of either variables ${\cal C}_{G}$ and $\zeta_{G}$ can take inside the PDF domain.

The mass fraction in Eqn. (\ref{massfraction}) is finally used to provide us the dark matter density fraction contained in the form of PBHs, or the PBH abundance, for a given $M_{\rm PBH}$ after numerically integrating over a range of horizon masses as \cite{Ferrante:2022mui, Franciolini:2023pbf}: 
\bea \label{pbhabundance} f_{\rm PBH} &=& \frac{1}{\Omega_{\rm DM}}\int d\ln{M_{H}}\left(\frac{M_{\odot}}{M_H}\right)^{1/2}\left(\frac{g_{*}}{106.75}\right)^{3/4}\left(\frac{g_{*s}}{106.75}\right)^{-1}\left(\frac{\beta_{\rm NG}(M_{H})}{7.9 \times 10^{-10}}\right),
\eea
where $M_{\odot}$ is the solar mass, \textcolor{black}{$\Omega_{\rm DM}=0.264$} is the current dark matter density. We will analyze this quantity in the coming sections, discussing various related numerical outcomes from our study by incorporating the PNGs.


\section{Scalar Induced Gravitational Waves from EFT of Bounce}\label{s5}

We briefly review the theory of scalar induced gravitational waves (SIGWs) and outline the necessary results needed to determine the GW spectrum from the regularized-renormalized-resummed scalar power spectrum reviewed in section \ref{s32}.  

The SIGWs present an interesting scenario studied abundantly in recent literature related to interpreting the latest pulsar timing array (PTA) signal \cite{NANOGrav:2023hvm, EPTA:2023fyk, Reardon:2023gzh, Xu:2023wog} of a stochastic gravitational wave background. Such induced gravitational waves have a much greater power distribution than the primordial gravitational waves and thus provide a worthy signal to measure. The basic idea behind generating SIGWs involves studying the equation of motion for the tensor modes using cosmological perturbation theory, where the scalar modes couple at second order and provide sourcing of these tensor perturbations. We will be interested in the scenario where, after their generation, the GWs re-enter into the RD epoch with the EoS, $w=1/3$. These GWs continue to explore the Universe since the Big Bang thus providing a valuable probe of the early Universe before inflation ends. Also, due to a large GW energy density associated with the wavenumbers of the NANOGrav15 signal, the SIGWs present a way to constrain the scalar power spectrum amplitude to provide an ample abundance of sub-solar mass PBHs, which we discuss in the spectrum results.

The important observable of interest here is called the spectral density of GWs, which is defined as the fractional GW energy density $(\rho_{\rm GW})$ per logarithmic interval of wavenumber \cite{Domenech:2021ztg}:
\bea \label{spectraldensity}
\Omega_{\rm GW}(\tau, k) &=& \frac{1}{3M_{pl}^{2}H^{2}}\frac{d\rho_{\rm GW}}{d\ln{k}} = \frac{1}{12}\bigg(\frac{k}{aH}\bigg)^{2}\overline{\Delta^{2}_{h}(\tau,k)},
\eea
where $\overline{\Delta^{2}_{h}(\tau,k)}$ refers to the tensor power spectrum of which an averaging over the oscillations is later performed and denoted by the overline. From the above Eqn. (\ref{spectraldensity}) we get the GW spectrum generated at some time in radiation domination. We can also write the expression for the spectral density measured at the current moment, if it was generated at some moment $t_{c}$, using the previous relation as:
\bea
\Omega_{\rm{GW},0}h^2 &=& \frac{h^{2}}{3M_{pl}^{2}H_{0}^{2}}\frac{d\rho_{\rm GW,0}}{d\ln{k}} = \Omega_{r,0}h^2\frac{1}{\rho_{r,0}}\frac{d\rho_{\rm GW,0}}{d\ln{k}} = \Omega_{r,0}h^2\frac{\rho_{r,c}}{\rho_{r,0}}\left(\frac{a_{0}}{a_{c}}\right)^{-4}\Omega_{\rm GW,c}.
\eea
where $\Omega_{r,0}h^{2}=\rho_{r,0}h^{2}/(3H_{0}^{2}M_{pl}^{2})$ denotes the current radiation energy density with $h=H_{0}/100\;{\rm kms^{-1}}/{\rm Mpc}$ (for the uncertainty in $H_{0}$). The last equality observes the use of redshift scaling of the radiation energy density. Eqn. (\ref{spectraldensity}) requires us to evaluate the tensor power spectrum which can be expressed as coming from taking the square of the scalar power spectrum, from Eqn. (\ref{RRRspectrum}), and an integration kernel such that the GW energy density finally reads of as:
\bea
\Omega_{\rm GW} &=& \int_{0}^{\infty}dv \int_{|{1-v}|}^{1+v} du \; {\cal F}(u,v)\times\Delta^{2}_{\zeta, {\bf RRR}}(ku) \times \Delta^{2}_{\zeta, {\bf RRR}}(kv),
\eea   
where the kernel is given by \cite{Kohri:2018awv, Domenech:2021ztg}:
\bea
{\cal F}(u,v) &=& 3y^{2}\bigg(\frac{4v^2 - (1-u^2 +v^2)^2}{4u^2v^2}\bigg)^2 \times \bigg[\frac{\pi ^2 y^2}{4}\Theta[(u+v)-\sqrt{3}]+\bigg(1-\frac{1}{2}y \ln \bigg|\frac{1+y}{1-y}\bigg|\bigg)^2\bigg],\eea
with
\bea y\equiv 1- \frac{3}{2uv}\bigg[1-\left(\frac{u-v}{\sqrt{3}}\right)^2\bigg].
\eea

After these GWs are generated, to measure their current energy density require implementing the entropy conservation principle, and after juggling with the energy density $(g(T))$ and entropy $(g_{s}(T))$ degrees of freedom within the Eqn. (\ref{spectraldensity}), one arrives at the following result \cite{Domenech:2021ztg}:
\bea
\Omega_{\rm{GW},0}h^2 &=& 1.62 \times 10^{-5}\;\bigg(\frac{\Omega_{r,0}h^2}{4.18 \times 10^{-5}}\bigg)\times\bigg(\frac{g(T_r)}{106.75}\bigg)\bigg(\frac{g_{s}(T_r)}{106.75}\bigg)^{-4/3}\Omega_{\rm GW},
\eea
where $T_{r}$ is the temperature at the instant of GW generation in the RD epoch. We are interested in the frequency regime that coincides with the one probed by the PTA signal, $10^{-9}{\rm Hz}\leq f \leq 10^{-8}{\rm Hz}$, and a simple conversion relation between the two variables used here follows:
\bea
f = 1.6 \times 10^{-9} {\rm Hz}\times \left(\frac{k}{10^{6}\;{\rm Mpc}^{-1}}\right).
\eea

In the later section, we will analyze the GW spectrum resulting from our use of the regularized-renormalized-resummed power spectrum and see that the RD case gives a strong signal within the PTA frequency regime.

\section{Numerical Analysis}
\label{s6}

In this section, we confer on the various aspects of our outcomes, owning to the compaction function approach to mitigate PBH overproduction. We devote the first subsection to a narrative explaining our choice of large negative NGs. Following this, we turn to the results associated with the GW spectrum from our regularized-renormalized-resummed version of the power spectrum. Proceeding ahead, we present a comprehensive analysis centered on the PBHs overproduction problem, where we also offer insights on the appropriate choice of the effective sound speed parameter $(c_{s})$ that can evade the overproduction problem while simultaneously being favored by the PTA data.

\subsection{Why large negative non-Gaussianity?}
\label{s6a}

At this point, it becomes imperative to unveil the rationale behind prominent negative NGs, and their efficacy in yielding improved results. Recall the parameterization relation in Eqn. (\ref{zetaPNG}). This definition represents a local form of NGs since the equation is local in real space. Now, the current observational constraint on the local NG comes to be $f_{\rm NL}=-0.9\pm5.1$ at $68\%$ confidence level \cite{Planck:2019kim}, implying that cosmologically significant deviations from gaussianity have not yet been observed.
For the specific case of single-field, slow-roll inflation, Maldacena presented his now-famous consistency condition \cite{Maldacena:2002vr} relating the spectral index $n_s$ for the two-point correlation function with the NG parameter $f_{\rm NL}$ from the three-point correlation, which led to a prediction of the value $f_{\rm NL}=(5/12)(1-n_{s})\sim {\cal O}(-0.1)$, based on current observational estimates of the spectral index. 

A significant influence of having PNGs in our analysis $(f_{\rm NL})$ lies in the fact that positive NGs can be quite harmful as they amplify the fluctuations, and that increases the PBH abundance. On the contrary, negative NGs can suppress the probability of even producing large fluctuations, thus lowering the abundance \cite{Young:2013oia}. As a result of such a property, one can expect to incorporate a negative and relatively large value of $f_{\rm NL}$ within their model to attempt an explanation for the reduction in PBH production. Since large NGs can result from the existence of a slow-roll violating USR phase, the Galileon inflation is a particular example where a significantly large and negative, $f_{\rm NL}\sim {\cal O}(-6)$ due to the USR is reported \cite{Choudhury:2023kdb}. In alternative inflationary frameworks, a more pronounced negative $f_{\rm NL}$ emerging from a USR phase, in contrast to the case of Galileon inflation, has not yet been studied. Having large deviations from gaussianity due to the USR is also intrinsically tied to the resulting strength of the curvature perturbations, which later present a noticeable impact on various physical phenomena of interest in the early universe. Our interest here lies in the production of PBHs and their SIGW counterpart. Utilizing the large and negative features of the NG, in \cite{Choudhury:2023fwk} it was shown that a value of $f_{\rm NL}\sim {\cal O}(-6)$ from Galileon inflation could suffice to circumvent the prevalent PBH overproduction issue in the recently observed PTA signal, see Refs. \cite{Franciolini:2023pbf,Franciolini:2023wun,Inomata:2023zup,LISACosmologyWorkingGroup:2023njw, Inui:2023qsd,Chang:2023aba, Gorji:2023ziy,Li:2023xtl, Li:2023qua,Firouzjahi:2023xke,Gorji:2023sil,Choudhury:2023fwk,Choudhury:2023fjs,Choudhury:2024one}. However, with such a negative amount of NG coming from Galileon inflation, the statistical concordance with the SIGW explanation of the latest SGWB signal is positioned just inside the $1\sigma$ confidence region of the NANOGrav15 and excluded at the $1\sigma$ regime of the EPTA posteriors for the signal. 

Our objective here is to resolve the PBH overproduction problem most efficiently by demonstrating the advantage of considering large negative NGs and the account for an ample power spectrum amplitude that best aligns with the SIGW interpretation of the PTA signal. To achieve this, we approach a setup featuring a non-singular bouncing cosmology integrated with the standard SR/USR/SR inflationary model employed to study PBH formation. Moreover, by generalizing this in the EFT framework allows us to encompass a wider class of models and throws out any model-specific considerations and their subsequent imprints on our results. The motivation to introduce bounce in our analysis stems from the  various studies in the past suggesting that bouncing scenarios can generate an order of magnitude greater NGs compared to the single-field inflationary models \cite{Cai:2009fn, Lehners:2008my}.

In standard slow-roll inflationary scenario, curvature perturbations originate directly from the quantum fluctuations of a scalar field. Due to the near-flatness of the inflaton potential, the slow roll parameter remains small and hence the fluctuations in the scalar field are approximately Gaussian. Consequently, the $f_{\rm NL}$ parameter is generally small. In sharp contrast, the ekpyrotic model features a steep, negative potential which leads to a large equation of state parameter or slow roll parameter. This steep potential forces the the scalar fields to experience significant non-linear interactions, which enhance the NGs. In the context of the ekpyrotic scenarios of contraction preceding the bounce, the authors in \cite{Lehners:2008my} demonstrated by solving the equation of motion for the perturbations produced during the ekpyrotic phase, and in \cite{Koyama:2007if} using the $\delta N$ formalism, that the NG parameter can achieve large and negative values starting from $f_{\rm NL}=-39.95$. We have adopted this specific value for our analysis due to its efficacy in producing elaborate results. Similarly, we incorporate the local NG value also as $f_{\rm NL}=-35/8$ for our analysis derived from the matter bounce scenario in \cite{Cai:2009fn} where they focus on the interaction terms in the cubic order action which remain suppressed during a slow-roll inflation case but instead provide substantial contributions when estimating the bispectrum under the conditions for a matter bounce. 
Therefore, our set-up featuring a contraction and a bouncing scenario, including a USR phase during inflation, unequivocally invites the inclusion of large negative NGs.

\begin{figure*}[htb!]
    \centering
    \subfigure[]{
    \includegraphics[height=8cm,width=8.5cm]{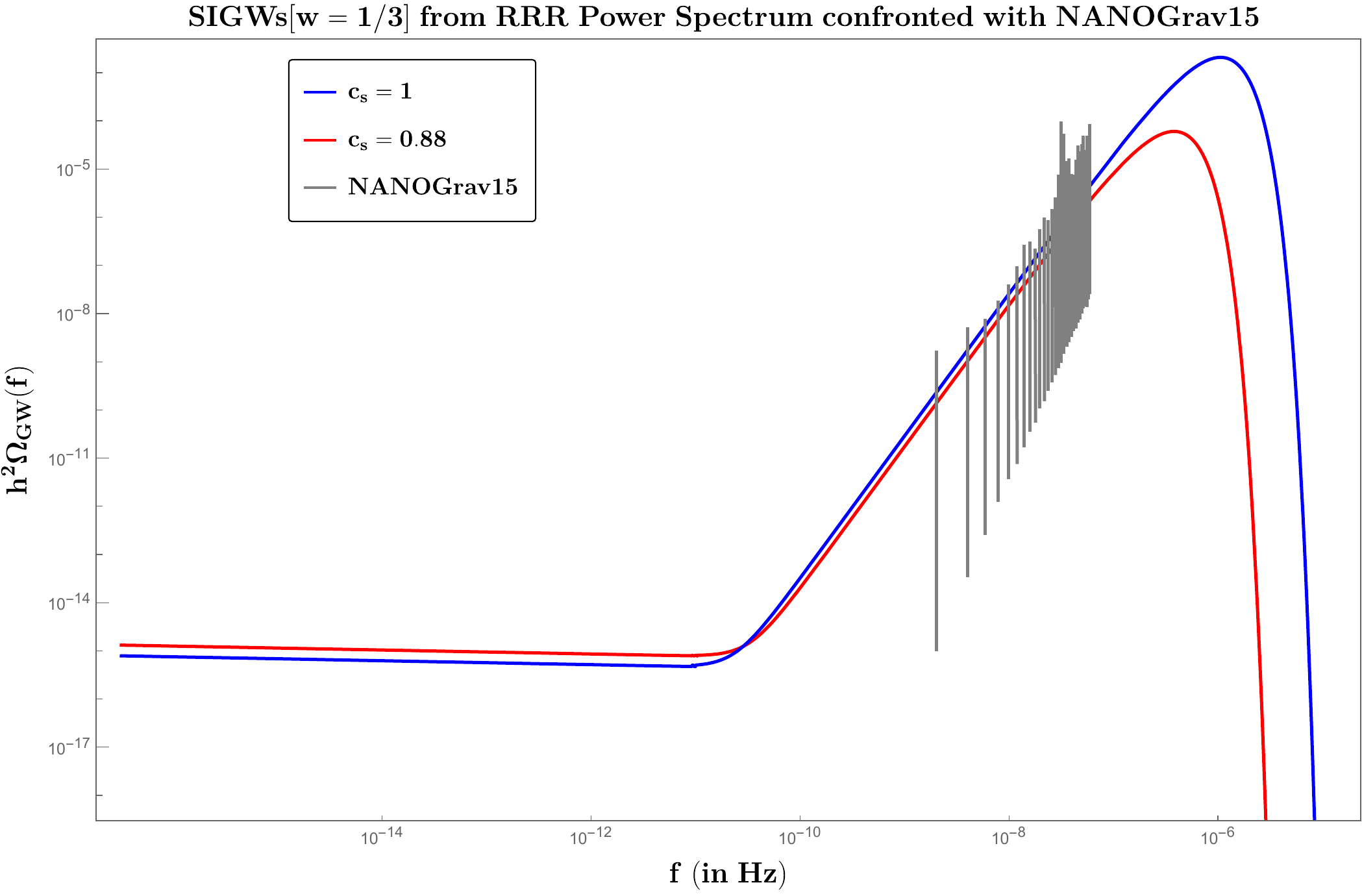}\label{DiffcsNG}
    }
    \subfigure[]{
    \includegraphics[height=8cm,width=8.5cm]{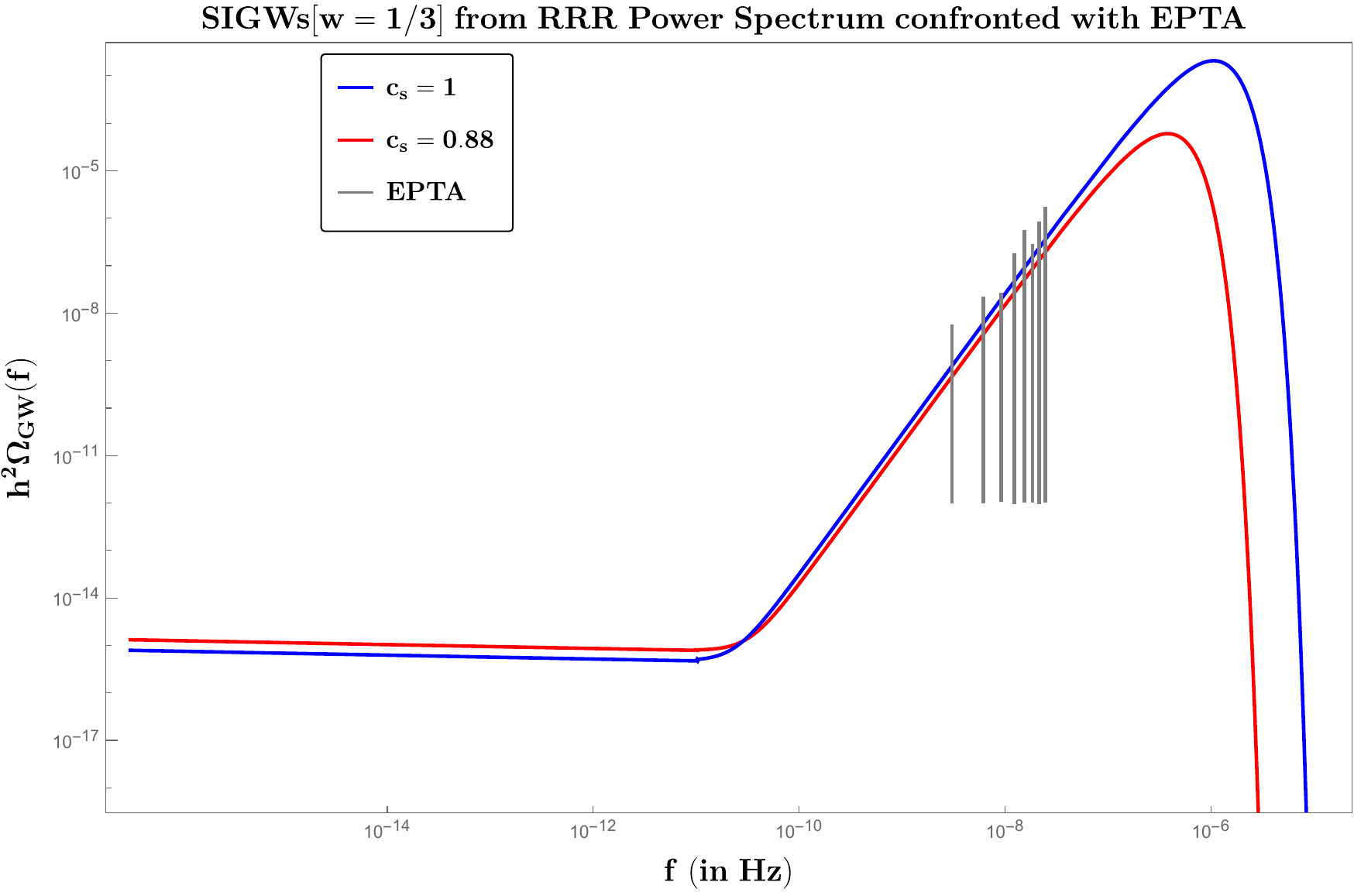}\label{DiffcsEPTA}
    }
    \caption{GW spectrum $h^2\Omega_{\rm GW}$ versus the frequency for different choices of the effective sound speed, $c_s=1$ (blue) and $c_s=0.88$ (red), while keeping the EoS $w=1/3$ fixed. The background consists of the experimental signal, shown here in black vertical lines, obtained by the NANOGrav15 (left) and EPTA (right) observations. The generated GW spectrum displays a growing trend consistent with the signals' features, and soon after, it falls sharply due to the exponentially decreasing amplitude feature present inside the regularized-renormalized-resummed scalar power spectrum in Eqn. (\ref{RRRspectrum}). }\label{csSIGW}
\end{figure*}

\begin{figure*}[htb!]
    \centering
    \subfigure[]{
    \includegraphics[height=6cm,width=18cm]{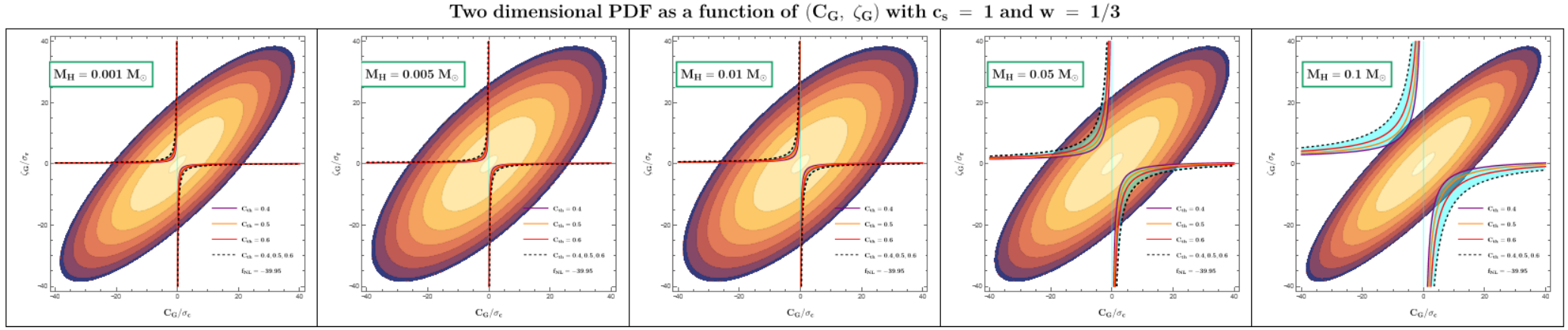}\label{fnl39cs1}
    }\\
    \subfigure[]{
    \includegraphics[height=6cm,width=18cm]{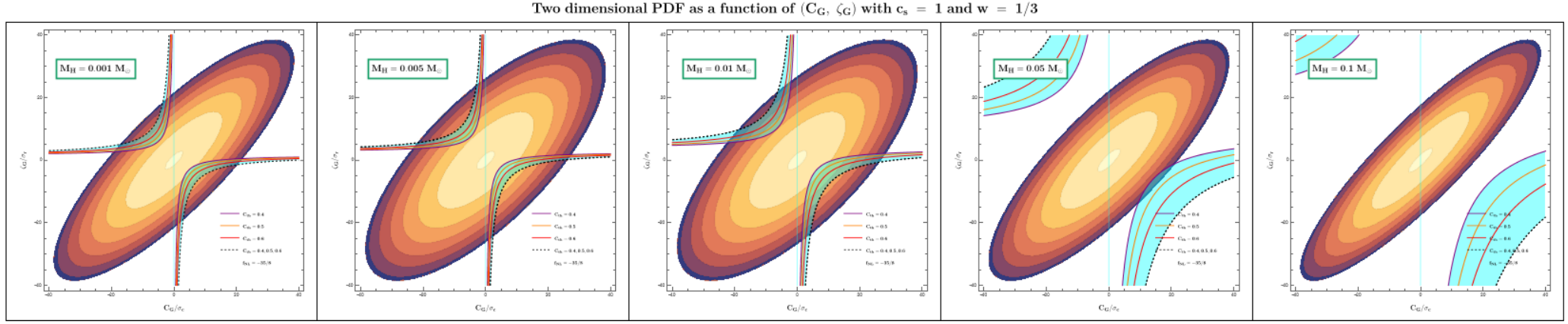}\label{fnl35by8cs1}
    }
    \caption{Joint probability distribution function for various values of PBH masses with a fixed amplitude $A=10^{-2}$, EoS parameter $w=1/3$, and effective sound speed $c_{s}=1$. The top fig. \ref{fnl39cs1} depicts results for $f_{\rm NL}=-39.95$ while in the bottom fig. \ref{fnl35by8cs1} corresponds to $f_{\rm NL}=-35/8$. The shaded cyan regions represent the permissible integration domains, from Eqn. (\ref{integraldomain}), for various compaction thresholds: ${\cal C}_{\rm th}=0.4$ (purple), ${\cal C}_{\rm th}=0.5$ (yellow), ${\cal C}_{\rm th}=0.6$ (red), respectively. Inside the PDFs, we observe increased support of the domains giving over-threshold perturbations for lower mass PBH while these lose support by moving outwards for higher mass PBH. This effect becomes pronounced in the bottom for $f_{\rm NL}=-35/8$. As we increase the PBH mass, the correlation coefficient $\gamma_{\rm cr}$ tends towards unity, which is evident here by the compression along the correlated direction of the PDF.  }\label{2dpdffnl}
\end{figure*}

\subsection{GW Spectrum}
\label{s6b}

\begin{figure*}[htb!]
    \centering
    \subfigure[]{
    \includegraphics[height=8cm,width=8.5cm]{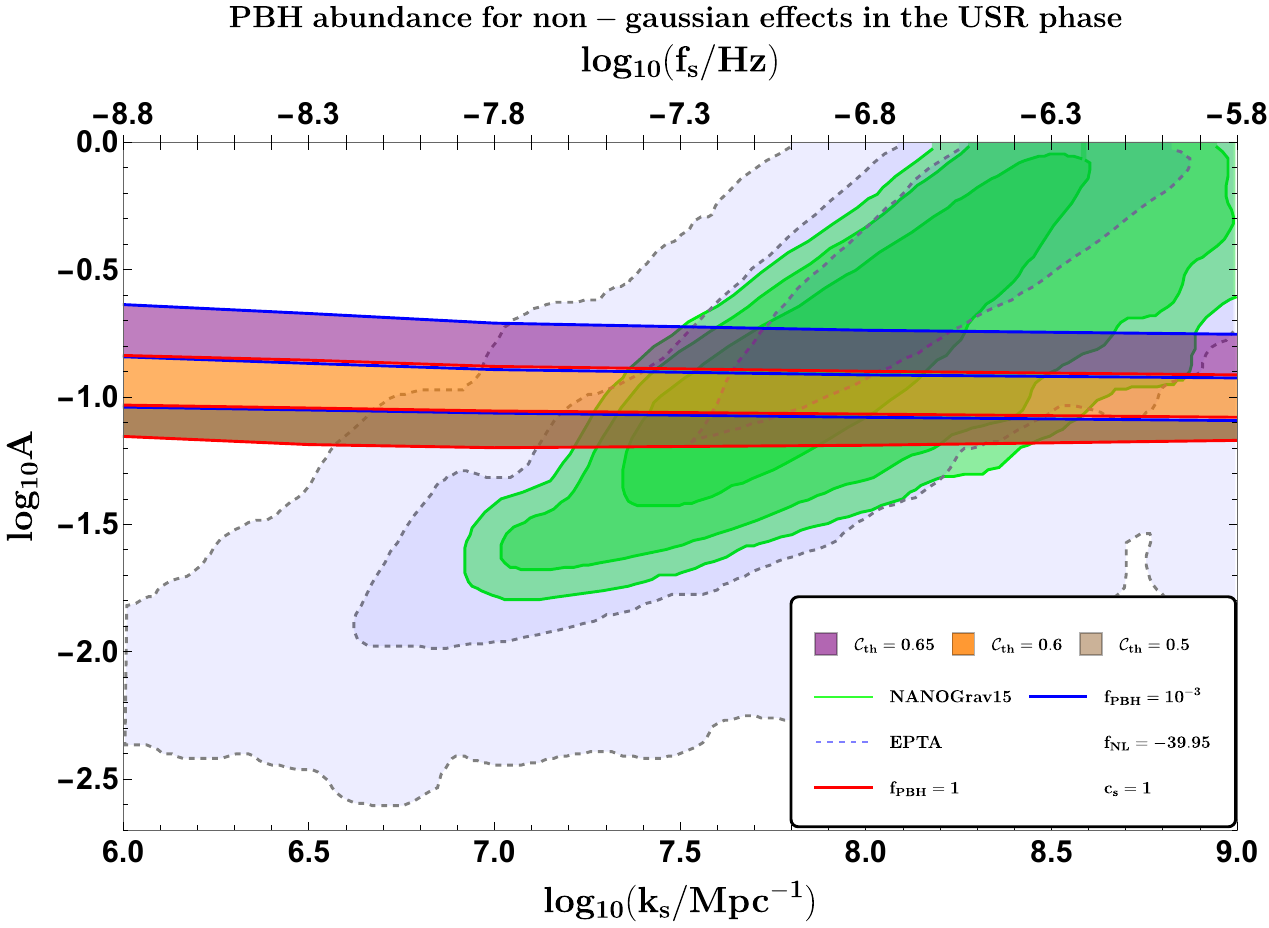}\label{fnl39cs1cth}
    }
    \subfigure[]{
    \includegraphics[height=8cm,width=8.5cm]{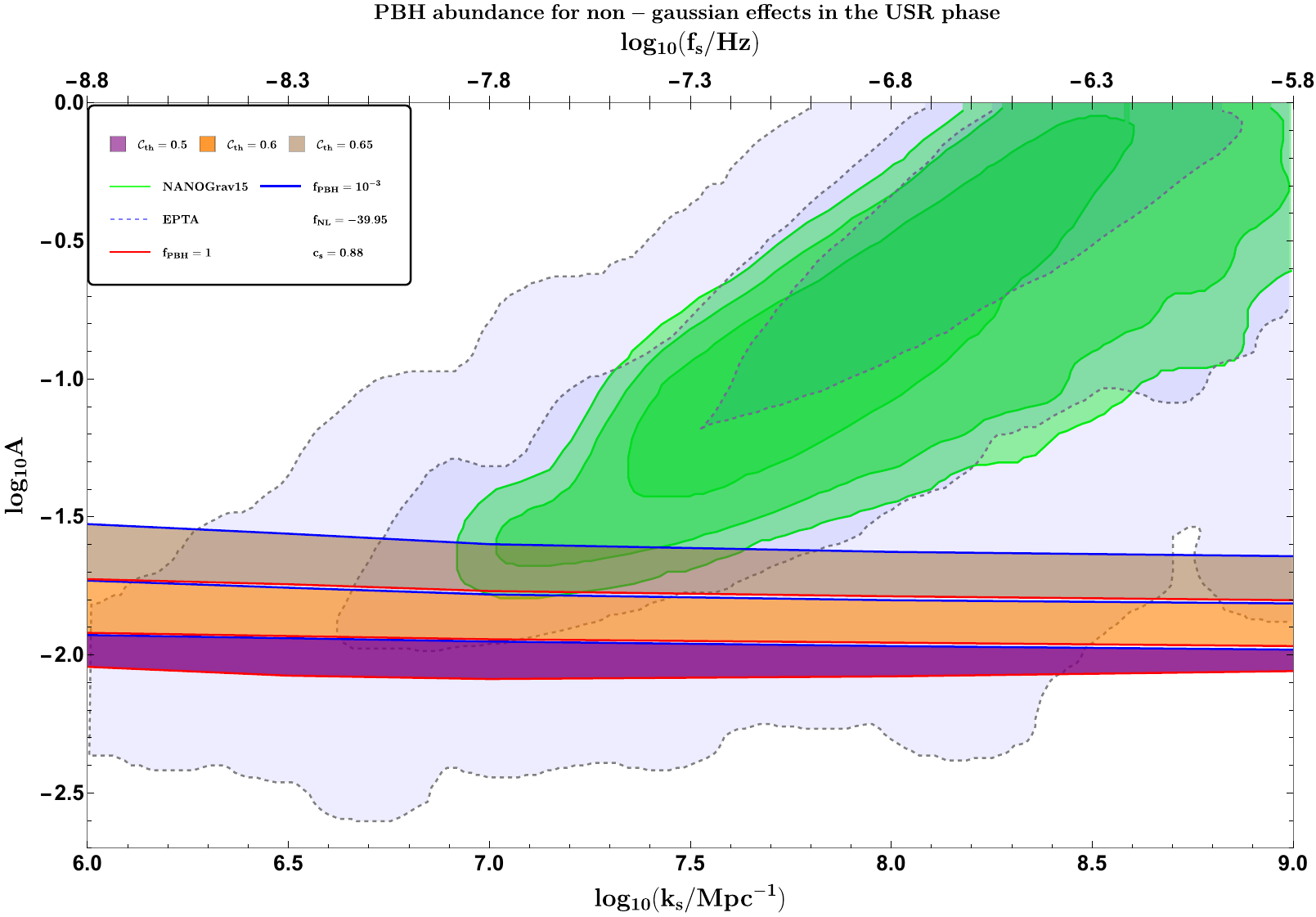}\label{fnl39cs0.88cth}
    }
     \subfigure[]{
    \includegraphics[height=8cm,width=8.5cm]{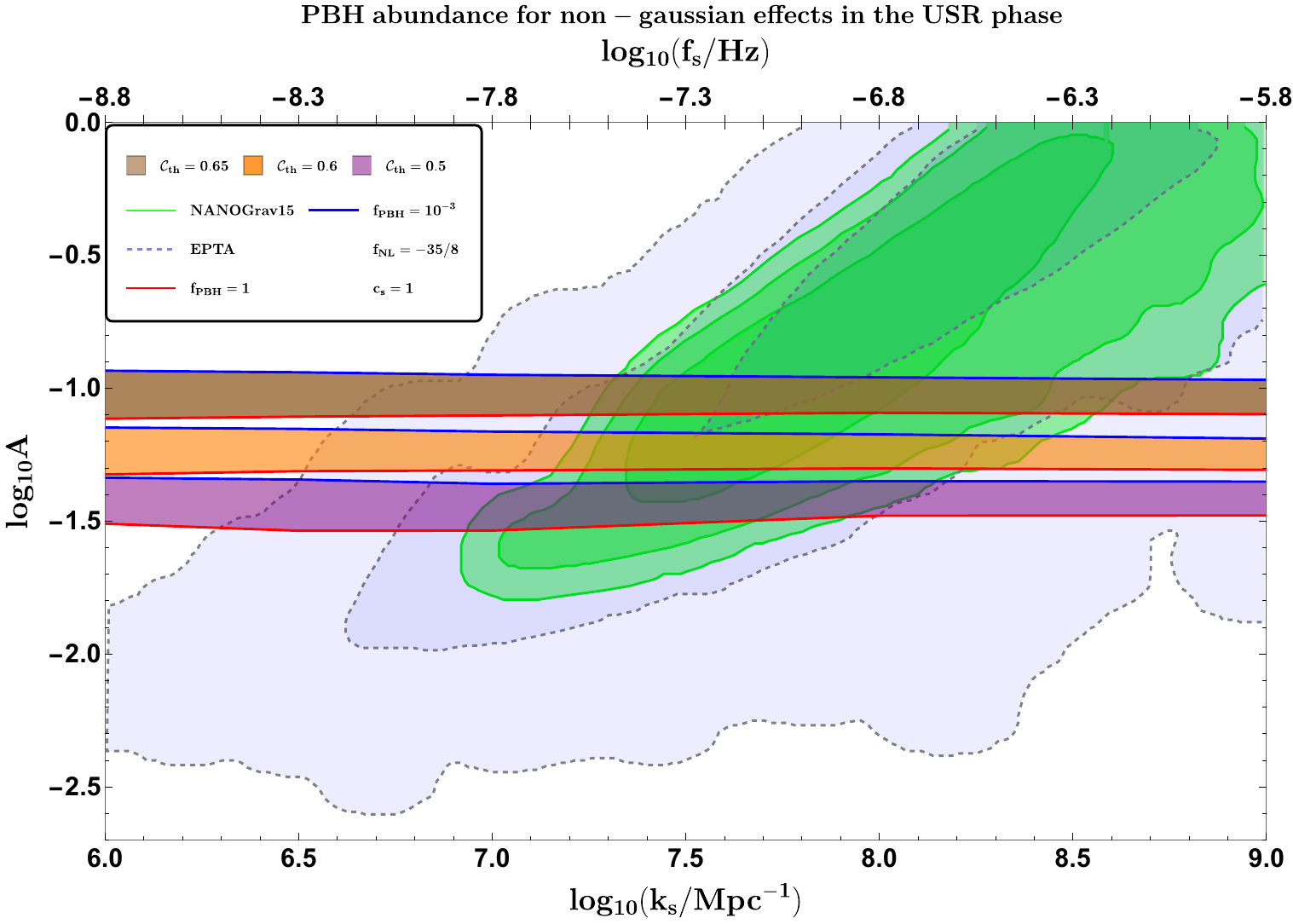}\label{fnl35by8cs1cth}
    }
     \subfigure[]{
    \includegraphics[height=8cm,width=8.5cm]{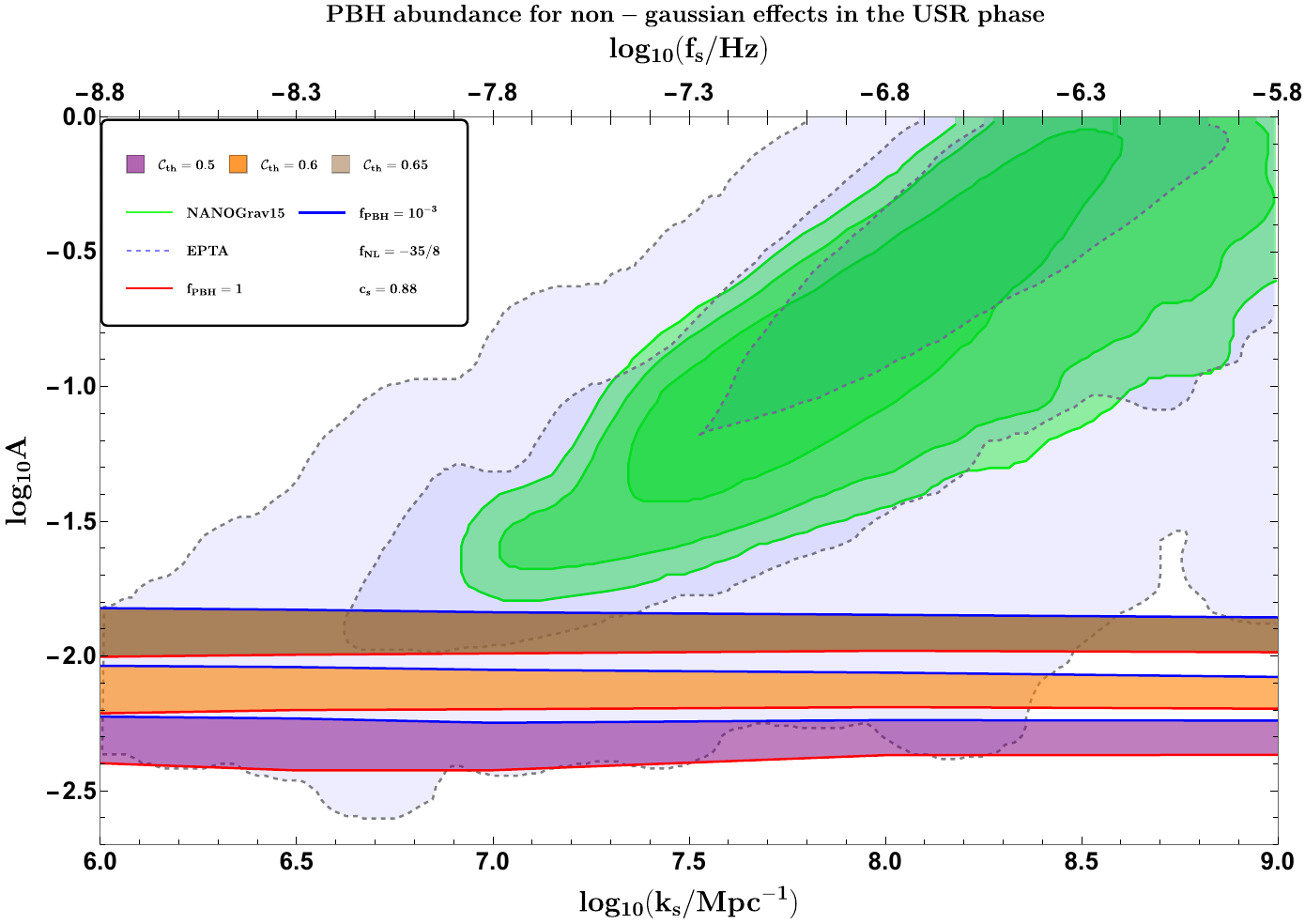}\label{fnl35by8cs0.88cth}
    }
    \caption{Variation in the USR peak amplitude $A$ of the regularized-renormalized-resummed power spectrum with the transition wavenumber $k_{s}$. The horizontal coloured bands highlight the region with abundance in the range $f_{\rm PBH} \in (10^{-3},1)$ relating to the distinct compaction thresholds, ${\cal C}_{\rm th}=0.65$ (light brown), ${\cal C}_{\rm th}=0.6$ (orange), ${\cal C}_{\rm th}=0.5$ (purple). The left plots of fig. \ref{fnl39cs1cth} and fig. \ref{fnl35by8cs1cth} utilize $c_s=1$,  while the right plots of fig. \ref{fnl39cs0.88cth} and fig. \ref{fnl35by8cs0.88cth} have $c_s=0.88$. In the top row $f_{\rm NL}=-39.395$ is fixed while in the bottom row it is $f_{\rm NL}=-38/5$. Both large negative PNG, $f_{\rm NL}=-39.395$, with greater sound speed, $c_{s}=1$, introduce a much significant difference in the peak amplitude than any other combination of the two values. The light blue and the green posteriors correspond to EPTA and the NANOGrav15 data, respectively, and underlay the bands to enable the SIGW interpretation \cite{Franciolini:2023pbf}.  }\label{NGfpbhA}
\end{figure*}

\begin{figure*}[htb!]
    \centering
    \subfigure[]{
    \includegraphics[height=8cm,width=8.5cm]{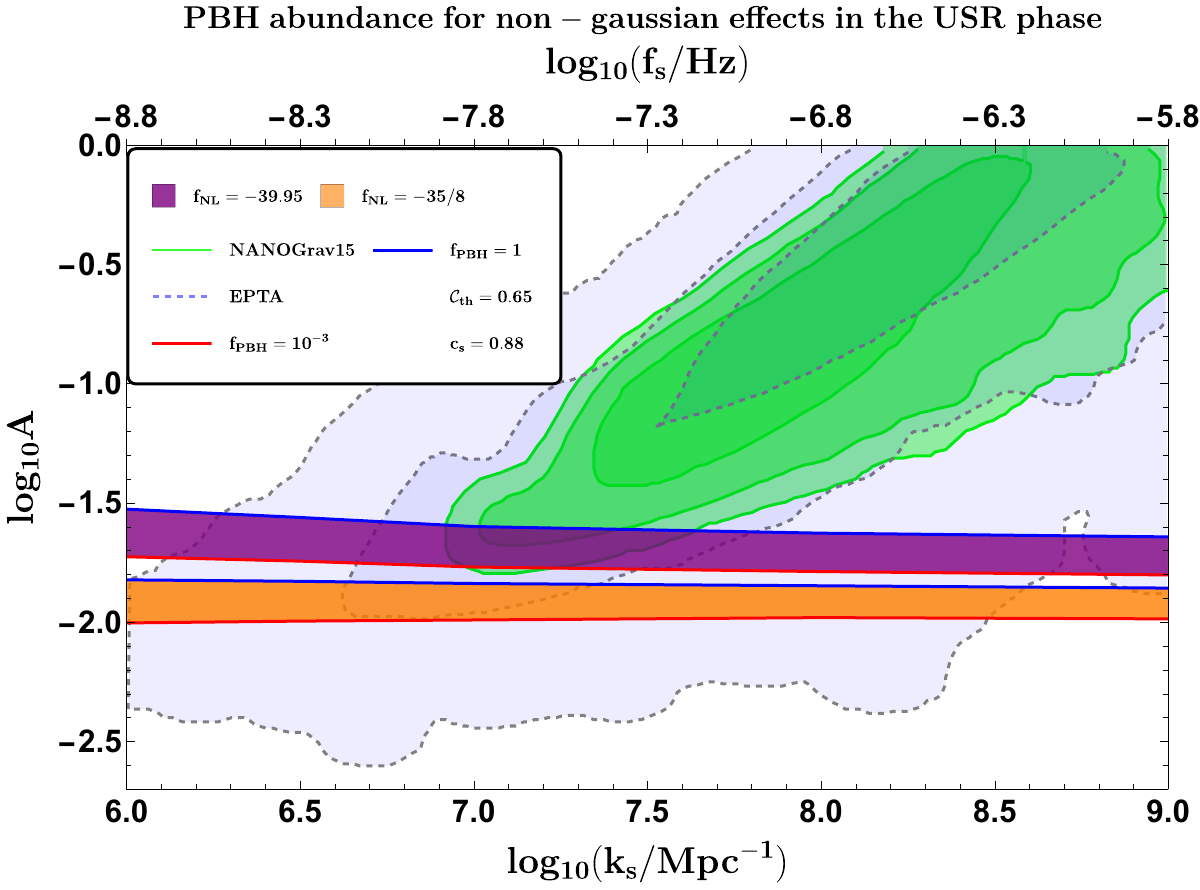}\label{Difffnlcs0.88cth0.65}
    }
    \subfigure[]{
    \includegraphics[height=8cm,width=8.5cm]{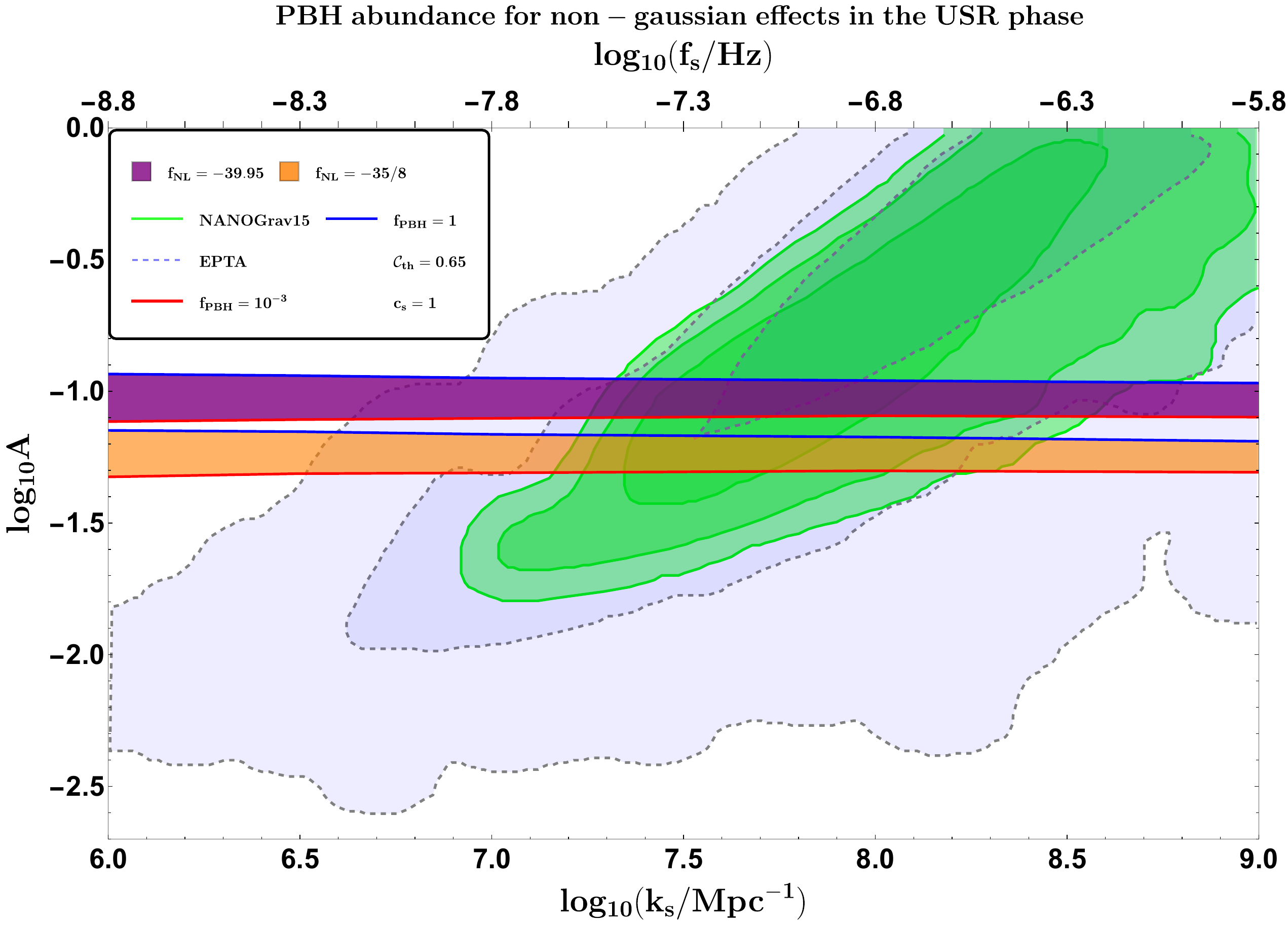}\label{Difffnlcs1cth0.65}
    }
    \caption{Plots demonstrating the impact of the effective sound speed parameter $c_s$ on the power spectrum amplitude $A$ in the USR phase. Both the figures contain comparisons between the two values of $f_{\rm NL}=(-39.95,-35/8)$ shown with blue and orange colours, respectively. In fig. \ref{Difffnlcs0.88cth0.65}, corresponding to $c_s=0.88$, the amplitude of both bands associated with two distinct $f_{\rm NL}$ parameters, is decisively excluded beyond the $1\sigma$ contours of the NANOGrav15 (green) and the EPTA (light blue) posteriors. In contrast, fig. \ref{Difffnlcs1cth0.65}, which employs $c_s=1$, provides a surge in the power spectra amplitude for both the bands and are well supported by the PTA data within the $1\sigma$ region. }\label{fnl358cs0.881}
\end{figure*}

\begin{figure*}[htb!]
    \centering
    \subfigure[]{
    \includegraphics[height=7.7cm,width=8.5cm]{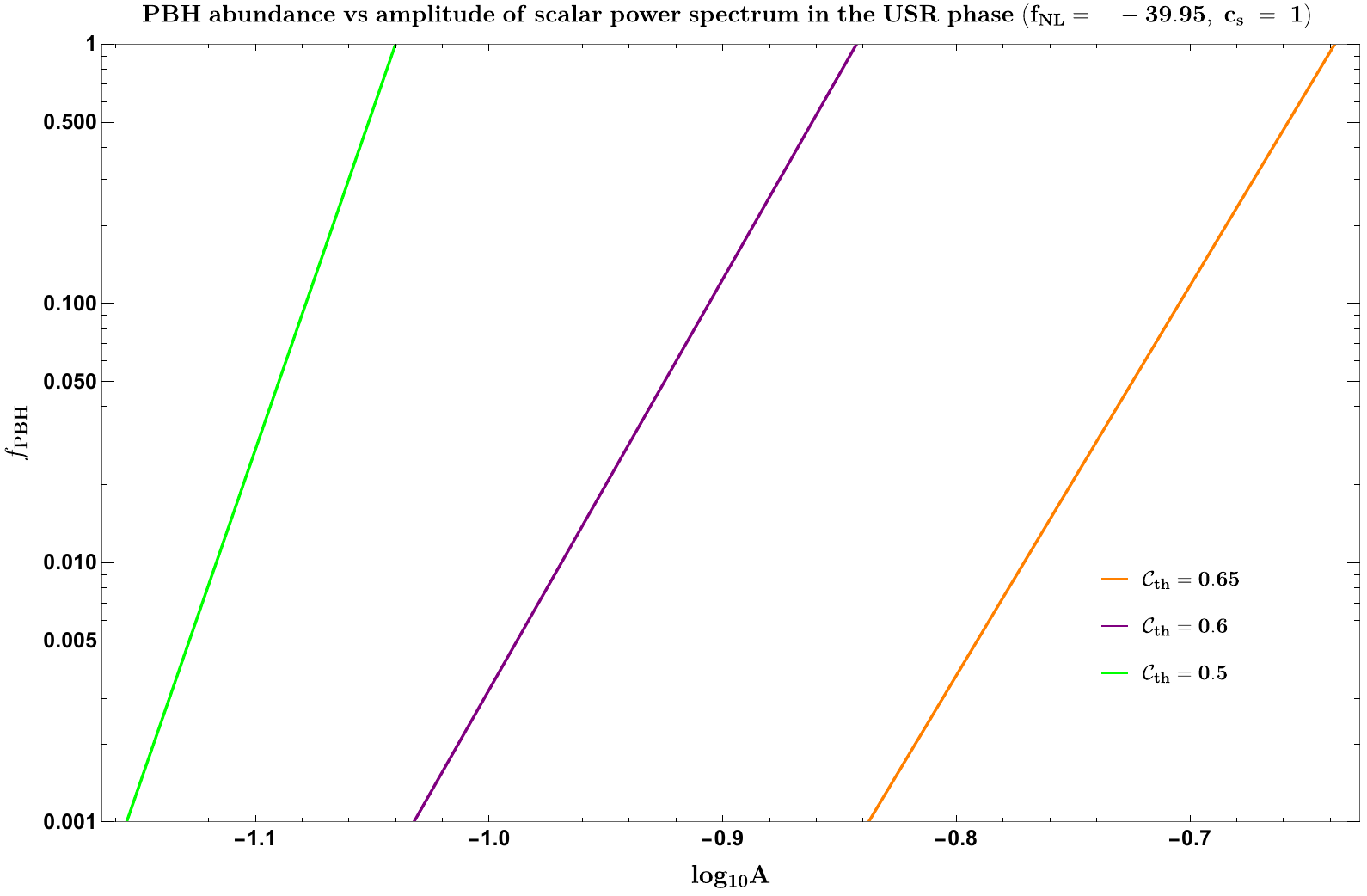}\label{fnl39cs1fPBH}
    }
    \subfigure[]{
    \includegraphics[height=7.7cm,width=8.5cm]{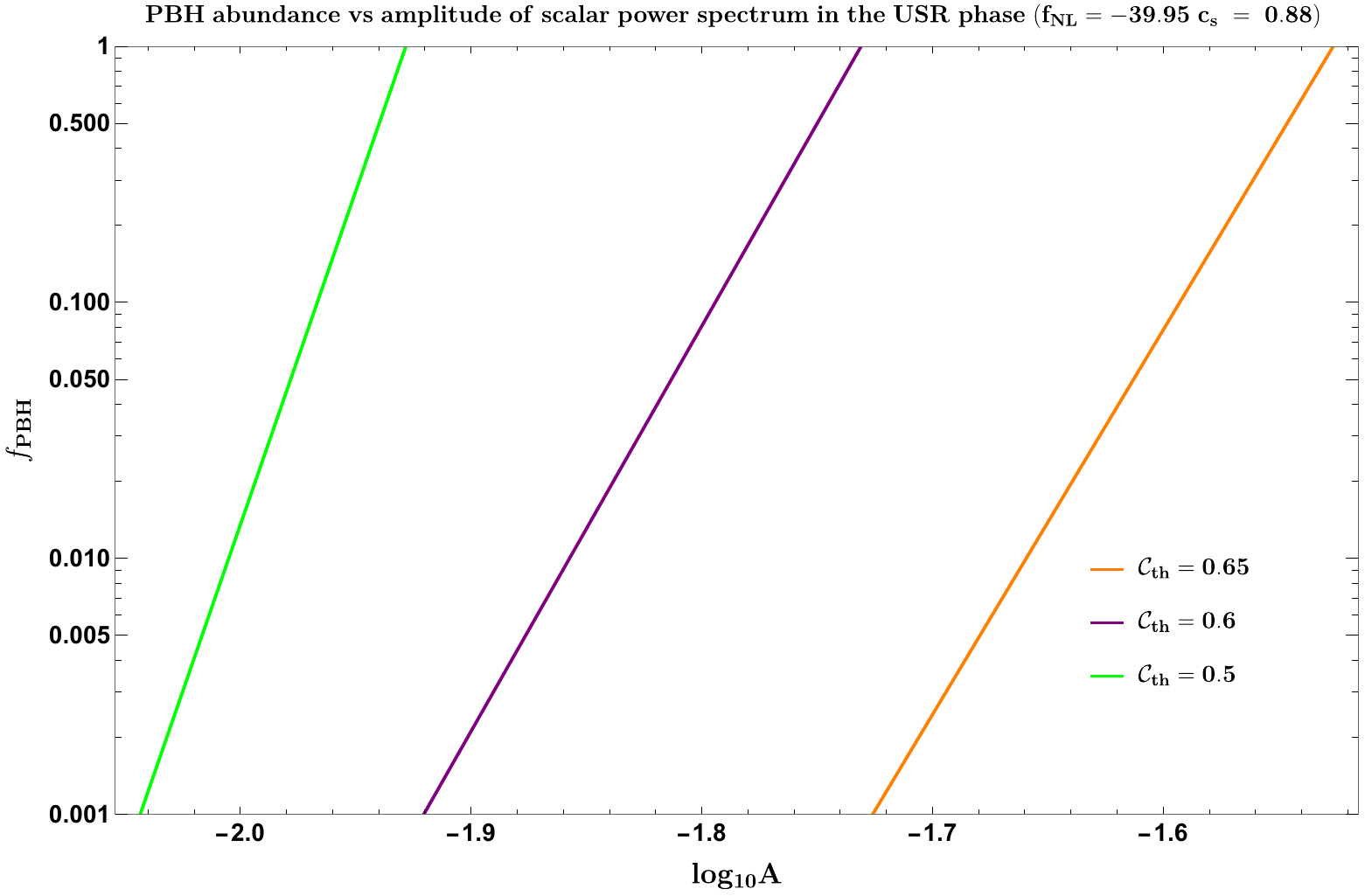}\label{fnl39cs0.88fPBH}
    }
    \subfigure[]{
    \includegraphics[height=7.7cm,width=8.5cm]{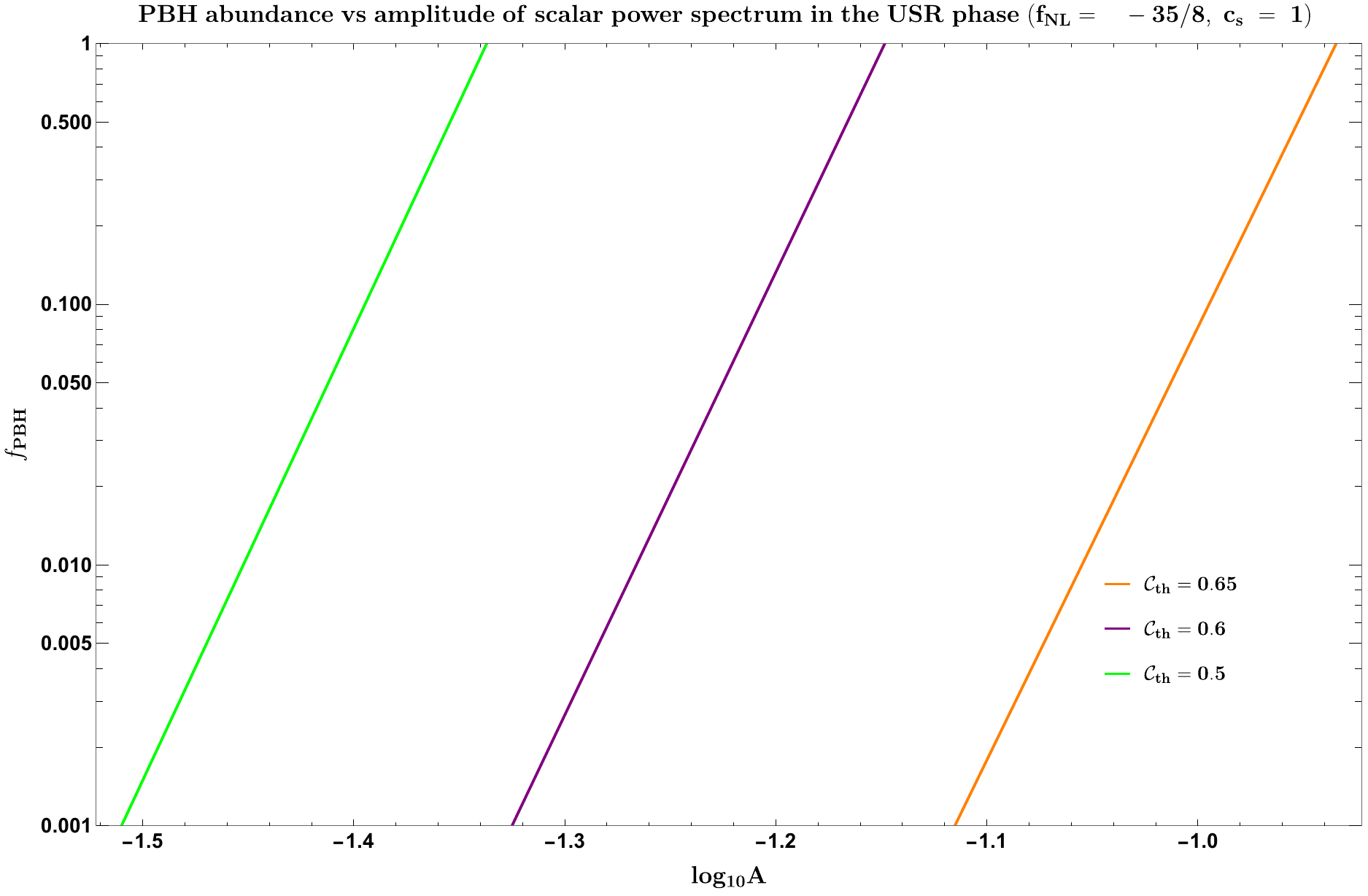}\label{fnl35by8cs1fPBH}
    }
    \subfigure[]{
    \includegraphics[height=7.7cm,width=8.5cm]{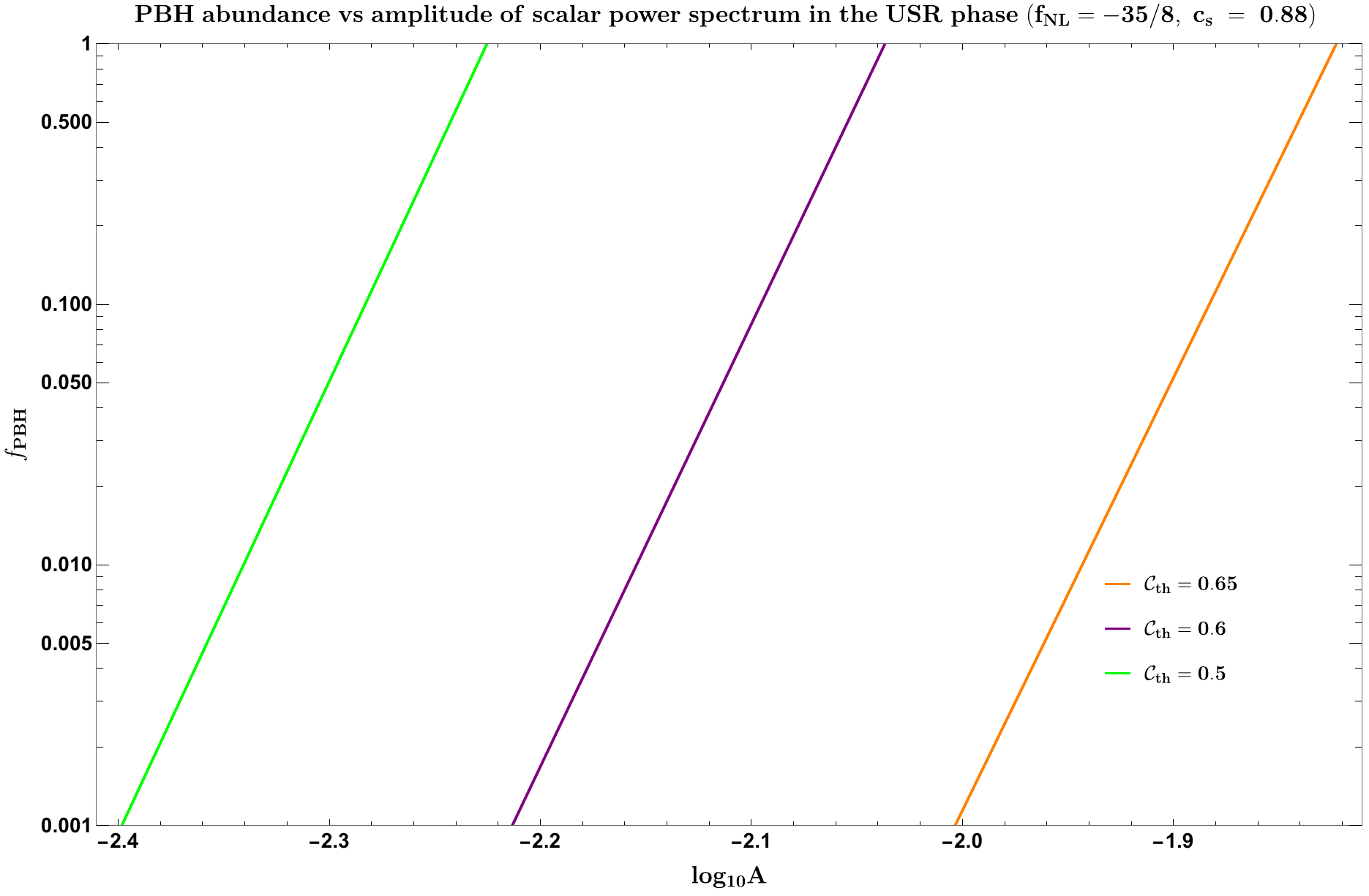}\label{fnl35by8cs0.88fPBH}
    }
    \caption{ The PBH fractional abundance $f_{\rm PBH}$ plotted against the change in peak amplitude $A$ of the regularized-renormalized-resummed scalar power spectrum from the EFT of bounce. The top-left (\ref{fnl39cs1fPBH}) and top-right (\ref{fnl39cs0.88fPBH}) figures exploits $f_{\rm NL}=-39.395$, with the effective sound speed $c_{s}=1$ and $c_{s}=0.88$, respectively. The bottom-left (\ref{fnl35by8cs1fPBH}) and bottom-right (\ref{fnl35by8cs0.88fPBH}) figures exploits $f_{\rm NL}=-35/8$, with $c_{s}=1$ and $c_{s}=0.88$, respectively. In each sub-figure, the impact of changing threshold ${\cal C}_{\rm th}$ between ${\cal C}_{\rm th}=0.65$ (orange), ${\cal C}_{\rm th}=0.6$ (purple), and ${\cal C}_{\rm th}=0.5$ (green), on the $f_{\rm PBH}$ is illustrated along with the transition number kept fixed at the same value $k_s=10^{7}{\rm Mpc^{-1}}$. Notice the sudden fall in the observed fraction of PBHs as the amplitude gets slightly altered for a given threshold. 
    }\label{PBHAlines}
\end{figure*}

We employ our EFT of bounce setup to derive the behaviour of the fractional GW energy density as a function of the frequency, with the results shown in figure \ref{csSIGW}. In Fig. \ref{DiffcsNG}, the spectrum crosses the NANOGrav15 signal, and in Fig. \ref{DiffcsEPTA}, it crosses the EPTA signal. Both images illustrates that the generated spectrum covers the frequency bins probed by the PTA observations. Following this, the spectrum rapidly drops to zero due exponential damping effect inherent in the quantum-loop corrected power spectrum of Eqn. (\ref{RRRspectrum}). In both the figures, we choose to highlight the difference coming from two different sound speed $(c_{s}=1,0.88)$ values which is motivated by the previous results on the total loop corrected power spectrum by the authors in \cite{Choudhury:2024dei}. After crossing the observed signals, the GW spectrum with $c_s=1$ exhibits larger amplitude by an order of magnitude relative to the spectrum with the choice of $c_s=0.88$. This analysis suggests that $c_s=1$ proves to be a more preferable choice to achieve a significant signal from SIGWs and its implication will also be reflected in our upcoming analysis of PBH formation.

\subsection{PBH Overproduction}
\label{s6c}

In this study, we specifically focus our analysis on two principal values of the NG parameter, namely $f_{\rm NL} \in (-39.95,-35/8)$, justifications as to why have been detailed before in Sec. \ref{s6c}. Having established that, we begin our discussion with the figure (\ref{2dpdffnl}), which depicts the 2D joint PDF with a uniform correlation between the variables ${\cal C}_{G}$ and ${\cal \zeta}_{G}$ for a spectrum of horizon masses in increasing order from left to right. The analysis has been carried out with EoS $w=1/3$, effective sound speed $c_s=1$, a fixed power spectrum amplitude $A=10^{-2}$ and two aforementioned values of the NG parameter, $f_{\rm NL}=-39.95$, and $f_{\rm NL}=-35/8$, as represented by figure \ref{fnl39cs1} and figure \ref{fnl35by8cs1} respectively. A noteworthy feature here is the non-uniqueness of the amplitude chosen. $A=10^{-2}$ denotes a baseline value beneath which the PBH abundance for all masses is significantly suppressed, rendering it unsuitable for meaningful analysis. This value is merely a critical minimum necessary to generate a substantial abundance of PBHs. Performing analysis with higher values of $A$ will also provide desirable results, provided the perturbation theory is kept intact. Furthermore, the amplitude value is chosen autonomously, without direct correspondence with the Gaussian variables ${\cal C}_{G}$, and $\zeta_{G}$. $A$ is anchored at $10^{-2}$ across all generated plots. Nevertheless, the PDF expression encapsulates $A$ through the different covariance elements. By modulating the Gaussian variables $({\cal C}_{G},\zeta_{G})$ for a particular $A$, we derive our PDF contour. The integration domains are permitted as per the Eqs. (\ref{integraldomain}-\ref{CGcriticalrange}) (details of which have been provided in the Sec. \ref{s4b2}), and delineated by the cyan bands in figure (\ref{2dpdffnl}) with different separation boundaries corresponding to distinct thresholds of the compaction function, namely ${\cal C}_{\rm th} \in (0.4,0.5,0.6)$. These benchmark thresholds reside within the interval $[2/5,2/3]$, that is obtained after not fully considering the non-linear effects to the perturbation profile at super-horizon scales in the gradient-expansion approach \cite{Musco:2020jjb}. 

Moving forward, it is crucial to emphasize that we follow the results of \cite{Musco:2020jjb}, where the entire analysis to determine the threshold ${\cal C}_{\rm th}$ has been performed with consideration of the NGs arising from the non-linear relationship between the density contrast and the curvature perturbation $\zeta$. However, PNGs have not been taken into account for calculating the threshold values. While several studies have examined the impact of PNGs on these threshold values \cite{Kehagias:2019eil, Escriva:2022pnz}, no deviations, from the linear case result, with a significant impact of greater than an order of percent level have been recorded. Nonetheless, incorporating PNGs in estimating the threshold itself is beyond the scope of this paper's current analysis. 

With this consideration in mind, we return to the analysis of figure (\ref{2dpdffnl}). Notably, all ${\cal C}_{\rm th}$ values share one common boundary, denoted by the dotted black line. The iso-contour lines for $f_{\rm NL}=-39.95$ (figure \ref{fnl39cs1}) show a pronounced overlap of the domain with the PDF for lower masses, indicating a greater probability of obtaining a sizeable abundance for these lower mass PBHs than for the higher mass. A parallel trend is evident for $f_{\rm NL}=-35/8$ (figure \ref{fnl35by8cs1}), where the domain gets drawn inwards for the low masses. However, a direct comparison between the two $f_{\rm NL}$ values reveals that the contour lines for $f_{\rm NL}=-39.95$ exhibit a greater compatibility with the domain within the distribution. This implies that $f_{\rm NL}=-39.95$ is more favourable for the scenario of ample PBHs production as compared to $f_{\rm NL}=-35/8$. 
The above observation is further corroborated by the illustrations in the plots of figure (\ref{NGfpbhA}), which demonstrates the influence of NGs on the power spectrum, which varies along with the transition wavenumber $k_s$ in the USR region, necessary to obtain a suitable fractional abundance range and subsequently avoid overproduction. The results are correlated with the PTA data to find observational constraints. The various coloured bands represent the PBHs abundance in the range $f_{\rm PBH} \in (10^{-3}-1)$ corresponding to thresholds ${\cal C}_{\rm th} \in (0.5,0.6,0.65)$. Before diving into the discussion of the figure, we like to present a detail about the suitability of our forthcoming analysis with the NANOGrav15 and EPTA posteriors. A careful examination will display a resemblance of our chosen scalar power spectrum with the log-normal power spectrum. This is indebted to the presence of a USR phase with a sharp peak. This resemblance justifies our use of the PTA log-normal-based posteriors for analysis.  Additionally, the SIGW spectrum displayed in fig.(\ref{csSIGW}) also exhibits a causality tail in the IR region, which is a distinctive feature when using the log-normal power spectrum. This further reinforces the idea that separate probability analysis is not required in our case to develop posteriors for deriving our outcomes.

Now, in figure \ref{fnl39cs1cth} with $f_{\rm NL}=-39.95$, and $c_s=1$, we observe that all the bands corresponding to the threshold values yield substantial abundance with amplitudes in the range of $A \in (10^{-1}-10^{-0.5})$. Consequently, they effectively circumvent the overproduction problem by residing within the $1\sigma$ region of the NANOGrav15 posteriors. Conversely, reducing the magnitude of the negative $f_{\rm NL}$ to $-35/8$, and keeping $c_s=1$, would alter the results as shown in figure \ref{fnl35by8cs1cth}. Here for the threshold ${\cal C}_{\rm th}=0.65$ (brown band), both the upper and lower limits of $f_{\rm PBH}$ fall within the $1\sigma$ region of the NANOGrav15 and EPTA contours, while for ${\cal C}_{\rm th}=0.6$ (orange band), the upper value of $f_{\rm PBH}$ merely grazes from outside of the $1\sigma$ contour of the EPTA. Additionally, for ${\cal C}_{\rm th}=0.5$ (purple band) in figure \ref{fnl35by8cs1cth}, the result is further constrained by being excluded over the $1\sigma$ contour of the EPTA and barely grazes the $1\sigma$ region of NANOGrav15. This observation aligns well with theoretical expectations since the perturbation amplitude is intricately linked with the compaction threshold, and a higher ${\cal C}_{\rm th}$ within the linear regime necessitates a greater amplitude to generate the same fractional abundance as compared to lower values of the same. A similar trend is visible if we take $c_s=0.88$ and examine their results from the figs. (\ref{fnl39cs0.88cth}, \ref{fnl35by8cs0.88cth}) for the two distinct $f_{\rm NL}$ cases. However, if we instead draw a direct comparison to inspect the impact of the effective sound speed $c_s$ on the amplitude, we can easily witness a significant difference separately captured in figure \ref{fnl358cs0.881}. 

The left \ref{Difffnlcs0.88cth0.65} and right \ref{Difffnlcs1cth0.65} figures represent the variation of the power spectra amplitude with $k_s$ for $c_s=0.88$, and $c_s=1$, respectively. The plots have been realized with ${\cal C}_{\rm th}=0.65$. Figure \ref{Difffnlcs0.88cth0.65} shows both the bands, i.e. the orange for $f_{\rm NL}=-39.95$, and purple for $f_{\rm NL}=-35/8$ remain largely excluded over the $1\sigma$ region of the PTA contours. Although the orange band circumvents overproduction and marginally favours the $2\sigma$ region of the NANOGrav15 contour, the scenario with $f_{\rm NL}=-35/8$ fails to generate sufficient amplitude to achieve a comparable abundance while simultaneously remaining consistent with the NANOGrav15. Attempts to amplify the amplitude to conform to the SIGW interpretation of the NANOGrav15 data would bring the risk of overproducing PBHs. Now, in contrast, figure \ref{Difffnlcs1cth0.65} temptingly highlights a dramatic surge in the amplitude for both the $f_{\rm NL}$ values. This remarkable shift prompted by the alteration of $c_s$ underscores the sensitivity of the setup to the effective sound speed parameter. Lowering $c_s$ from $1$ to $0.88$ drops the power spectrum amplitude to $A=10^{-3}$, an order of magnitude less when compared to $c_s=1$. This reduction in amplitude is consequential because it directly impacts the conditions requisite for PBH formation, as amplitude is integrated into the $f_{\rm PBH}$ through the Eqn. (\ref{pbhabundance}). The likelihood of PBH production drops further, drastically dropping for choices below $c_s=0.88$, making such scenarios irrelevant for analysis. Moreover, restraining to $c_s \leq 1$ is essential to uphold causality constraints. Exceeding $c_s=1$ would jeopardize the perturbativity conditions and therefore can be ruled out.

Additionally, we notice the appearance of features in the width of the bands in the figure (\ref{NGfpbhA}). Notably, they are not uniform across the transition wavenumbers $k_s$. These bands are seen to exhibit a greater thickness near $k_s\sim {\cal O}(10^6{\rm  Mpc^{-1}})$ and become narrower toward the right side of the plots. Note that the transition wavenumber is linked with the PBHs mass by Eqn. (\ref{HMass}). This is indicative of the idea that with increased wavenumber, PBHs with lower masses are poised to produce substantial abundance with a slightly lower amplitude as compared to those with higher masses. 
 
We now examine the results for the fractional abundance of PBHs represented by the figure (\ref{PBHAlines}). The three inclined coloured lines on each of these plots correspond to distinct compaction thresholds ${\cal C}_{\rm th}=0.65$ (orange), ${\cal C}_{\rm th}=0.6$ (purple), and ${\cal C}_{\rm th}=0.5$ (green). These lines show the variation of the PBH abundance with the regularized-renormalized-resummed power spectrum amplitude for the two principal values of $f_{\rm NL}$ and $c_s$, recurrently mentioned in this paper. The transition wavenumber is kept fixed here at $k_{s}=10^{7}{\rm Mpc^{-1}}$. As clearly evident from the graphs, opting for a higher value of ${\cal C}_{\rm th}$ calls for a larger amplitude across all the generated plots, and vice versa. Further, to obtain a suitable $f_{\rm PBH}$ within the range $(10^{-3}, 1)$ with $(f_{\rm NL}, c_s) = (-39.95,1)$, we require a peak amplitude $A\sim {\cal O}(0.1)$. The abundance is extremely susceptible to the change in peak amplitude irrespective of the chosen threshold condition and any specific combination of the two $f_{\rm NL}$ and $c_s$ values. The scenario with $f_{\rm NL}=-35/8$ and $c_s=0.88$ (non-canonical) leads to the minimal amount of peak amplitude, while the negatively large  $f_{\rm NL}=-39.95$ and $c_s=1$ (canonical) scenario leads to a huge amplification of $A$ that comes extremely close to breaching the perturbativity conditions. Now, having presented the results of our setup, let us look at how it compares to alternative frameworks and models previously studied to address the recent crucial overproduction problem.

\section{Comparative analysis of previous studies}
\label{s7}

This section aims to address previous analyses conducted in line with studying the overproduction of PBHs and compares them through their efficiency and limitations in contrast to the approach adopted for this work and its results. Quite recently, since the definitive announcement of an SGWB signal via the PTA collaborations, several groups have highlighted a dramatic issue concerning the abundant production of PBHs. This issue directly relates to the SIGWs being a leading interpretation of the PTA signal. Upon confrontation with the observed signal, it suggests a highly amplified scalar power spectrum, which implies that the PBHs can be overproduced. Below we discuss some of the major advances specific to this problem and supply the table\footnote{Note that the table serves to display a certain parameters of few chosen models and no attempt has been made to present a comprehensive list of all the models studied so far.}\ref{tab1:comparison} summarizing the same.

\begin{table}[ht]
\centering
\renewcommand{\arraystretch}{4}
\setlength{\tabcolsep}{11pt} 
\begin{tabular}{|>{\columncolor{pink!40}}c|>{\centering\arraybackslash}p{2.0cm}|>{\centering\arraybackslash}p{2.0cm}|>{\centering\arraybackslash}p{2.0cm}|>{\centering\arraybackslash}p{2.0cm}|>{\centering\arraybackslash}p{2.0cm}|>
{\centering\arraybackslash}p{2.0cm}|}
\specialrule{1.5pt}{0pt}{0pt} 
\hline
\textbf{Parameter} & \cellcolor{green!20} \textbf{Ekpyrotic Bounce-USR} & \cellcolor{green!20} \textbf{Matter Bounce-USR} &  \cellcolor{green!20} \textbf{Curvaton \cite{Franciolini:2023pbf}} & \cellcolor{green!20} \textbf{Galileon \cite{Choudhury:2023kdb,Choudhury:2023hfm,Choudhury:2023fwk,Choudhury:2024one}} & \cellcolor{green!20} \textbf{MST \cite{Choudhury:2023fjs,Bhattacharya:2023ysp}} \\ 
\hline
\textbf{f$_{\rm \bf NL}$} & ${\cal O}(-39.95)$ & ${\cal O}(-35/8)$ & ${\cal O}(-2)$ & ${\cal O}(-6)$ & NYS \\
\hline
\textbf{SIGW} & Consistent with NANOGrav$15$ and EPTA & Consistent with NANOGrav$15$ and EPTA & Consistent with NANOGrav$15$ and EPTA  & Consistent with NANOGrav$15$ and EPTA & Consistent with NANOGrav$15$ and EPTA \\
\hline
\textbf{\hspace{-3mm} Over-production \hspace{-3mm}} & Sharply within $1\sigma$ of NANOGrav$15$ and EPTA contours & Barely touches $1\sigma$ of EPTA contours & Within $2\sigma$ of NANOGrav$15$ and EPTA contours & Just enters inside the $1\sigma$ of NANOGrav$15$ & Within $2\sigma$ of EPTA and just touches the $2\sigma$ of NANOGrav$15$  \\
\hline
\textbf{PBHs Mass ($M_{\odot}$)} & $(10^{-6}-0.1)$ & $(10^{-6}-0.1)$ & $(10^{-6}-0.01)$ & $(10^{-6}-0.1)$ & $(10^{-31}-10^4)$ \\
\hline
\textbf{EoS parameter} & $(0.31,1/3)$ & $(0.31,1/3)$ & $1/3$ & $(0.24,1/3)$ & ($-0.01,1/3$) \\
\hline
\textbf{QCD effects} & NSY & NSY & Studied & Studied & NSY \\
\hline
\textbf{Quantum loop effects} & $\bullet$ Consistent perturbative treatment \quad\quad $\bullet$ Evades the no-go bound of $M_{\rm PBH}\sim {\cal O}(10^{2}{\rm gms})$ & $\bullet$ Consistent perturbative treatment \quad\quad $\bullet$ Evades the no-go bound of $M_{\rm PBH}\sim {\cal O}(10^{2}{\rm gms})$ & NSY & $\bullet$ Consistent perturbative treatment \quad\quad $\bullet$ Evades the no-go bound of $M_{\rm PBH}\sim {\cal O}(10^{2}{\rm gms})$ & $\bullet$ Consistent perturbative treatment \quad\quad $\bullet$ Evades the no-go bound of $M_{\rm PBH}\sim {\cal O}(10^{2}{\rm gms})$ \\
\hline
\specialrule{1.5pt}{0pt}{0pt} 
\end{tabular}

\caption{An overview on the results from studies on various models associated with the formation of curvature perturbations in the early universe. The models are compared based on various parameters that they have addressed. The abbreviation "NSY" indicates "Not Studied Yet" to denote parameters that have not been explored as of yet within a given model. Different PBH mass ranges investigated by the models have been expressed in the solar mass $M_{\odot}$ units. The values quoted for EoS parameter $w$ does not denote the best favourable outcomes of analysis, but rather the range of values utilized for investigation.}
\label{tab1:comparison}

\end{table}

Among the many attempts to avoid this unfortunate outcome, some include the presence of an additional field that does not back-reacts to the dynamics of the background metric perturbations (a spectator field) like the curvaton model \cite{Ferrante:2023bgz,Franciolini:2023pbf,Gow:2023zzp}. Much like the model discussed in this paper, the curvaton scenario in \cite{Inomata:2023drn} also invites NGs into the PDF of $\zeta$ which is brought into effect by expansion of the curvature perturbation in terms of the leading order NG parameter $f_{\rm NL}$.  In \cite{Inomata:2023drn}, using small and negative NGs, $f_{\rm NL}\sim {\cal O}(-1)$, the curvaton model is shown to provide a significant decrease in the abundance $f_{\rm PBH}$ thus avoiding overproduction while simultaneously explaining the SIGW from the PTA data. It is also able to generate large, solar mass, PBHs with $M_{\rm PBH} \sim {\cal}{O}(0.1-1)M_{\odot}$. Similarly, the analysis with curvaton in \cite{Ferrante:2023bgz} shows that $f_{\rm NL}\sim{\cal O}(-1)$ generates the best possible outcomes for PBH formation while again conforming to the data from experiments (NANOGrav$15$ and EPTA). It was shown that by assuming a broken-power law power spectrum, and after using the similar quadratic curvature perturbation ansatz, the power spectrum peaked near $f_{\rm NL} =-2$ and on diving further below this value caused a substantial decrement in the amplitude required for PBHs formation. A similar effect has also been displayed considering positive $f_{\rm NL}$. Another notable example from these attempts involves using a spectator tensor field in \cite{Gorji:2023sil}, where the authors propose that it is this particular field responsible for the generation of the observed PTA signal and contributes sub-dominantly to the actual energy density fraction that is finally shared by the PBHs. 

Using the equation of state parameter $(w)$, right before the RD era following inflation, is also shown as favorable by the authors in \cite{Choudhury:2023fjs} to address the PBH overproduction issue when employed within the framework that contains multiple sharp transitions (MSTs) across a wide range of frequencies covering the PTA sensitivity and beyond. The authors here consider an EFT set-up with multiple SR-USR-SR phases in succession with a sufficient number of e-foldings $\Delta {\cal N} \sim (60-70)$ to complete inflation. This set-up is shown to generate a wide range of PBH masses ranging from sub-solar to beyond solar mass. Moreover, the SIGW spectrum generated from the one-loop corrected power spectrum also conforms to the NANOGrav15 data and other experimental data from LISA \cite{LISA:2017pwj}, DECIGO \cite{Kawamura:2011zz}, Cosmic explorer (CE) \cite{Reitze:2019iox}, HLVK (LIGO, VIRGO, and KAGRA) \cite{LIGOScientific:2014pky, VIRGO:2014yos, KAGRA:2018plz}, Einstein telescope (ET) \cite{Punturo:2010zz}, BBO \cite{Crowder:2005nr}. In \cite{Choudhury:2023fjs}, after integrating the effect of varying $w$ within the MST set-up, a major conclusion drawn was to demonstrate that $w=1/3$ produces the best possible results to derive all the aforementioned outcomes. However, the mentioned studies do not include the effect of consideration of either non-linearity between the density contrast and the curvature perturbation or the PNGs, which is yet to be studied. Numerous studies have focused on employing local PNGs, and it is this particular feature of the curvature perturbations that has been the most effective in doing away with PBH overproduction. Previously, in section \ref{s6a}, we elaborated on the impact of having this feature within a theory to study PBH production, and many of the latest studies in \cite{Franciolini:2023pbf,Firouzjahi:2023xke,Franciolini:2023wun,Wang:2023ost,Zhu:2023gmx,Li:2023xtl,Li:2023qua,Choudhury:2024one} illustrate this viewpoint thoroughly to examine the overproduction of PBHs. For the case of Galileon inflation, inferring from the results of \cite{Choudhury:2023fwk}, we observe that, with $f_{\rm NL}\sim {\cal O}(-6)$, it is able to cure the overproduction and strengthen the support for a SIGW explanation of the signal.

The emergence of QCD effects in the PTA signal stems from the sudden decrease in the energy and entropy degrees of freedom around the temperature, $T\gtrsim 100{\rm MeV}$, which aligns perfectly within the frequency regime probed by the PTA and thus the corresponding decrease in the equation of state can noticeably alter the low-frequency tail of the signal \cite{Franciolini:2023wjm}. Using such conditions for Galileon inflation, \cite{Choudhury:2024one} shows the softening of the IR tail of the GW spectrum to $f\simeq 5\times 10^{-9}{\rm Hz}$, that is, before we reach the peak of the spectrum. However, in the present study, QCD effects have not been taken into account for the formation of PBHs.

In the last row, we consider the case where quantum loop effects are considered while estimating the results on PBH production from a model of interest. In light of the \textit{no-go theorem} \cite{Choudhury:2023vuj, Choudhury:2023jlt, Choudhury:2023rks}, all the studies mentioned, except for the curvaton, come as examples of how to successfully evade the stringent bound, $M_{\rm PBH}\sim {\cal O}(10^{2}{\rm gms})$, as a consequence of this theorem.


The present work applies the same spirit of using the PNGs to our advantage but goes a step further by taking into account enormous values for the same, $f_{\rm NL}\ll {\cal O}(-10)$, compared to previous studies. This work also exploits the compaction function approach rather than the standard Press-Schechter formalism or the method from the peaks theory to study the PBH formation process. And finally, the underlying setup of this paper goes beyond the choice of working just within the inflationary picture that numerous other studies above prefer to address this issue, but also combines the features and consequences of having a non-singular bouncing scenario before the usual slow-roll commences. To conduct the analysis for the overproduction problem, with section \ref{s6c} devoted towards its end results, we utilize the theoretical set-up that is built upon the works \cite{Choudhury:2023vuj,Choudhury:2023jlt,Choudhury:2023rks,Choudhury:2024dei} and by integrating the aforementioned improvements we aim to achieve a stronger and broader understanding of curing the overproduction along with shedding some light in an upcoming section on the acceptable amounts of such type of NG.

\section{Confrontation with experiments}
\label{s8}

After conducting an elaborate discussion including various models and their relationship with multiple parameters encountered during PBH formation, we briefly discuss in this section their results when confronted with different gravitational wave experiments and, if possible, microlensing experiments for searching signatures of dark matter. The summary of these results can be seen in Table \ref{experimentstable}.

\begin{table}[ht]
\footnotesize
\centering
\renewcommand{\arraystretch}{2} 
\setlength{\tabcolsep}{15pt} 
\begin{tabular}{
|>{\columncolor{pink!40}}c
|>{\centering\arraybackslash}m{2cm}
|>{\centering\arraybackslash}m{1.8cm}
|>{\centering\arraybackslash}m{1.8cm}
|>{\centering\arraybackslash}m{1.8cm}
|>{\centering\arraybackslash}m{1.8cm}|}
\specialrule{1.5pt}{0pt}{0pt} 
\hline
\textbf{Experiment} & \cellcolor{green!20} \textbf{Ekpyrotic Bounce-USR} & \cellcolor{green!20} \textbf{Matter Bounce-USR} & \cellcolor{green!20} \textbf{Galileon} \cite{Choudhury:2023fwk,Choudhury:2023fjs} & \cellcolor{green!20} \textbf{Curvaton \cite{Ferrante:2023bgz,Gow:2023zzp,Franciolini:2023pbf}} & \cellcolor{green!20} \textbf{MST \cite{Bhattacharya:2023ysp}} \\
\hline 
\textbf{EPTA} & Consistent SIGW signal generation & Consistent SIGW signal generation & Consistent SIGW signal generation & Consistent SIGW signal generation & Consistent SIGW signal generation \\
\hline
\textbf{NANOGrav15} & Consistent SIGW signal generation & Consistent SIGW signal generation & Consistent SIGW signal generation & Consistent SIGW signal generation & Consistent SIGW signal generation \\
\hline
\textbf{LVK} & NSY & NSY & Similar strength of $\Omega_{\rm GW}h^{2}$ inside the LVK constraints as for the PTA & NSY & Uniform strength of  $\Omega_{\rm GW}h^{2}$ across all transitions into the LVK  \\
\hline
\textbf{Microlensing} & Possible to generate $\log_{10}{(M_{\rm PBH})}\sim {\cal O}(-6,-2)$ with $f_{\rm PBH}\in (10^{-3}-1)$ and consistent with the PTA  & Similar result to the Ekpyrotic section & Maximum possible value $f_{\rm PBH}\sim 10^{-1.6}$ for $M_{\rm PBH}\sim 10^{-2}M_{\odot}$ and $w=1/3$ after constraints from EROS/MACHO  & Generates asteroid mass PBHs with $f_{\rm PBH} \lesssim 1$, after constraints from SUBARU HSC. & Maximum possible value $f_{\rm PBH}\sim 10^{-1.7}$ for $M_{\rm PBH}\sim {\cal O}(10^{-4}M_{\odot}$) and $w=1/3$ after constraints from OGLE \\
\specialrule{1.5pt}{0pt}{0pt} 
\hline
\end{tabular}
\caption{Various experiments and their conclusions have been highlighted in the light of PBH formation, and SIGW interpretation. "NSY" denotes that the parameter is "Not Studied Yet" inside that specific model.} 
\label{experimentstable}
\end{table}

Among the list of experiments, we focus here on the PTA (NANOGrav and EPTA) observations that cover the nanohertz (nHz) and beyond frequency range of $f\sim {\cal O}(10^{-9}-10^{-7}){\rm Hz}$ and corresponds roughly to probing PBHs with masses, $M_{\rm PBH}\sim {\cal O}(10^{-4}-10^{-2})M_{\odot}$. Then there are the microlensing experiments, namely the Subaru HSC \cite{Niikura:2017zjd}, OGLE \cite{Niikura:2019kqi}, EROS/MACHO \cite{EROS-2:2006ryy}, that can provide tight constraints covering the PBH mass range, $M_{\rm PBH}\sim {\cal O}(10^{-10}-10^{-1})M_{\odot}$. And following this, at frequencies $f\sim {\cal O}(1-10^3){\rm Hz}$, we have constraints coming from the LIGO, VIRGO, and KAGRA (LVK) gravitational wave observatories that can probe PBH masses with $M_{\rm PBH}\sim {\cal O}(0.1)M_{\odot}$. We then collect results on PBH formation from various models as listed along the top row of the table \ref{experimentstable} and comment on them if any study utilizing that model concerns the regime probed by the above experiments. 

Beginning with the model focused on in the present work, our EFT treatment helps us to combine the contraction and bouncing scenarios, regardless of their kind, either ekpyrotic or matter, with the inflationary framework that contains a USR phase for PBH generation. In the figure (\ref{csSIGW}), we show that under the choice of $w=1/3$ for the RD era, and either choice of $c_s$, the resulting SIGW spectrum follows consistently both the PTA (NANOGrav$15$ and EPTA) signals in their respective frequency regimes. This conclusion does not alter whether choosing the ekpyrotic case $(\epsilon_{c/b}>3)$ or the matter case $(\epsilon_{c/b}=3/2)$ before inflation, thereby giving us a positive signature upon confrontation with the PTA experiments. A dedicated analysis focusing on the high-frequency LVK regimes is yet to be performed for this model. For the case of Galileon inflation, the analysis in \cite{Choudhury:2023hfm} suggests the conclusive feature of the resulting SIGW spectrum with the PTA signals. Due to the benefits coming from the non-renormalization theorem, as elaborated in \cite{Choudhury:2023hvf, Choudhury:2023hfm}, the spectrum can provide enough signature inside the frequency regime probed by the LVK observations while preserving its similar features as in the low-frequency PTA regime. While in \cite{Gow:2023zzp}, asteroid mass PBHs with $M_{\rm PBH} \sim {\cal O}({10^{-16}-10^{-10}})M_{\odot}$, probed by SUBARU HSC \cite{Niikura:2017zjd}, and produced in the curvaton models can explain the totality of dark matter depending on the duration of the curvaton decay, the curvaton mass and Hubble scale at the end of inflation. In the case of having multiple sharp transitions \cite{Bhattacharya:2023ysp}, the SIGW spectrum, once generated, has the feature, owning to the setup, to provide for a signature consistent with both the observed nHz regime of the PTA signal and the high-frequency regime probed by the LVK.

\begin{figure*}[htb!]
    \centering
    \subfigure[]{
    \includegraphics[height=8cm,width=8.5cm]{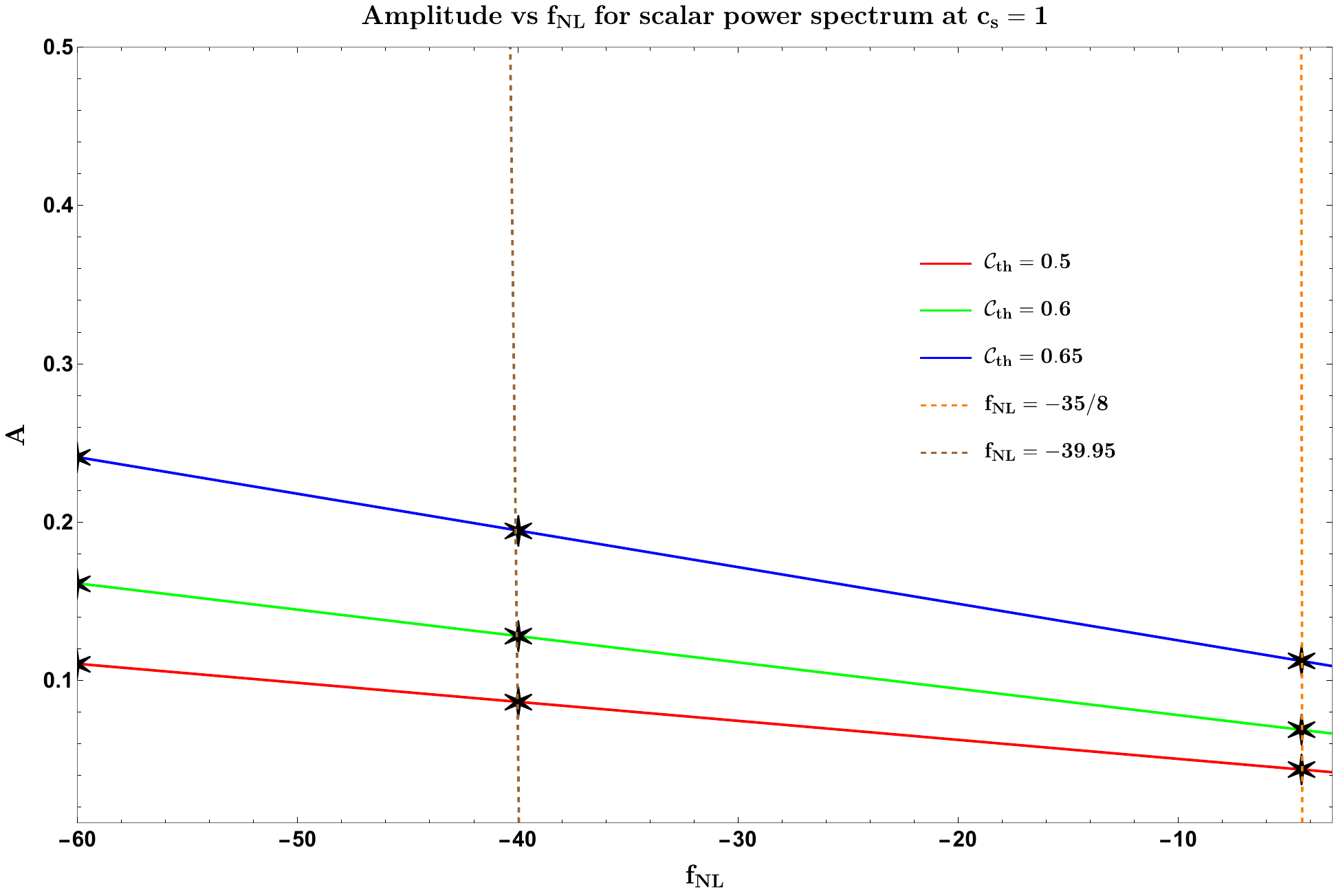}\label{ampfnlcs1}
    }
    \subfigure[]{
    \includegraphics[height=8cm,width=8.5cm]{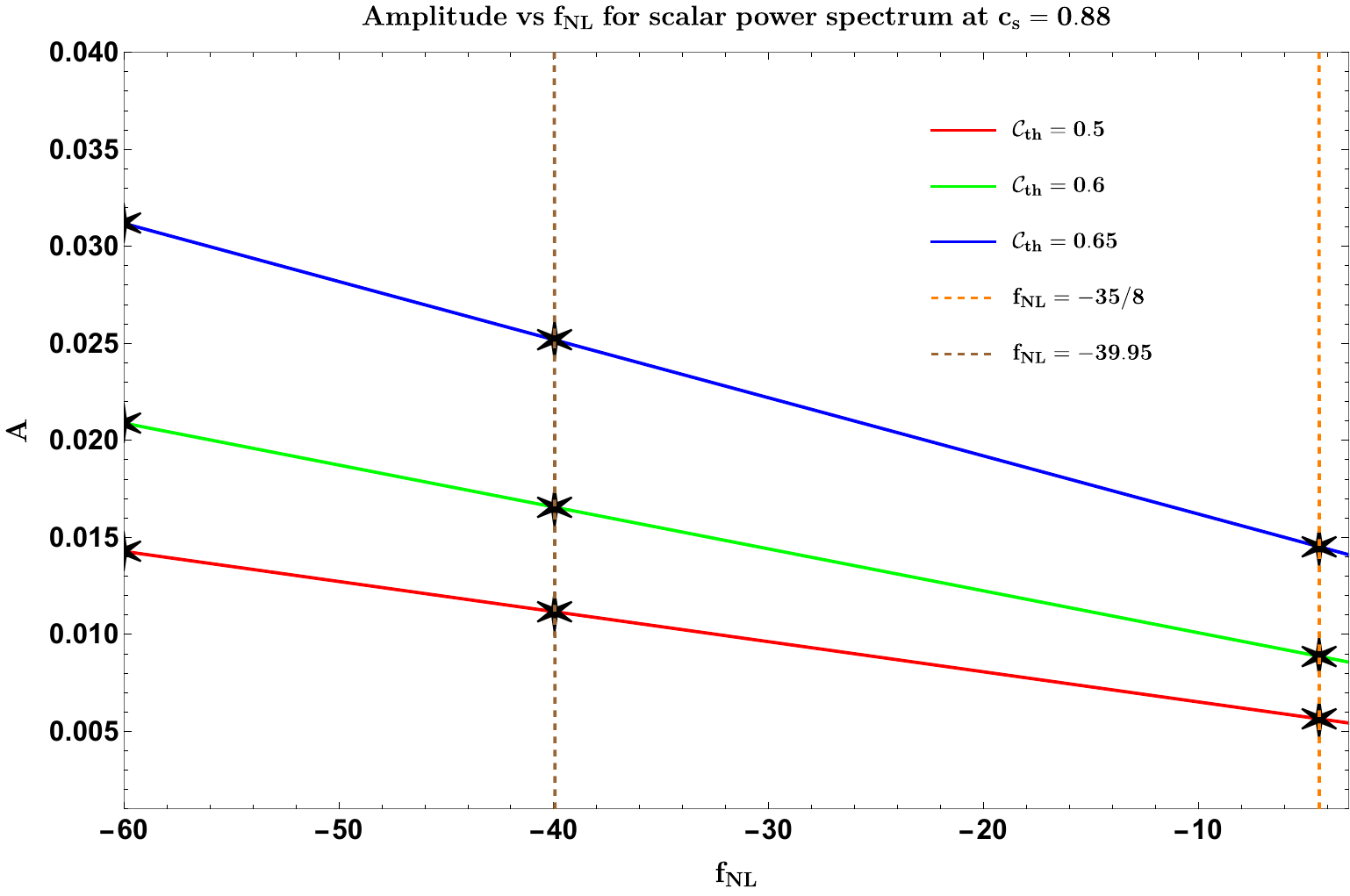}\label{ampfnlcs0.88}
    }
    \caption{ Plots highlighting the variation of peak amplitude $A$ of the power spectrum with the non-Gaussianity parameter $f_{\rm NL}$ for separate cases of the effective sound speed $c_s$. The left figure (\ref{ampfnlcs1}) corresponds to $c_s=1$, where the amplitude achieves a noticeable amplification, standing around an order of magnitude higher than the right figure (\ref{ampfnlcs0.88}) which employs $c_s=0.88$. Both figures incorporate distinct compaction thresholds, specifically ${\cal C}_{\rm th}\in (0.5,0.6,0.65)$, denoted by red, green, and blue lines respectively. The red and green dotted vertical lines mark the position of the values $f_{\rm NL}=(-39.95,-35/8)$ that we incorporate continuously within our analysis, and these intersect with the various coloured inclined lines for separate values of $A$ marked using a solid star. We establish an upper bound on the value of $f_{\rm NL}$ through this observation.  }\label{ampvsfnl}
\end{figure*}

\section{Establishment of strong upper bound on $f_{\rm NL}$}
\label{s9}

This work has delved into the significance of having large and negative primordial non-Gaussianity in our theory to evade the issue of PBH overproduction. In section \ref{s6c}, we show through our results how large NGs, with value $f_{\rm NL}\sim {\cal O}(-40)$, can not only suppress the overproduction of near-solar mass PBHs but the corresponding curvature perturbations can lead to an induced GW spectrum that successfully aligns with the latest PTA observations. We now recall the discussions in Sec. \ref{s6a} about having such large magnitude of NGs within our theory as a direct signature of violating the famous consistency condition for single-field, slow-roll inflation proposed by Maldacena in \cite{Maldacena:2002vr} which suggests $f_{\rm NL}\sim {\cal O}(-0.1)$. Instead, if one assumes a contracting phase followed by a matter bounce scenario, then, as pointed out in \cite{Cai:2009fn}, the strength of NGs increases due to their amplitude getting unaffected by the slow-roll parameters, and the curvature perturbations in the contracting phase constantly experience growth in the super-horizon scales until the matter bounce initiates. The other interesting aspect is the scenario of an ekpyrotic contraction before the bouncing phase. During this phase, the potential well becomes negative, extremely steep, and increases the kinetic energy density in the scalar fields to dominate, leading to highly non-linear interactions and thus resulting in an increase in the non-Gaussian signature from such a phase. The corresponding ekpyrotic equation of state controls the largeness of the interactions. We work with the results in \cite{Lehners:2008my}, where they discuss the production of NGs in the ekpyrotic(and cyclic) models in great detail. During the contraction phase, the scalar field fluctuations produce the entropic perturbations, which later get converted into the curvature perturbations where the EoS in this intervening instant of conversion, right before going into the bounce, significantly affects the non-Gaussianities. If such a conversion occurs during the ekpyrotic phase, it can produce a favorable amount of negative and large NGs of order $f_{\rm NL}\ll {\cal O}(-10)$.

In our results, it is evident from figures (\ref{NGfpbhA}-\ref{PBHAlines}) that the requisite peak scalar power spectrum amplitude is quite substantial, $A\sim {\cal O}(0.1)$, which approaches to the verge of breaking the underlying perturbative approximation. At this point, we ask ourselves, how large can we expect the magnitude of NG to grow before it breaches this approximation? To address this, we examine the figure (\ref{ampvsfnl}).

The need for amplified strength of inflationary fluctuations for PBH formation does not necessarily warrant the validation of perturbativity constraints. From the figure (\ref{ampvsfnl}), we observe a monotonic trend for the peak amplitude $A$ with increasing magnitude of the NG parameter $f_{\rm NL}$. Our previous results have shown how helpful large negative NGs are in dealing with PBH overproduction. Presently, in both figures \ref{ampfnlcs1} and \ref{ampfnlcs0.88}, we can see that the amplitude $A$ required to achieve $f_{\rm PBH}\sim 1$, increases with negative $f_{\rm NL}$ as well as with higher values of the compaction threshold ${\cal C}_{\rm th}$. However, there comes a moment where the required amplitude crosses over $A\sim {\cal O}(0.1)$ for both $f_{\rm NL}=(-39.95,-35/8)$ and $c_s=1$ and over $A\sim {\cal O}(0.01)$ for the same set of $f_{\rm NL}$ but now with $c_s=0.88$. Following our repeated emphasis over choosing the canonical case where $c_s=1$, we notice from figure \ref{ampfnlcs1} that going beyond $f_{\rm NL}\sim -60$ will surely keep pushing the amplitude close to eventually hitting $A\sim {\cal O}(1)$ and putting the analysis at risk of breaking perturbativity. In a similar spirit, with $c_s=0.88$ in figure \ref{ampfnlcs0.88}, one observes the condition $A> 0.01$ for $f_{\rm NL}\lesssim -60$. Furthermore, since we aim for the interval, $f_{\rm PBH}\in (10^{-3},1)$, which quantifies as having a sizeable fraction, the maximum amplitude difference tolerable for a particular wavenumber comes to, $\log_{10}(\Delta A)=-0.2$, which is an estimate that we can expect even when we consider $f_{\rm NL}\sim {\cal O}(-60)$.

For the space of allowed compaction thresholds, the values we have considered here for ${\cal C}_{\rm th}$ lie between $(0.5,0.65)$, with ${\cal C}_{\rm th}=0.5$ demanding that the amplitude come just close to $A\sim 0.1$. So, any value of this threshold lower than ${\cal C}_{\rm th}\sim 0.5$ will accordingly require a low enough amplitude even if we consider $f_{\rm NL}\sim -60$. Thus, we conclude that for $0.4< {\cal C}_{\rm th}< 0.5$, where the lower limit arises from our choice to work with the linear regime result, ${\cal C}_{\rm th}\in [2/5,2/6]$ \cite{Musco:2020jjb}, the strength of the scalar power spectrum is still sufficient to incorporate further large and negative NG, $f_{\rm NL}<-60$. The significance of $f_{\rm NL}=-60$ also extends to the fact that in figure (\ref{ampvsfnl}) we choose to show the analysis for the same set of compaction thresholds that we previously incorporate in figures (\ref{NGfpbhA}-\ref{PBHAlines}) to highlight the removal of PBH overproduction with the choice of ${\cal C}_{\rm th}\in (0.5, 0.65)$, which gives us the best agreement with the data.  

\textcolor{black}{ Some models also predict extremely large but positive NGs closer to the wavenumber regime where the power spectrum gets enhanced due to modifications in the inflationary potential, with one such study addressing this feature present in \cite{Rezazadeh:2021clf}. The model there supposes modifying the $\alpha-$attractor potential via a tiny bump to enhance the primordial power spectrum \cite{Mishra:2019pzq}, which promotes the formation of PBHs. In \cite{Rezazadeh:2021clf}, they show that at small scales, close to the wavenumbers that lead to PBHs, the value of the equilateral NG configuration reaches $f^{\rm equil}_{\rm NL}\sim {\cal O}(10^{3})$. In contrast to this feature, our study focuses on the implications of local type or squeezed NGs that are proven, as stated at the beginning of this section, to dominate models that have an ekpyrotic or matter-type contraction or bouncing phase. After choosing a set of $f_{\rm NL}$ values obtained within models with such pre-inflationary features, we conduct our study of curing PBH overproduction and finally propose the aforementioned upper bound on negatively large NGs.   }

Considering the pre-inflationary dynamics of the perturbations, studies have pointed out that, in principle, one great advantage of such conditions is the generation of a large amount of NG that reaches $f_{\rm NL}\sim {\cal O}(-100)$ \cite{Lehners:2008my, Koyama:2007if}. In such cases, we can, however, comment about large NGs as our use of an EFT framework encompasses both the scenarios before the beginning of inflation and hereafter, thus allowing for a generalized study. Hence, upon conclusion from the observations, we provide a strict upper bound on the possible magnitude of NG, $f_{\rm NL}\sim -60$, which does not endanger the validity of our analysis within our EFT setup.

\section{Conclusion}
\label{s10}

In this section, we summarize our crucial findings regarding the cure for PBH overproduction and its connection with exhibiting large and negative primordial non-Gaussianities. Our underlying set-up takes advantage of the features coming from pre-inflationary physics, with the introduction of the contracting and bouncing phases, into the generation of extremely large non-Gaussianities in the curvature perturbations. The benefit of bearing such large NGs is reflected in controlling the abundance of PBHs, which are formed during inflation, and they determine whether their abundance gets either suppressed or enhanced. We began with a brief introduction on the EFT of bounce framework built on the works of \cite{Choudhury:2023vuj,Choudhury:2023jlt,Choudhury:2023rks,Choudhury:2024dei}, and described how the scalar power spectrum after contributions from each phase is constructed including the one-loop quantum corrections. In order to provide a lucid procedure of handling the quantum-loop corrections, we summarize its main components of regularization, renormalization, and resummation which altogether assist in removing any quadratic (or power-law) UV and smoothen the logarithmic IR divergences resulting from each phase to the tree-level result. At the end of the procedures, we obtained the crucial one-loop renormalized and DRG resummed scalar power spectrum through which we later discuss the production of PBHs. 

Keeping in line with the latest surge in the theoretical advancements of PBH formation, we preferred to employ the compaction function criterion for the same and elaborate on its formalism, which provides a means to calculate the PBH mass fraction. We further integrate the inclusion of the quadratic form of NGs within the compaction function approach, to conduct a qualitative study of the impact NGs have on PBH formation. We follow this with a discussion on how we go about calculating the necessary abundance and the effect of having PNGs in such calculations. After this, we take a brief interlude to discuss the energy density spectrum of scalar-induced gravity waves in the radiation-dominated era where $w=1/3$. Throughout this work, we do not entertain any free equation of state and proceed with the fact that the scales contributing most to the PBH collapse do so after their re-entry into the radiation era, for a general equation of state analysis within the EFT of bounce set-up we refer to \cite{Choudhury:2024dzw}. 

In the latter half of this paper, we focused primarily on the outcomes after getting the help of large and negative PNGs to tackle the overproduction of PBHs. To better comprehend the crucial insights gained from the rest of our analysis, we illustrate them using bullet points as follows:
\begin{itemize}
    \item[$\bullet$]  
        We selected two distinct benchmark values for investigating PNGs namely, $f_{\rm NL}=(-39.95,-35/8)$, which result from independent analysis focusing on either an ekpyrotic contraction \cite{Lehners:2008my} or an expanding period of matter bounce \cite{Cai:2009fn}, respectively. Within this analysis, we added another variable, the effective sound speed component, and considered its two separate values, $c_s=(0.88,1)$ based on a previous work \cite{Choudhury:2024dei}.
    \item[$\bullet$]
        We found from our set-up that the probability of producing near solar mass PBHs, $M_{\rm PBH}\geq {\cal O}(0.01)M_{\odot}$, kept decreasing at a faster rate when $f_{\rm NL}=-35/8$ than in the case where $f_{\rm NL}=-39.95$, signalling the importance of large and negative PNGs in the PBH production probability for a given amplitude of the scalar power spectrum.
    \item[$\bullet$]
        To highlight our best result, we established that larger negative values such as $f_{\rm NL}=-39.95$, coupled with the canonical case $c_{s}=1$, proved utmost beneficial in increasing the abundance of PBHs and simultaneously establishing an excellent agreement of the requisite peak power spectrum amplitude within $1\sigma$ of having a SIGW explanation of the recent PTA signal.
    \item[$\bullet$]
        Utilizing a lower value of $c_s=0.88$ leads to an order-of-magnitude suppression in the amplitude, pushing the agreement outside the $2\sigma$ regime when $f_{\rm NL}=-39.95$ and even outside by $3\sigma$ for the choice of $f_{\rm NL}=-38/5$.
    \item[$\bullet$]
        After this, we conducted a comparative study highlighting the latest attempts towards addressing overproduction through various models that include only the PNGs to their advantage or incorporate other new parameters. We followed this by confronting these models' results with a few observational experiments for detecting gravitational wave signals and candidates for dark matter via microlensing.
    \item[$\bullet$]
        Later, we scrutinized the acceptance of negatively large NGs in our EFT model and concluded this discussion by extracting an upper bound of $f_{\rm NL}\sim -60$ as the most allowable negative value for the case of large compaction thresholds, ${\cal C}_{\rm th}\in (0.5,0.65)$, utilizing which we previously strongly mitigated the overproduction of PBHs.
    \item[$\bullet$]
         When going more negative beyond $f_{\rm NL}\sim -60$, we found the power spectrum amplitude to cross over the condition $A\sim {\cal O}(0.1)$ for $f_{\rm NL}=-39.95$, and also over $A\sim {\cal O}(0.01)$ for $f_{\rm NL}=-35/8$, signalling a breach in the underlying perturbativity arguments.  
\end{itemize}

\section*{Acknowledgement}

SC would like to thank The North American Nanohertz Observatory for Gravitational Waves (NANOGrav) collaboration and the National Academy of Sciences (NASI), Prayagraj, India, for being elected as an associate member and the member of the academy respectively. 
SC would also like to thank all the members of Quantum Aspects of the Space-Time \& Matter
(QASTM) for elaborative discussions. Last but not least, we acknowledge our debt to the people
belonging to the various parts of the world for their generous and steady support for research in natural sciences.
\newpage
\section*{Appendix}
\appendix

\section{Evolution of slow-roll parameters in the EFT of bounce framework}
\label{appA0}

\textcolor{black}{In this section, we discuss the evolution of the two slow-roll parameters, $\epsilon(N)$ and $\eta(N)$, depending on the nature of the pre-inflationary setup via its corresponding scale factor, $a(t)$, and how such parameters then evolve throughout the complete EFT of bounce framework studied in this work. We use the two scale-factor definitions as stated previously in Eqs. (\ref{contractionscale}-\ref{bouncescale}). }

\textcolor{black}{In order to derive the behaviour of these slow-roll parameters, we would use the help of the simple relation, $dN=Hdt={\cal H}d\tau$, for the e-folding elapsed in a conformal time interval, $d\tau$, and where, ${\cal H}=aH$, is the conformal Hubble parameter. Let us first analyze the slow-roll parameters during the pre-inflationary phases, after which we will then enter the SR/USR/SR setup. The final result for the evolution is shown in the figures (\ref{epsilonplots}-\ref{etaplots}) for the complete setup. }

\begin{enumerate}
    \item \underline{\rm \bf Phase-I~ (Contraction):} \textcolor{black}{This phase can be an ekpyrotic or matter-type contraction phase. Using the definition in Eqn. (\ref{contractionscale}) we can write the relation, ${\cal H}=a'/a=1/(\tau(\epsilon-1))$. Once we know this, we can now proceed with the following relation for the conformal time as:}
    \bea \label{timeC}
\int_{N_{0}}^{N}dN = \int_{\tau_{0}}^{\tau}\frac{d\tau}{\tau(\epsilon-1)}\implies \tau = \tau_{0}\exp{[(N-N_{0})(\epsilon-1)]},
\eea
\textcolor{black}{for $N< N_c$ where $N=N_c$ signals the end of the contraction phase, and $N_0$ and $\tau_0$ are reference variables for the e-folding and conformal time such that $a(\tau_0)=a_0$ in Eqn. (\ref{contractionscale}). The above result allows us to obtain the conformal Hubble parameter as a function of the e-folds $N$ during contraction. Now we are ready to write the relation for $\epsilon(N)$ using its following definition:}
\bea
\epsilon(N) &=& 1-\frac{{\cal H}'}{{\cal H}^2} = 1+\frac{\tau^2(\epsilon-1)^2}{\tau^2(\epsilon-1)} = \epsilon,
\eea
\textcolor{black}{where $\epsilon$ in the RHS takes the value $\epsilon=3/2$ during matter contraction or $\epsilon> 3$ during the ekpyrotic contraction phase.
In a similar fashion, we can derive the relation with e-folding for the second slow-roll parameter $\eta(N)$. We use its corresponding definition to arrive at the following relation:}
\bea
\eta = \frac{\epsilon'}{\epsilon{\cal H}} = 0,
\eea
\textcolor{black}{since $\epsilon$ does not change during this phase.}

\begin{figure*}[htb!]
    	\centering
    \subfigure[]{
      	\includegraphics[width=8.5cm,height=7.5cm]{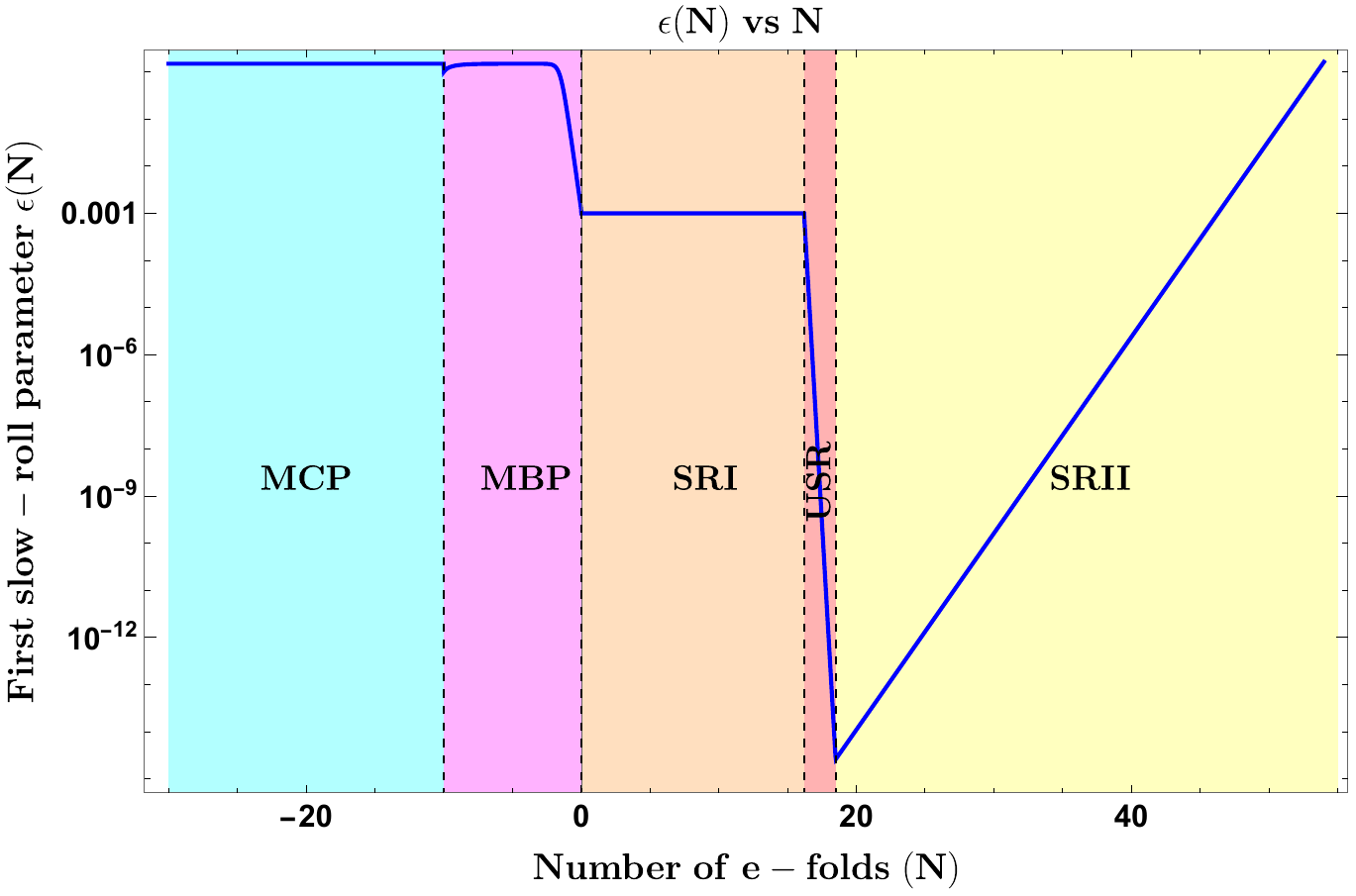}
        \label{epsilonMatter}
    }
    \subfigure[]{
       \includegraphics[width=8.5cm,height=7.5cm]{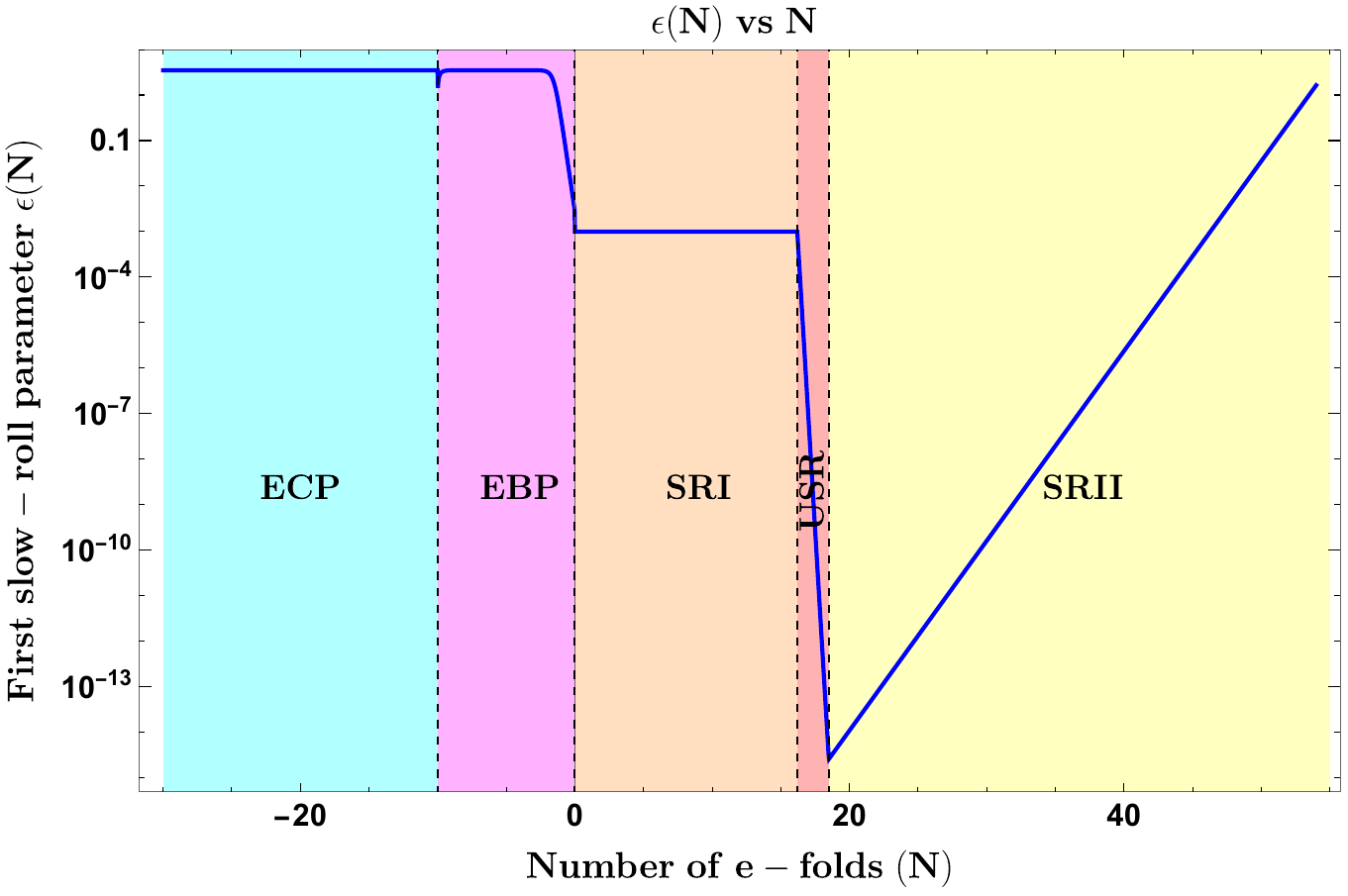}
        \label{epsilonEkpy}
    }
    	\caption[Optional caption for list of figures]{\textcolor{black}{The first slow-roll parameter $\epsilon(N)$ as a function of the e-folds $N$ in \ref{epsilonMatter} with the matter contraction (MCP) and matter bounce (MBP) phases and in \ref{epsilonEkpy} with the ekpyrotic contraction (ECP) and ekpkyrotic bounce (EBP) phases followed by the SRI, USR, and SRII phases.} } 
    	\label{epsilonplots}
    \end{figure*}

\begin{figure*}[htb!]
    	\centering
    \subfigure[]{
      	\includegraphics[width=8.5cm,height=7.5cm]{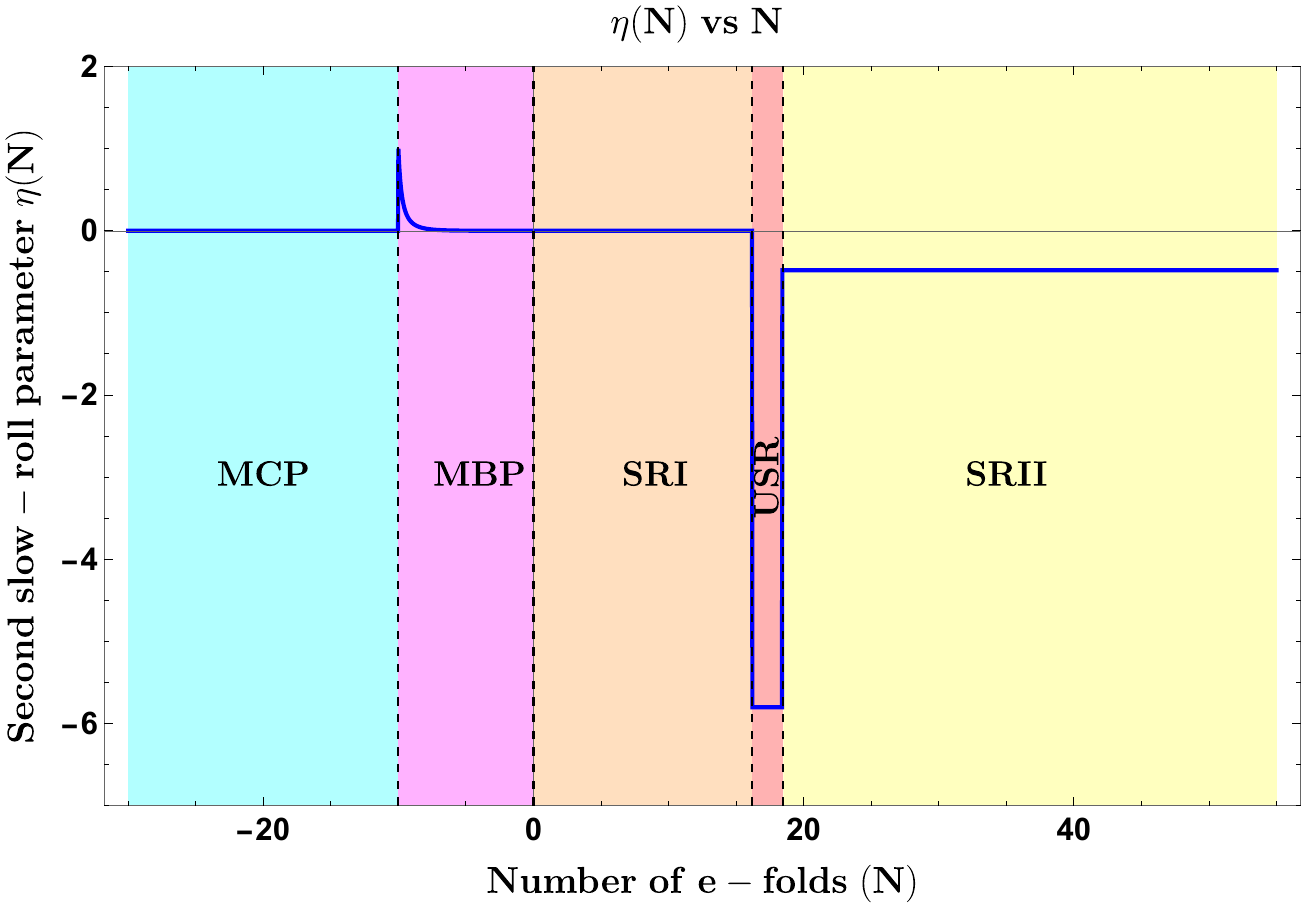}
        \label{etaMatter}
    }
    \subfigure[]{
       \includegraphics[width=8.5cm,height=7.5cm]{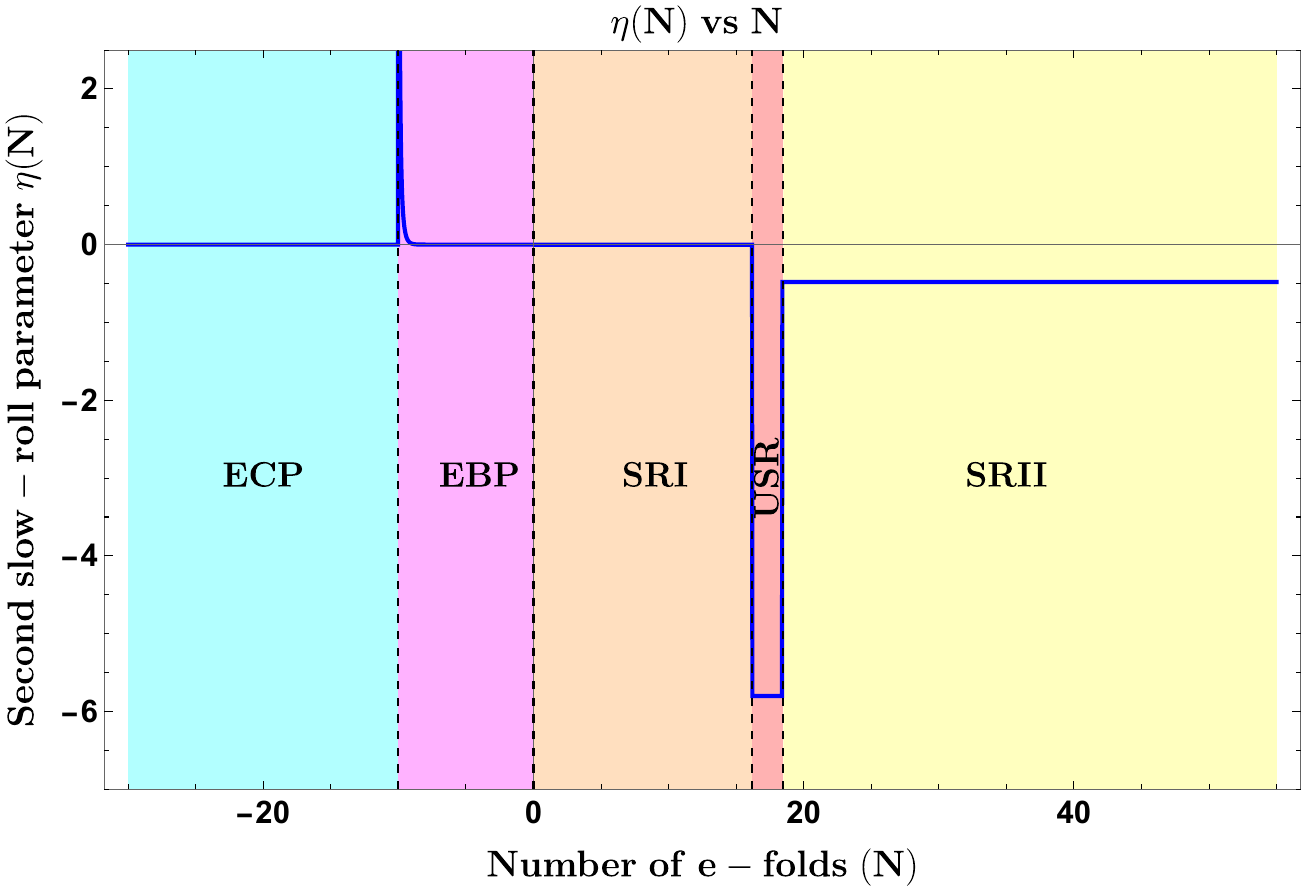}
        \label{etaEkpy}
    }
    	\caption[Optional caption for list of figures]{\textcolor{black}{The second slow-roll parameter $\eta(N)$ as a function of the e-folds $N$ in \ref{etaMatter} with the matter contraction (MCP) and matter bounce (MBP) phases and in \ref{etaEkpy} with the ekpyrotic contraction (ECP) and ekpkyrotic bounce (EBP) phases followed by the SRI, USR, and SRII phases.} } 
    	\label{etaplots}
    \end{figure*}

\item \underline{\rm \bf Phase-II~ (Bounce):} \textcolor{black}{This phase can also be an ekpyrotic or matter-type bouncing phase. This time after utilising the definition of the scale factor in Eqn. (\ref{bouncescale}) we get the following relation for the conformal time as:}
\bea \label{timeB}
\int_{N_{0}}^{N}dN &=& \int_{\tau_{0}}^{\tau}d\tau\frac{\tau}{\tau_{0}^2 (\epsilon-1)}\bigg(1+\bigg(\frac{\tau}{\tau_0}\bigg)^2 \bigg)^{-1},\nonumber\\
\implies N-N_{0} &=& \frac{\ln{[\tau^2 + \tau_{0}^2]}}{2(\epsilon-1)}\Biggr|_{\tau_{0}}^{\tau} = \frac{1}{2(\epsilon-1)}\ln{\bigg[\frac{1}{2}\bigg(1+\bigg(\frac{\tau}{\tau_0}\bigg)^2\bigg)\bigg]},
\eea
\textcolor{black}{for $N_c<N<N_b$ where the bouncing phase ends at $N=N_b$, and $N_0$ and $\tau_0$ are reference variables for the e-folding and conformal time such that $a(\tau_0)=a_0$ in Eqn. (\ref{bouncescale}). Using this we can express the conformal Hubble parameter in terms of the e-folds as follows:}
\bea
{\cal H}(N) &=& \frac{1}{\tau_{0}(\epsilon-1)}\frac{\tau}{\tau_{0}}\bigg(1+\bigg(\frac{\tau}{\tau_0}\bigg)^2 \bigg)^{-1},\nonumber\\
&=& \frac{1}{2\tau_{0}(\epsilon-1)}\exp{[-2(N-N_{0})(\epsilon-1)]}\sqrt{2\exp{[2(N-N_{0})(\epsilon-1)]}-1}.
\eea
\textcolor{black}{Continuing from here we can get the relation for $\epsilon(N)$ using its definition, and the relation in Eqn (\ref{timeB}), as follows:}
\bea
\epsilon(N) = 1-\frac{{\cal H}'}{{\cal H}^2} = 1- \frac{(\epsilon-1)(\tau_{0}^2-\tau^2)}{\tau^2} = \epsilon + \frac{(1-\epsilon)}{2\exp{[2(N-N_{0})(\epsilon-1)]}-1},
\eea
\textcolor{black}{which tells us that for large enough $N$ the exponential in the denominator becomes large enough so that $\epsilon$ again becomes constant in the respective phase. The $\epsilon$ values on the RHS take again similar values during the ekpyrotic $(\epsilon>3)$ or matter $(\epsilon=3/2)$ bouncing phase. Finally, from the definition of the parameter $\eta(N)$ and Eqn. (\ref{timeB}), we have the following relation:}
\bea
\eta(N) = \frac{\epsilon'}{\epsilon{\cal H}} &=&  2(\epsilon-1)^{2}\times \frac{\tau_{0}^2}{\tau^2}\frac{1+\big(\frac{\tau}{\tau_0}\big)^2}{1+ \epsilon\big(\frac{\tau^2}{\tau_0^2}-1\big)},\nonumber\\
&=& 2(\epsilon-1)^{2}\times \frac{1}{(2\exp{[2(N-N_{0})(\epsilon-1)]}-1)}\frac{2\exp{[2(N-N_{0})(\epsilon-1)]}}{1+2\epsilon(\exp{[2(N-N_{0})(\epsilon-1)]}-1)}.
\eea

\item \underline{\rm \bf Phase-III~ (First Slow-Roll):} \textcolor{black}{We can now begin analysing the slow-roll parameters in the phases during inflation starting with the first slow-roll. We assume that throughout the phase the SR parameter $\eta(N)$ remains a small and negative almost constant value. This further leads us to conclude from its definition that the first SR parameter $\epsilon (N)$ is also a constant since:}
\bea
\eta &=& \frac{\epsilon'}{\epsilon{\cal H}} = \frac{1}{\epsilon}\frac{d\epsilon}{dN},
\eea
\textcolor{black}{and that it also remains an almost constant value during $N_{*}<N<N_s$ to satisfy the assumption of $\eta(N)$. The value $N_*$ here refers to a reference e-folding value corresponding to the start of SRI, which we choose to be $\epsilon(N=N_{*})={\cal O}(10^{-3})$. $N_s$ refers to the point where SRI ends and the next USR phase begins.}

\item \underline{\rm \bf Phase-IV~ (Ultra Slow-Roll):}
\textcolor{black}{Here we assume that the transition from the SRI to the USR occurs sharply such that the value of the second SR parameter suddenly changes to become, $\eta(N)\sim {\cal O}(-6)$, which is a negatively large value. It stays the same throughout the interval of this phase, $N_s<N<N_e$, and, using its definition, we can write the following relation for the first SR parameter:}
\bea
\eta &=& \frac{1}{\epsilon}\frac{d\epsilon}{dN} \implies \int_{N_s}^{N}\eta dN =\int_{\epsilon(N_s)}^{\epsilon}d\ln \epsilon \implies \epsilon(N) = \epsilon(N_s)\exp{(\eta (N-N_{s}))},
\eea
\textcolor{black}{which tells us that $\epsilon(N)$ suffers an exponential decrease during the USR phase. The value $\epsilon(N_s)$ matches its value during the end of the SRI phase.}

\item \underline{\rm \bf Phase-V~ (Second Slow-Roll):}
\textcolor{black}{The previous assumption regarding the nature of the transition between the SRI and USR phases follows again after we exit the USR and enter the final SRII phase. The second SR parameter, this time, increases from $\eta(N)\sim {\cal O}(-6)$ to $\eta(N)\sim {\cal O}(-1)$, indicating the end of inflation. This value remains the same throughout this phase until the first SR parameter climbs back close to unity. Its behaviour can be understood in a similar fashion done before as follows:}
\bea
\eta &=& \frac{1}{\epsilon}\frac{d\epsilon}{dN}\implies  \int_{N}^{N_{\rm end}}\eta dN = \int_{\epsilon}^{\epsilon(N_{\rm end})}d\ln \epsilon \implies \epsilon(N) = \epsilon(N_{\rm end})\exp{(-\eta (N_{\rm end}-N))},
\eea
\textcolor{black}{where we take $\epsilon(N_{\rm end})\sim {\cal O}(1)$ as the moment when inflation truly ends with final e-foldings being $N_{\rm end}$. As usual, continuity demands that the parameters remain the same value at the transition instant $N=N_e$.}

\end{enumerate}

\section{Comoving curvature perturbation mode solutions}
\label{appA}

In this Appendix, we underline the explicit curvature perturbation mode solutions for each phase in our EFT of bounce set-up. These solutions will later help us construct the tree-level scalar power spectrum and are needed further to evaluate the various momentum and temporal integrals for the quantum loop corrections. The discussions related to one-loop quantum corrections are part of the next section. The upcoming results are basically solutions of the Fourier space MS equation, see Eqn. (\ref{MSEq}), for any wavenumber $k$. 
Our complete setup consists of $5$ phases, starting with the contraction, followed by a bounce into an expanding phase, which joins the SRI as inflation begins, a sharp transition into the USR, and a sharp transition into the final SRII phase where inflation ends. For both contracting and bouncing phases, we do not restrict to any particular matter or ekpyrotic kind but generalize the study to suit either.

We begin below from the contracting phase and mention the mode solutions until the SRII phase and illuminate along the way the properties of each general solution.

\begin{enumerate}
    \item \underline{\rm \bf Phase-I~ (Contraction):}
    The contraction phase of our set-up is characterized using the scale factor solution present in Eqn. (\ref{contractionscale}) which involves the use of its equation of state parameter $\epsilon_c$ and determines whether its a matter $(\epsilon_c=3/2)$ or an ekpyrotic $(\epsilon_c> 3)$ kind of phase. This phase operates during the conformal time interval, $\tau_c < \tau < \tau_b$, at the end of which we transition into the next bouncing phase. The curvature mode solution during contraction has the following structure: 
        \bea
        \label{contraction}
        {\bf \zeta}_{\bf C}&=&\frac{2^{\nu-\frac{3}{2}} c_s (-k c_s \tau )^{\frac{3}{2}-\nu}}{ia_0\tau \sqrt{2 \epsilon_*}(k c_s)^{\frac{3}{2}}\sqrt{2}  M_{pl}}\left(\frac{\tau}{\tau_0}\right)^{-\frac{1}{(\epsilon-1)}}\sqrt{\left(\frac{\epsilon_*}{\epsilon_c}\right)}\Bigg|\frac{\Gamma(\nu)}{\Gamma(\frac{3}{2})}\Bigg|\nonumber\\
&&\quad\quad\quad\quad\quad\quad\quad\quad\quad\quad\quad\quad \times\Bigg\{\alpha_1 (1+i k c_s\tau) e^{-i\left(k c_s\tau+\frac{\pi}{2}\left(\nu+\frac{1}{2}\right)\right)}-\beta_1(1-i k c_s \tau)e^{i\left(k c_s\tau+\frac{\pi}{2}\left(\nu+\frac{1}{2}\right)\right)}\Bigg\},
        \eea
        where variable $\nu$ denotes the effective mass parameter from Eqn. (\ref{MSvariableprops}). The above solution consists of a linear combination which comes with two Bogoliubov coefficients namely $\alpha_1$ and $\beta_1$ which provide information about the underlying quantum vacuum state and must satisfy the normalized condition $|\alpha_1|^2 - |\beta_1|^2 = 1$. We must note here that in this work the choice of the initial vacuum structure is set to that of the Bunch-Davies vacuum which determines the coefficients with $\alpha_1=1,\;\beta_1=0$. 
    \item \underline{\rm \bf Phase-II~ (Bounce):}
    The expanding bounce phase is the second phase of our set-up which begins at $\tau_b$ and operates within the conformal time interval, $\tau_b < \tau < \tau_i$. This phase is characterized by its corresponding scale factor in Eqn. (\ref{bouncescale}) and which also includes the equation of state parameter $\epsilon_b$ through which we are able to capture the matter $(\epsilon_c=3/2)$ or ekpyrotic $(\epsilon_c>3)$ bouncing scenario. The curvature mode solution during the bounce takes the following form:
        \bea
        \label{bounce}
        &&{\bf \zeta}_{\bf B}=\frac{2^{\nu-\frac{3}{2}} c_s (-k c_s \tau )^{\frac{3}{2}-\nu}}{ia_0\tau \sqrt{2 \epsilon_*}(k c_s)^{\frac{3}{2}}\sqrt{2}  M_{pl}}\left[1+\left(\frac{\tau}{\tau_0}\right)^2\right]^{-\frac{1}{2(\epsilon-1)}}\sqrt{\left(\frac{\epsilon_*}{\epsilon_b}\right)}\Bigg|\frac{\Gamma(\nu)}{\Gamma(\frac{3}{2})}\Bigg |\nonumber\\
&&\quad\quad\quad\quad\quad\quad\quad\quad\quad\quad\quad\quad\quad\quad\times\Bigg\{\alpha_1 (1+i k c_s\tau) e^{-i\left(k c_s\tau+\frac{\pi}{2}\left(\nu+\frac{1}{2}\right)\right)}-\beta_1(1-i k c_s \tau)e^{i\left(k c_s\tau+\frac{\pi}{2}\left(\nu+\frac{1}{2}\right)\right)}\Bigg\}.
        \eea
        Throughout this phase, the choice of underlying quantum vacuum remains the same as our initial Bunch-Davies scenario at the beginning of the contraction. We choose this kind of smooth transitory behaviour between these phases, till we enter inside the inflationary epoch, for the sake of ease in our calculations. The Bogoliubov coefficients therefore follow with $\alpha_1=1,\;\beta_1=0$, still keeping the normalization condition intact.  
    \item \underline{\rm \bf Phase-III~ (First~Slow-Roll):} After the bouncing phase ends, we enter into inflation starting with the usual SR phase which lasts for the interval, $\tau_i < \tau < \tau_s$. Here the crucial assumption is that of the similar vacuum structure as before for the contraction and bounce phases. The general mode solution for this phase has the following form:
        \bea
        \label{SRI}
        {\bf \zeta}_{\bf SRI}&=&\frac{2^{\nu-\frac{3}{2}} c_s H (-k c_s \tau )^{\frac{3}{2}-\nu}}{i \sqrt{2 \epsilon_*}(k c_s)^{\frac{3}{2}}\sqrt{2} M_{pl}}\Bigg|\frac{\Gamma(\nu)}{\Gamma(\frac{3}{2})}\Bigg | \Bigg\{\alpha_1 (1+i k c_s\tau) e^{-i\left(k c_s\tau+\frac{\pi}{2}\left(\nu+\frac{1}{2}\right)\right)} -\beta_1(1-i k c_s \tau)e^{i\left(k c_s\tau+\frac{\pi}{2}\left(\nu+\frac{1}{2}\right)\right)}\Bigg\}.
        \eea
        where the Bogoliubov coefficients remain the same as stated, $\alpha_1=1,\;\beta_1=0$. During the previous phases we did not assume any special kind of transitory behaviour when connecting the phases, but this changes soon as the SRI phases comes to an end at $\tau=\tau_s$.
    \item \underline{\rm \bf Phase-IV~ (Ultra~Slow-Roll):} The USR phase continues for the conformal time interval, $\tau_s < \tau < \tau_e$, where the assumption is that, at the beginning, the transition happens sharply without any relaxation. As a result of this transition nature, the underlying vacuum also undergoes a shift in its properties from the usual Bunch-Davies. The curvature mode solution now takes the following form: 
        \bea
        \label{USR}
        {\bf \zeta}_{\bf USR}&=&\frac{2^{\nu-\frac{3}{2}} c_s   H (-k c_s \tau )^{\frac{3}{2}-\nu}}{i \sqrt{2 \epsilon_*}(k c_s)^{\frac{3}{2}}\sqrt{2} M_{pl}}\bigg(\frac{\tau_s}{\tau}\bigg)^3\Bigg|\frac{\Gamma(\nu)}{\Gamma(\frac{3}{2})}\Bigg |\Bigg\{\alpha_2 (1+i k c_s\tau) e^{-i\left(k c_s\tau+\frac{\pi}{2}\left(\nu+\frac{1}{2}\right)\right)}\nonumber\\
&&\quad\quad\quad\quad\quad\quad\quad\quad\quad\quad\quad\quad\quad\quad\quad\quad\quad\quad\quad\quad\quad\quad\quad\quad\quad -\beta_2(1-i k c_s \tau)e^{i\left(k c_s\tau+\frac{\pi}{2}\left(\nu+\frac{1}{2}\right)\right)}\Bigg\}.
        \eea
        where we notice the appearance of two new sets of Bogoliubov coefficients namely, $\alpha_2,\;\beta_2$. These coefficients can be calculated keeping in mind the above two formulas in Eqn. (\ref{SRI}-\ref{USR}) and applying the continuity and differentiability conditions for the modes at the instant $\tau=\tau_s$ of the transition. Solving these boundary conditions would yield the following solutions for the coefficients:
        \bea \label{alpha2b}
    \alpha_{{2}} &=&  \frac{1}{2 k^3 \tau_s^3 c_s^3}  \Bigg(3 i + 3 i  k^2 c_s^2 \tau_s ^2  + 2  k^3 c_s^3  \tau_s ^3  \Bigg  ),\\ \label{beta2b}  \beta_{{2}} &=& \frac{1}{2 k ^3 c_s ^3 \tau_s ^3}  \Bigg( 3i -6 k c_s \tau_s -3i k^2 c_s ^2 \tau_s^2 \Bigg) e^{-i\left(\pi\left(\nu+\frac{1}{2}\right)+ 2k c_s \tau_s\right)}, \eea
    which describe the general solution in terms of the new shifted vacuum structure. Following this phase for a brief interval of time, we finally move to the last slow-roll phase at $\tau=\tau_e$.
    \item \underline{\rm \bf Phase-V~ (Second~Slow-Roll):} The USR phase further observes a sharp transition when passing over to the SRII phase and that operates within, $\tau_e < \tau < \tau_{0}$, after which inflation officially terminates. The vacuum, as a consequence of another sharp transition, witnesses further changes in its structure and a new set of Bogoliubov coefficients, $\alpha_3,\;\beta_3$, are needed to describe the following general mode solution in this phase:
        \bea
        \label{SRII}
        {\bf \zeta}_{\bf SRII}&=&\frac{2^{\nu-\frac{3}{2}} c_s   H (-k c_s \tau )^{\frac{3}{2}-\nu}}{i \sqrt{2 \epsilon_*}(k c_s)^{\frac{3}{2}}\sqrt{2} M_p}\left(\frac{\tau_s}{\tau_e}\right)^3\Bigg|\frac{\Gamma(\nu)}{\Gamma(\frac{3}{2})}\Bigg |\Bigg\{\alpha_3 (1+i k c_s\tau) e^{-i\left (kc_s\tau+\frac{\pi}{2}(\nu+\frac{1}{2})\right )} \nonumber\\
&&\quad\quad\quad\quad\quad\quad\quad\quad\quad\quad\quad\quad\quad\quad\quad\quad\quad\quad\quad\quad\quad\quad\quad\quad\quad -\beta_3(1-i k c_s \tau)e^{i\left(k c_s\tau+\frac{\pi}{2}(\nu+\frac{1}{2})\right)}\Bigg\}.
        \eea
        where, via the similar continuity and differentiabilty condition applied to the solution in Eqn. (\ref{USR}-\ref{SRII}), provides with the following new set of coefficients:
        \bea  \alpha _{3} &=& \frac{1}{(2 k^3 \tau_e^3 c_s^3)(2 k^3 \tau_s^3 c_s^3)}\Bigg\{\left(-3 i -3 i  k^2 \tau_e^2 c_s^2 +2  k^3 \tau_e^3 c_s^3 \right)\times\left(3 i + 3 i  k^2 c_s^2 \tau_s ^2  + 2  k^3 c_s^3  \tau_s ^3  \right  ) \nonumber\\ && \quad\quad\quad\quad- \left(-3 i -6  k \tau_e c_s   +3 i  k^2 \tau_e^2 c_s^2 \right)\times
         \left( 3i -6 k c_s \tau_s -3i k^2 c_s ^2 \tau_s^2 \right) e^{2 i k c_s( \tau_e -\tau_s) }\Bigg\},\\
        \beta _{3} &=&  \frac{1}{(2 k^3 \tau_e^3 c_s^3)(2 k ^3 c_s ^3 \tau_s ^3)}\Bigg\{ \left(-3 i  +6  k \tau_e c_s +3 i k^2 \tau_e^2 c_s^2\right) \times\left(3 i + 3 i  k^2 c_s^2 \tau_s ^2  + 2  k^3 c_s^3  \tau_s ^3  \right) e^{-\left(2 i k \tau_e c_s + i \pi  \left(\nu +\frac{1}{2}\right)\right)} \nn\\&&\quad\quad\quad\quad +\left(2  k^3 \tau_e^3 c_s^3 + 3 i  k^2 \tau_e^2 c_s^2 +3 i \right )  \times\left( 3i -6 k c_s \tau_s -3i k^2 c_s ^2 \tau_s^2 \right) e^{-i\left(\pi(\nu+\frac{1}{2})+ 2k c_s \tau_s \right)}\Bigg\}. \eea
\end{enumerate}

\section{One-loop corrections using In-In formalism}\label{appC}

In this Appendix we describe the approach towards calculating the one-loop quantum corrections to the $2$-point correlation function using the well-known in-in formalism in cosmology.

For this aim, we require the cubic order expansion \cite{Maldacena:2002vr} of the EFT action for the curvature perturbation whose final form reads as:
\begin{widetext}
    \bea &&S^{(3)}_{\zeta}=\int d\tau\;{\cal L}_3=\int d\tau\;  d^3x\;  M^2_{ pl}a^2\; \bigg[\left(3\left(c^2_s-1\right)\epsilon+\epsilon^2-\frac{1}{2}\epsilon^3\right)\zeta^{'2}\zeta+\frac{\epsilon}{c^2_s}\bigg(\epsilon-2s+1-c^2_s\bigg)\left(\partial_i\zeta\right)^2\zeta\nonumber\\ 
&&\quad\quad\quad\quad\quad\quad\quad\quad\quad\quad\quad-\frac{2\epsilon}{c^2_s}\zeta^{'}\left(\partial_i\zeta\right)\left(\partial_i\partial^{-2}\left(\frac{\epsilon\zeta^{'}}{c^2_s}\right)\right)-\frac{1}{aH}\left(1-\frac{1}{c^2_{s}}\right)\epsilon \bigg(\zeta^{'3}+\zeta^{'}(\partial_{i}\zeta)^2\bigg)
     \nonumber\\
&& \quad\quad\quad\quad\quad\quad\quad\quad\quad\quad\quad+\frac{1}{2}\epsilon\zeta\left(\partial_i\partial_j\partial^{-2}\left(\frac{\epsilon\zeta^{'}}{c^2_s}\right)\right)^2
+\underbrace{\frac{1}{2c^2_s}\epsilon\partial_{\tau}\left(\frac{\eta}{c^2_s}\right)\zeta^{'}\zeta^{2}}_{\bf Dominant ~in~USR}+\cdots
  \bigg],\quad\quad\eea
\end{widetext}
where at the end we ignore the suppressed higher-order contributions in the current analysis. The last term with the operator $\zeta'\zeta^{2}$ represents the term most dominant when contributing in the USR phase, at an order of ${\cal O}(\epsilon)$, and thus constitutes a crucial part in the quantum loop corrections. Also important regarding the same term is the fact that its contribution diminishes when talking about the contraction and more so in the bouncing phases. We deal with the correlations involving this particular interaction operator in order to derive the one-loop corrections from each phase.

Here we briefly demonstrate the techniques used to compute the necessary correlations with the explicit calculations available in \cite{Choudhury:2024dei, Choudhury:2024aji}. To evaluate these late-time correlation functions we implement the Schwinger-Keldysh or in-in formalism in whose language the leading order contribution to the two-point correlation with the interaction Hamiltonian, $H_{\rm int}=-{\cal L}_{3}$ (only at the cubic order), is written as:
\bea \label{inin2pt}\langle\hat{\zeta}_{\bf k}\hat{\zeta}_{-{\bf k}}\rangle:&=&
    \lim_{\tau\rightarrow 0}\left\langle\bigg[\overline{T}\exp\bigg(i\int^{\tau}_{-\infty(1-i\epsilon)}d\tau'\;H_{\rm int}(\tau')\bigg)\bigg]\;\;\hat{\zeta}_{\bf k}(\tau)\hat{\zeta}_{-{\bf k}}(\tau)
\;\;\bigg[{T}\exp\bigg(-i\int^{\tau}_{-\infty(1+i\epsilon)}d\tau''\;H_{\rm int}(\tau'')\bigg)\bigg]\right\rangle.
\eea
where the $(\bar{T})T$ denote the (Anti-)time ordering of the time-dependent operators inside the Dyson series expansion, and using the small deformation of the temporal axis in the complex plane, $\tau \rightarrow \tau(1\pm i\epsilon)$, in the far past $(\tau\rightarrow -\infty)$ allows one to project out the vacuum state of the free theory. Following this, the only non-trivial contributions to the $2$-point correlation at the one-loop level are calculated via the relations: 
\bea
    &&\label{loop1}\langle\hat{\zeta}_{\bf k}\hat{\zeta}_{-{\bf k}}\rangle_{(0,2)}=\lim_{\tau\rightarrow 0}\int^{\tau}_{-\infty}d\tau_1\;\int^{\tau}_{-\infty}d\tau_2\;\langle \hat{\zeta}_{\bf k}(\tau)\hat{\zeta}_{-{\bf k}}(\tau)H_{\rm int}(\tau_1)H_{\rm int}(\tau_2)\rangle,\\
    &&\label{loop2}\langle\hat{\zeta}_{\bf k}\hat{\zeta}_{-{\bf k}}\rangle^{\dagger}_{(0,2)}=\lim_{\tau\rightarrow 0}\int^{\tau}_{-\infty}d\tau_1\;\int^{\tau}_{-\infty}d\tau_2\;\langle \hat{\zeta}_{\bf k}(\tau)\hat{\zeta}_{-{\bf k}}(\tau)H_{\rm int}(\tau_1)H_{\rm int}(\tau_2)\rangle^{\dagger},\\
    &&\label{loop3}\langle\hat{\zeta}_{\bf k}\hat{\zeta}_{-{\bf k}}\rangle_{(1,1)}=\lim_{\tau\rightarrow 0}\int^{\tau}_{-\infty}d\tau_1\;\int^{\tau}_{-\infty}d\tau_2\;\langle H_{\rm int}(\tau_1)\hat{\zeta}_{\bf k}(\tau)\hat{\zeta}_{-{\bf k}}(\tau)H_{\rm int}(\tau_2)\rangle.
\eea
The above relations combine, after using the mode solutions for each phase in our set-up, to give the corresponding total correction at the one-loop and thus we can write this sum from Eqs. (\ref{loop1},\ref{loop2},\ref{loop3}) as:
\bea
\langle\hat{\zeta}_{\bf k}\hat{\zeta}_{-{\bf k}}\rangle_{\bf One-loop} &=&\langle\hat{\zeta}_{\bf k}\hat{\zeta}_{-{\bf k}}\rangle_{(0,2)} + \langle\hat{\zeta}_{\bf k}\hat{\zeta}_{-{\bf k}}\rangle^{\dagger}_{(0,2)} + \langle\hat{\zeta}_{\bf k}\hat{\zeta}_{-{\bf k}}\rangle_{(1,1)}.
\eea
Once we obtain the above result, we can combine this with the result for the $2$-point correlation function at the tree-level for each phase, as expressed in Eqn. (\ref{treecorrl}), and our final result can be now written as:
\bea
\langle\hat{\zeta}_{\bf k}\hat{\zeta}_{-{\bf k}}\rangle &=& \langle\hat{\zeta}_{\bf k}\hat{\zeta}_{-{\bf k}}\rangle_{\bf Tree} + \langle\hat{\zeta}_{\bf k}\hat{\zeta}_{-{\bf k}}\rangle_{\bf One-loop}.
\eea

\section{ Regularized-Renormalized-Resummed formulas }\label{app:RRR}

This Appendix is dedicated to underline the key results after performing the procedures of regularization and renormalization and followed by the DRG resummation. We do not, however, aim to delve into the intricacies of the steps involved in each said procedure and for which we recommend the reader to look into \cite{Choudhury:2024dei} for a full analysis. 

When treating the one-loop corrections to the $2$-point correlation function, we encounter as a result that these corrections diverge both in the UV and the IR limits and thus must be dealt with to end up with a finite result for our observables at the super-horizon scales. As a first step, we begin by implementing the cut-off regularization method (see \ref{s32a}) for the momentum integrals in the one-loop correlators that involve introducing two separate momentum cut-off scales, $k_{\rm IR}$ and $k_{\rm UV}$, while working with each phase of our set-up. Next, we invoke the renormalization prescription (see \ref{s32b}) to remove the UV divergent contributions dominated by the sub-Horizon physics. As shown in \cite{Choudhury:2023rks,Choudhury:2024dei}, irrespective of the prescription choice, we get the same result, as we must, for the final regularized and renormalized scalar power spectrum: 
\bea \Delta^{2}_{\zeta, {\bf RR}}(k)&=&\bigg[\Delta^{2}_{\zeta,{\bf Tree}}(k)\bigg]_{\bf SRI}\times\bigg(1+\overline{{\bf T}}_{\bf C}+\overline{{\bf T}}_{\bf B}+\overline{{\bf T}}_{\bf SRI}+\overline{{\bf T}}_{\bf USR}+\overline{{\bf T}}_{\bf SRII}\bigg), \eea 
where the above contains various contributions coming from the contraction, bounce, SRI, USR, and SRII phase, respectively, and which we now list below explicitly:
\bea 
    \label{t1}\overline{\bf T}_{\bf C}&=&-\frac{4}{3}\bigg[\Delta^{2}_{\zeta,{\bf Tree}}(k)\bigg]_{\bf SRI}\times\Bigg(1-\frac{2}{15\pi^2}\frac{1}{c^2_{s}k^2_c}\left(1-\frac{1}{c^2_{s}}\right)\epsilon_c\Bigg)\times \left(\frac{\epsilon_*}{\epsilon_c}\right)\times\bigg[\frac{1}{\delta_{\bf C}}\bigg\{\left(\frac{k_{b}}{k_*}\right)^{\delta_{\bf C}}-\left(\frac{k_{c}}{k_*}\right)^{\delta_{\bf C}}\bigg\}\nonumber\\   &&\quad\quad\quad\quad\quad\quad\quad\quad\quad\quad\quad\quad\quad\quad\quad\quad\quad\quad\quad\quad\quad\quad\quad\quad\quad+\frac{1}{\left(\delta_{\bf C}+2\right)}\bigg\{\left(\frac{k_{b}}{k_*}\right)^{\delta_{\bf C}+2}-\left(\frac{k_{c}}{k_*}\right)^{\delta_{\bf C}+2}\bigg\}\bigg],\\
    \label{t2}\overline{\bf T}_{\bf B}&=&-\frac{4}{3}\bigg[\Delta^{2}_{\zeta,{\bf Tree}}(k)\bigg]_{\bf SRI}\times\Bigg(1-\frac{2}{15\pi^2}\frac{1}{c^2_{s}k^2_b}\left(1-\frac{1}{c^2_{s}}\right)\epsilon_b\Bigg)\times \left(\frac{\epsilon_*}{\epsilon_b}\right)\times\frac{1}{\delta_{\bf B} +2}\bigg[\, _2F_1\left(\frac{\delta_{\bf B}+2}{2},\frac{1}{\epsilon_b-1}-1;\frac{\delta_{\bf B}+4}{2};-1\right)\nonumber\\
    &&\quad\quad\quad\quad\quad\quad\quad\quad\quad\quad\quad\quad\quad\quad\nonumber\\
    &&\quad\quad\quad\quad\quad\quad\quad\quad\quad\quad\quad\quad\quad\quad\quad\quad-\left(\frac{k_{b}}{k_{*}}\right)^{\delta_{\bf B}+2}\, _2F_1\left(\frac{\delta_{\bf B}+2}{2},\frac{1}{\epsilon_b-1}-1;\frac{\delta_{\bf B}+4}{2};-\left(\frac{k_{b}}{k_{*}}\right)^{2}\right)\bigg],\\
    \label{t3}\overline{\bf T}_{\bf SRI}&=&-\frac{4}{3}\bigg[\Delta^{2}_{\zeta,{\bf Tree}}(k)\bigg]_{\bf SRI}\times\Bigg(1-\frac{2}{15\pi^2}\frac{1}{c^2_{s}k^2_*}\left(1-\frac{1}{c^2_{s}}\right)\epsilon_*\Bigg)\times\ln\left(\frac{k_s}{k_*}\right),\\
    \label{t4}\overline{\bf T}_{\bf USR}&=&\frac{1}{4}\bigg[\Delta^{2}_{\zeta,{\bf Tree}}(k)\bigg]_{\bf SRI}\times\bigg[\bigg(\frac{\Delta\eta(\tau_{e})}{{c}^{4}_{s}}\bigg)^{2}{\bigg(\frac{k_{e}}{k_{s}}\bigg)^6} - \left(\frac{\Delta\eta(\tau_{s})}{{c}^{4}_{s}}\right)^{2}\bigg]\times\ln\left(\frac{k_e}{k_s}\right),\\
    \label{t5}\overline{\bf T}_{\bf SRII}&=&\bigg[\Delta^{2}_{\zeta,{\bf Tree}}(k)\bigg]^2_{\bf SRI}\times\Bigg(1-\frac{2}{15\pi^2}\frac{1}{c^2_{s}k^2_*}\left(1-\frac{1}{c^2_{s}}\right)\epsilon_*\Bigg)\times\ln\left(\frac{k_{\rm end}}{k_e}\right).
\eea
After the renormalization, the late-time secular growth of the remaining IR divergences, coming in the form of Logarithms of different wavenumbers from each phase, can be softened further via the DRG resummation procedure (see \ref{s32c}). The final result uses the terms in Eqn. (\ref{t1}-\ref{t5}) after evaluating them at the pivot scale $(*)$ and they package the effect of IR divergences in the form of quantum corrections which reads:
\bea &&{\cal Q}_{c}=-\left\{\frac{\bigg[  \Delta_{\zeta,\textbf{Tree}}^{2}(k)\bigg]_{\textbf{SRI}}}{\bigg[  \Delta_{\zeta,\textbf{Tree}}^{2}(k_{*})\bigg]_{\textbf{SRI}}}\right\}\times\bigg[\bigg(\overline{{\bf T}}^2_{\bf C,*}+\overline{{\bf T}}^2_{\bf B,*}+\overline{{\bf T}}^2_{\bf SRI,*}+\overline{{\bf T}}^2_{\bf USR,*}+\overline{{\bf T}}^2_{\bf SRII,*}\bigg)+\cdots\bigg],\eea
and contributes to the regularized-renormalized-resummed power spectrum that is our desired finite result of importance here: 
\bea
\Delta^{2}_{\zeta, {\bf RRR}}(k)
=\bigg[\Delta^{2}_{\zeta,{\bf Tree}}(k_*)\bigg]_{\bf SRI} \left(\frac{k_e}{k_s}\right)^6\times\left(1+\left(\frac{k}{k_s}\right)^2\right)\exp\bigg(6\ln\left(\frac{k_s}{k_e}\right)+{\cal Q}_{c}\bigg).\eea

The two quantities $\delta_{\bf C}$ and $\delta_{\bf B}$ are also defined as follows:
\bea
\delta_{\bf C}&\equiv&\left(3-2\nu+\frac{2\epsilon_c}{\epsilon_c-1}\right),\\
\delta_{\bf B}&\equiv&\left(3-2\nu+\frac{2}{\epsilon_b-1}\right).
\eea
Here $\epsilon_c$ and $\epsilon_b$ indicate the equation of state for the contraction and bounce phases in our set-up, and $\epsilon_*$ denotes the equation of state at the earlier stages of inflation corresponding to the pivot scale.

\bibliography{RefsGWB}
\bibliographystyle{utphys}
\end{document}